\documentclass[aps,twocolumn,groupedaddress,showpacs]{revtex4-1}
\usepackage{amsmath}
\usepackage{amssymb}
\usepackage{mathrsfs}
\usepackage[pdftex,colorlinks=true,urlcolor=blue,linkcolor=blue,citecolor=blue]{hyperref}
\usepackage{graphicx}
\usepackage{float}
\usepackage[abs]{overpic}
\usepackage[authormarkup=none,final]{changes}
\definechangesauthor[name={ShovanA},color=blue]{SA}
\definechangesauthor[name={ShovanD},color=red]{SD}

\begin{document}

\title{Collective Modes of a Soliton Train in a Fermi Superfluid}

\author{Shovan Dutta}
\email[E-mail: ]{sd632@cornell.edu}
\author{Erich J. Mueller}
\email[E-mail: ]{em256@cornell.edu}
\affiliation{Laboratory of Atomic and Solid State Physics, Cornell University, Ithaca, New York 14853, USA}

\date{\today}

\begin{abstract}
We characterize the collective modes of a soliton train in a quasi-one-dimensional Fermi superfluid, using a mean-field formalism. In addition to the expected Goldstone and Higgs modes, we find novel long-lived gapped modes associated with oscillations of the soliton cores. The soliton train has an instability that depends strongly on the interaction strength and the spacing of solitons. It can be stabilized by filling each soliton with an unpaired fermion, thus forming a commensurate Fulde-Ferrell-Larkin-Ovchinnikov (FFLO) phase. We find that such a state is always dynamically stable, which paves the way for realizing long-lived FFLO states in experiments via phase imprinting.
\end{abstract}


\maketitle

A unifying theme of contemporary physics is understanding emergent dynamics of many-particle systems. One motif is the appearance of persistent nonlinear structures such as solitons \cite{dauxois2006physics}. Solitons arise naturally in diverse physical systems, including water waves \cite{zabusky1965interaction, *chabchoub2013experimental, *trillo2016experimental, *sisan2014solitons}, plasmas \cite{kuznetsov1986soliton, *shukla2006formation}, optical fibers \cite{kivshar2003optical, *kivshar1998dark, *mollenauer2006solitons, *mollenauer1980experimental, *mitschke1987experimental, *hasegawa1984generation, *drummond1993quantum}, conducting polymers \cite{lu1988solitons, *heeger1988solitons, *su1979solitons, takayama1980continuum, [{}] [{ [\href{http://www.jetpletters.ac.ru/ps/1354/article_20458.shtml}{JETP Lett. {\bf 31}, 456 (1980)}]; }]brazovskii1980exact, *horovitz1981soliton, *mertsching1981incommensurate, *[{}] [{ [\href{http://www.jetp.ac.ru/cgi-bin/e/index/e/59/2/p434?a=list}{Sov. Phys. JETP {\bf 59}, 434 (1984)}].}]brazovskii1984peierls}, superconductors \cite{[{}] [{ [\href{http://www.jetp.ac.ru/cgi-bin/e/index/e/58/2/p428?a=list}{Sov. Phys. JETP {\bf 58}, 428 (1983)}]; }]buzdin1983phase, *[{}] [{ [\href{http://www.jetp.ac.ru/cgi-bin/e/index/e/66/2/p422?a=list}{Sov. Phys. JETP {\bf 66}, 422 (1987)}]. }]buzdin1987nonuniform, machida1984superconductivity, kivelson1987topology, *tanaka2001soliton, dienst2013optical}, Bose-Einstein condensates \cite{frantzeskakis2010dark, *kevrekidis2007emergent, dutton2001observation, *donadello2014observation, burger1999dark, *anderson2001watching, denschlag2000generating, *becker2008oscillations, *stellmer2008collisions, khaykovich2002formation, *strecker2002formation, *nguyen2017formation, *[{}] [{ (2017).}]everitt2017observation, lamporesi2013spontaneous, weller2008experimental, *shomroni2009evidence, bongs2003spectroscopy, *scott2003creation, *theocharis2010multiple, *becker2013inelastic}, DNA dynamics \cite{prohofsky1988solitons, *muto1989thermally, *scott1992davydov, *yakushevich2002nonlinear, *vanitha2012internal, *peyrard2004nonlinear}, quantum field theory \cite{rajaraman1982solitons, *bacsar2008self, *dunne2013time, *dunne2014full, *schon2000emergence, thies2006relativistic, *thies2004analytical, *campbell1982soliton}, and early-universe cosmology \cite{kibble1976topology, *sikivie1982axions, *frieman1988primordial, *kusenko1997solitons, *lozanov2014end, *weinberg2012classical}. They are technologically important, with applications in telecommunications \cite{haus1996solitons, *nakazawa1994soliton, *hasegawa1997recent, *zakharov2013optical, *nakazawa2000recent, *hasegawa2000soliton, *turitsyn2012dispersion, *rohrmann2012solitons}, information processing \cite{dienst2013optical, [{}] [{ and references therein; }]scheuer2005all, *al2016all, *steiglitz2009photon, *steiglitz2010soliton, *steiglitz2010making, *jakubowski2001computing, *yang2008ultraslow, *pang2016all, *bonetti2015direct, *janutka2008quantum}, and matter-wave interferometry \cite{mcdonald2014bright, *gertjerenken2013generating, *helm2015sagnac, *scott2008exploiting, *polo2013soliton, *negretti2004enhanced, *veretenov2007interferometric, *cuevas2013interactions, *martin2012quantum, *sakaguchi2016matter}. Moreover, cold-atom experiments can now engineer matter-wave solitons in atomic superfluids, and directly observe their motion \cite{burger1999dark, anderson2001watching, dutton2001observation, donadello2014observation, denschlag2000generating, becker2008oscillations, stellmer2008collisions, khaykovich2002formation, strecker2002formation, nguyen2017formation, [{}] [{ (2017).}]everitt2017observation, lamporesi2013spontaneous, weller2008experimental, shomroni2009evidence, bongs2003spectroscopy, scott2003creation, theocharis2010multiple, becker2013inelastic, yefsah2013heavy, *ku2014motion, ku2016cascade}. Understanding the behavior of these collective objects is vital to the larger problem of forming a cohesive theory of nonequilibrium dynamics \cite{takahashi2016bogoliubov}. In particular, the next generation of Fermi gas experiments will be creating clouds with many of these nonlinear defects \cite{ku2016cascade}. While past theoretical studies have shed light on the behavior of individual \cite{dziarmaga2005gap, *antezza2007dark, *liao2011traveling, *scott2011dynamics, *spuntarelli2011gray, efimkin2015moving, hammer2016bound, cetoli2013snake, *wen2013dark, *reichl2013vortex, *scherpelz2014phase, *mateo2014chladni, bulgac2014quantized, reichl2017core, klimin2014finite, *lombardi2016soliton} or pairs of solitons \cite{bulgac2012quantum, scott2012decay, wen2009dynamics}, the behavior and even stability of soliton trains are not understood. Here we study the linearized dynamics of a soliton train in a one-dimensional (1D) Fermi gas, finding a rich set of collective modes. We characterize these modes, finding distinct differences from Bose superfluids, which may generalize to nonlinear excitations of other systems.

We consider a two-component Fermi gas in an elongated trap with tight radial confinement so that the dynamics is effectively 1D \cite{guan2013fermi}. The strong radial confinement suppresses the snake instability by which solitons decay into vortices and sound waves in three dimensions \cite{ku2016cascade, cetoli2013snake, mateo2014chladni, wen2013dark, reichl2013vortex, bulgac2014quantized, reichl2017core, scherpelz2014phase, burger1999dark, anderson2001watching, dutton2001observation, donadello2014observation, muryshev1999stability, *feder2000dark, *brand2002solitonic, *komineas2003solitons, *kevrekidis2004avoiding, *kamchatnov2008stabilization, *mamaev1996propagation, *tikhonenko1996observation, *[{}] [{ [\href{http://www.jetp.ac.ru/cgi-bin/e/index/e/67/8/p1583?a=list}{Sov. Phys. JETP {\bf 67}, 1583 (1988)}]; }]kuznetsov1988instability, *josserand1995generation}. To avoid the idiosyncracies of strictly 1D systems, we envision a weakly-coupled array of such tubes which have long-range superfluid order. Past experiments have studied fermionic superfluids in such geometries \cite{liao2010spin, *revelle20161d}. Using phase-imprinting techniques \cite{ku2016cascade, burger1999dark, anderson2001watching, yefsah2013heavy, ku2014motion, sacha2014proper, *karpiuk2002solitons, *denschlag2000generating, *becker2008oscillations, *stellmer2008collisions, *gajda1999optical, *burger2002generation, *law2003dynamic, *wu2002controlled}, one can generate a train of solitons in the superfluid. The collective modes of the soliton train would show up as pronounced peaks in spectroscopic measurements of the pairing susceptibility, or in the density response of the system \cite{moritz2003exciting, lutchyn2011spectroscopy, edge2009signature, edge2010collective, heikkinen2011collective}. Here we extract the collective modes by linearizing the self-consistent Bogoliubov-de Gennes (BdG) equations governing the fermion fields.

The soliton train has two gapless Goldstone modes which originate from the spontaneous breaking of gauge- and translational symmetry: a ``phonon'' mode describing phase twists and an ``elastic'' mode describing oscillations in the spacing between the domain walls. The elastic mode is only well defined for wave vectors smaller than the inverse separation of the solitons, but we find a second gapped branch of oscillations which persists to large wave vectors [Fig.~\ref{figure2}(a)]. This branch is the remnant of the ``Higgs'' mode in a uniform superfluid \cite{pekker2014amplitude, *littlewood1982amplitude, *matsunaga2013higgs, anderson2015superconductivity, *matsunaga2014light}.

In addition, we find a twofold degenerate gapped mode which, at small wave vectors, describes oscillations in the width and grayness of each soliton [Fig.~\ref{figure2}(d)-(e)]. To our knowledge, this ``core'' mode hasn't appeared before in the literature. It lies outside the particle-hole continua, and should therefore be long-lived, hence easier to detect in experiments than those embedded in a continuum.

However, we also find that the soliton train has two kinds of instabilities toward a uniform superfluid state: in one, pairs of neighboring solitons approach and annihilate each other [Fig.~\ref{figure2}(f)], whereas in the other, the order parameter moves off into the complex plane [Fig.~\ref{figure2}(g)]. Both instabilities grow at the same rate, which depends on the degree of overlap between adjacent solitons. This overlap can be reduced by creating solitons farther apart, or by increasing the attractive interaction strength to produce sharper solitons [Fig.~\ref{figure3}(a)-(b)]. Using either approach, one can make the instability rate much smaller compared to the frequency of the ``core'' modes, thus allowing them to be resolved.

One can also stabilize the train by filling each soliton with unpaired fermions, i.e., by polarizing the Fermi gas. Such a state constitutes a realization of the long-sought-after Fulde-Ferrell-Larkin-Ovchinnikov (FFLO) phase \cite{fulde1964superconductivity, *[{}] [{ [Sov. Phys. JETP {\bf 20}, 762 (1965)]; }]larkin1965inhomogeneous, *casalbuoni2004inhomogeneous, *matsuda2007fulde, *zwicknagl2010breaking, [{}] [{ and references therein.}]dutta2015dimensional, liu2007fulde, *liu2008finite, *parish2007quasi, *mizushima2005direct, *sun2012oscillatory, *radzihovsky2010imbalanced, *baksmaty2011bogoliubov}, whose experimental evidence in solid state
\cite{[{}] [{ and references therein. Also see references in [51].}]koutroulakis2016microscopic}
 and cold gas \cite{liao2010spin, *revelle20161d} systems has so far been indirect. We find that the instability rate falls with increasing polarization, vanishing for the ``commensurate FFLO'' (C-FFLO) phase with one excess fermion per soliton [Fig.~\ref{figure3}(c)]. Thus, a C-FFLO phase is always dynamically stable, even when energetics favor a different state (Fig.~\ref{figure4}). This means one can directly engineer stable FFLO states by phase imprinting, as opposed to searching for the one that minimizes the free energy. This enlarged parameter space will facilitate more direct probes of the exotic state. In the Supplemental Material \footnote{See Supplemental Material, which includes Refs. \cite{sagi2015breakdown, belokolos2001exact, cooper1995supersymmetry, novikov1984theory, orso2007attractive, askey1978q, plischke2006equilibrium} for an outline of a protocol to generate C-FFLO states, analytic results, collective-mode spectra at different interactions and spin imbalance, simulations showing instability, and collective modes of a soliton train in the Gross-Pitaevskii and a nonlinear Klein-Grodon equation.}, we briefly outline an experimental protocol for creating such states. A detailed analysis of the protocol can be found in \cite{[{}] [{ (2017).}]dutta2017protocol}.

Our results are based on a mean-field BdG formalism. Such a mean-field treatment gives a reasonably accurate description of quasi-1D Fermi gases for moderate to weak interactions, becoming quantitative in the weak-coupling limit \cite{lutchyn2011spectroscopy, edge2009signature, xu2014dark, liu2007fulde, *liu2008finite, *parish2007quasi, *mizushima2005direct, *sun2012oscillatory, *radzihovsky2010imbalanced, *baksmaty2011bogoliubov}. Further, past theoretical work has shown that the 1D BdG equations accurately describe the equilibrium properties of an array of tubes \cite{lutchyn2011spectroscopy, [{}] [{ [\href{http://www.jetp.ac.ru/cgi-bin/e/index/e/58/2/p428?a=list}{Sov. Phys. JETP {\bf 58}, 428 (1983)}]; }]buzdin1983phase, *[{}] [{ [\href{http://www.jetp.ac.ru/cgi-bin/e/index/e/66/2/p422?a=list}{Sov. Phys. JETP {\bf 66}, 422 (1987)}]. }]buzdin1987nonuniform}.

We start with the many-body Hamiltonian
\begin{equation}
\hat{H} = \int dx \Big[\sum\nolimits_{\sigma=\uparrow,\downarrow} \hat{\Psi}_{\sigma}^{\dagger} \hat{H}^{\mbox{\tiny{(0)}}}_{\sigma} \hat{\Psi}_{\sigma} + g_{\mbox{\tiny{1D}}} \hat{\Psi}_{\uparrow}^{\dagger} \hat{\Psi}_{\downarrow}^{\dagger} \hat{\Psi}_{\downarrow} \hat{\Psi}_{\uparrow}\Big],
\label{hamiltonian}
\end{equation}
where $\hat{\Psi}_{\sigma} \equiv \hat{\Psi}_{\sigma} (x,t)$ denote the fermion field operators in the Heisenberg picture, and $g_{\mbox{\tiny{1D}}}$ is the 1D coupling constant whose relationship to the 3D scattering length is well-studied \cite{olshanii1998atomic, *bergeman2003atom, *haller2010confinement, [{}] [{ and references therein.}]dutta2015dimensional}. The single-particle Hamiltonian is $\hat{H}^{\mbox{\tiny{(0)}}}_{\uparrow,\downarrow} = -\partial_x^2/2 - \epsilon_{\mbox{\tiny{F}}} \pm h$, where $\epsilon_{\mbox{\tiny{F}}}$ is the Fermi energy, and $h$ is an effective magnetic field which controls the polarization. We have set $\hbar = m = 1$, where $m$ is the mass of each fermion. Attractive interactions ($g_{\mbox{\tiny{1D}}}<0$) lead to Cooper pairing, which we encode in the superfluid order parameter $\Delta(x,t) = g_{\mbox{\tiny{1D}}} \langle \hat{\Psi}_{\downarrow} (x,t) \hat{\Psi}_{\uparrow} (x,t) \rangle$. Ignoring quadratic fluctuations about $\Delta$ yields mean-field equations of motion for $\hat{\Psi} \equiv (\hat{\Psi}_{\uparrow} \;\; \hat{\Psi}_{\downarrow}^{\dagger})^{T}$. The many-body state is formed by occupying fermionic quasiparticle modes $\hat{\gamma}_j^s$, defined by $\smash{\hat{\Psi} = \sum_{s,j} e^{i s k_{\text{F}} x} (U^{s}_j(x,t)\;\;V^{s}_j(x,t))^{T} \hat{\gamma}^{s}_j}$, where $k_{\text{F}}$ is the Fermi momentum, $(U,V)$ are coherence factors, and $s=\pm$ breaks modes into right moving and left moving. For weak interactions, only the modes near the Fermi points contribute significantly to pairing. Thus we write $(-\partial_x^2/2 - \epsilon_{\mbox{\tiny{F}}}) [e^{\pm i  k_{\mbox{\tiny{F}}} x} (U^{\pm}_j\hspace{-0.02cm}, \hspace{-0.02cm}V^{\pm}_j)] \approx e^{\pm i k_{\mbox{\tiny{F}}} x} [\mp i k_{\mbox{\tiny{F}}} \partial_x (U^{\pm}_j\hspace{-0.02cm},\hspace{-0.02cm} V^{\pm}_j)]$ (the Andreev approximation \cite{[{}] [{ [\href{http://www.jetp.ac.ru/cgi-bin/e/index/e/19/5/p1228?a=list}{Sov. Phys. JETP {\bf 19}, 1228 (1964)}].}]andreev1964thermal}), obtaining \cite{Note1}
\begin{flalign}
&& \hspace{-2cm}i \partial_t \left(\hspace{-0.1cm}\begin{array}{c} U^{\pm}_j \\ V^{\pm}_j \end{array}\hspace{-0.1cm}\right) = 
\begin{pmatrix}  \mp i k_{\mbox{\tiny{F}}} \partial_x + h & \Delta(x,t) \\ \Delta^*(x,t) & \pm i k_{\mbox{\tiny{F}}} \partial_x + h\end{pmatrix} 
\left(\hspace{-0.1cm}\begin{array}{c} U^{\pm}_j \\ V^{\pm}_j \end{array}\hspace{-0.1cm}\right) ,&
\label{BdG}\\
\text{where}&&\textstyle{\Delta(x,t) = g_{\mbox{\tiny{1D}}} \sum\nolimits_{s,j} \langle \hat{\gamma}^{s \dagger}_j \hat{\gamma}^{s}_j \rangle\hspace{0.05cm} U^{s}_j V^{s *}_j}.\hspace{1cm}&
\label{timedependentselfconsistency}
\end{flalign}
For a real stationary solution, $\Delta(x,t)=\Delta_0(x)$, the coherence factors are of the form $\smash{(U_j^+\hspace{-0.05cm},\hspace{-0.05cm} V_j^+) \hspace{-0.05cm}=\hspace{-0.05cm} (u_j(x),v_j(x))}$ $\smash{e^{-i(\epsilon_j + h)t}}$ and  $\smash{(U_j^-\hspace{-0.05cm},\hspace{-0.05cm} V_j^-) \hspace{-0.05cm}=\hspace{-0.05cm} (u_j^*(x),v^*_j(x))\hspace{0.02cm} e^{-i(\epsilon_j + h)t}}$, where $\epsilon_j$ represents the quasiparticle spectrum.
\begin{figure}[t]
\includegraphics[scale=1]{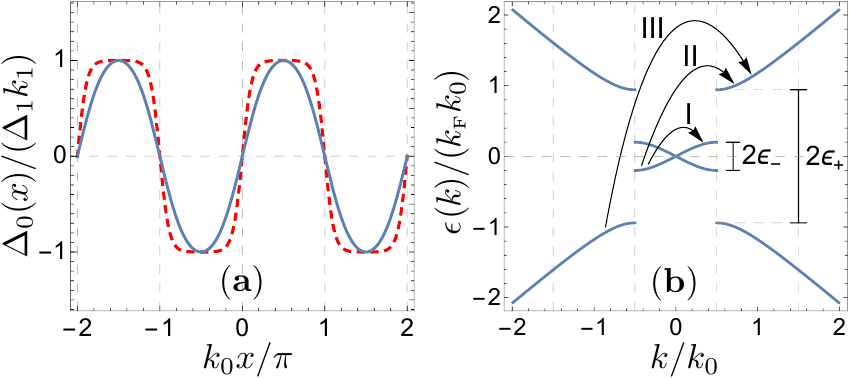}
\caption{\label{figure1}(Color online.) ({\bf a}) Stationary soliton train profile of the order parameter with wave vector $k_0$ for different values of the sharpness parameter $k_1$. Solid: $k_1 = 0.65$, dashed: $k_1 = 0.999$. The sharpness is set by the soliton spacing, interaction strength, and spin imbalance. ({\bf b}) BdG single-particle spectrum of the soliton train in the extended zone, for $k_1 = 0.65$. The arrows show three types of particle-hole excitations, which give rise to disconnected continua in the collective excitation spectrum [gray regions in Fig.~\ref{figure2}(a)].}
\end{figure}

Prior studies have found \cite{[{}] [{ [\href{http://www.jetpletters.ac.ru/ps/1354/article_20458.shtml}{JETP Lett. {\bf 31}, 456 (1980)}]; }]brazovskii1980exact, *horovitz1981soliton, *mertsching1981incommensurate, *[{}] [{ [\href{http://www.jetp.ac.ru/cgi-bin/e/index/e/59/2/p434?a=list}{Sov. Phys. JETP {\bf 59}, 434 (1984)}].}]brazovskii1984peierls, [{}] [{ [\href{http://www.jetp.ac.ru/cgi-bin/e/index/e/58/2/p428?a=list}{Sov. Phys. JETP {\bf 58}, 428 (1983)}]; }]buzdin1983phase, *[{}] [{ [\href{http://www.jetp.ac.ru/cgi-bin/e/index/e/66/2/p422?a=list}{Sov. Phys. JETP {\bf 66}, 422 (1987)}]. }]buzdin1987nonuniform, machida1984superconductivity} stationary soliton train solutions of the form $\Delta_0(x) = \Delta_1 k_1 \hspace{0.05cm}\mbox{sn} (\Delta_1 x/k_{\mbox{\tiny{F}}}, k_1)$ with $\Delta_1 \equiv 2 k_{\mbox{\tiny{F}}} k_0 K(k_1)/\pi$, where $2\pi/k_0$ is the period of the train, $\mbox{sn}$ is a Jacobi elliptic function \cite{whittaker1996course}, $K$ is the complete elliptic integral of the first kind, and $k_1 \in (0,1)$ is a parameter controlling the sharpness of solitons, which is set by imposing self-consistency [Eq.~\eqref{timedependentselfconsistency}]. The quasiparticle spectrum has a continuum of free states for $|\epsilon|>\epsilon_+$, and a band of midgap states for $|\epsilon|<\epsilon_-$, where \linebreak $\epsilon_{\pm} = \Delta_1 (1\pm k_1) /2$ (Fig.~\ref{figure1}). The midgap band describes Andreev bound states localized at the soliton cores.\deleted[id=SD]{ Note that the band gap equals $\Delta_1 k_1$, the maximum value of $\Delta_0(x)$, which increases with stronger interactions \cite{Note1}.}

To find the collective modes, we linearize small fluctuations about the stationary solution. Thus we write $\Delta = \Delta_0(x) + \delta \Delta(x,t)$, $\smash{U^{+}_j} \hspace{-0.05cm}= (u_j(x) + \delta \smash{u^{+}_j}(x,t))\hspace{0.04cm} e^{-i(\epsilon_j+h)t}$, $\smash{U^{-}_j}\hspace{-0.05cm}= (u_j^*(x) + \delta u^{-}_j(x,t))\hspace{0.04cm} e^{-i(\epsilon_j+h)t}$, and similar expressions for $\smash{V_j^{\pm}}$ in Eqs.~\eqref{BdG} and \eqref{timedependentselfconsistency}, yielding a set of coupled equations relating $\delta u_j^{\pm}$, $\delta v_j^{\pm}$, and $\delta \Delta$. Next, we decompose the fluctuations into frequency components, and use the completeness of the stationary wave functions to eliminate $\smash{\delta u_j^{\pm}}$ and $\smash{\delta v_j^{\pm}}$, thus arriving at an integral equation for $\delta\Delta$. In particular, we write $\delta\Delta = \mbox{Re} (\delta_a(x) e^{i\Omega t}) + i\hspace{0.05cm} \mbox{Im} (\delta_p(x) e^{i\Omega t})$ where $\delta_a$ and $\delta_p$ describe the amplitude and phase fluctuations respectively, and find (full derivation in Supplemental Material \cite{Note1}),
\begin{equation}
\delta_{p,a}(x) = -g_{\mbox{\tiny{1D}}} \hspace{-0.05cm}\int\hspace{-0.05cm} dx^{\prime} \mathcal{M}^{\pm}(x,x^{\prime};\Omega) \hspace{0.05cm}\delta_{p,a}(x^{\prime})\hspace{0.05cm},
\label{integraleqn}
\end{equation}
where, at zero temperature,
\begin{equation}
\mathcal{M}^{\pm} \hspace{-0.05cm}=\hspace{-0.08cm} \sideset{}{'}\sum_{j,j^{\prime}} \hspace{-0.05cm}\frac{2(\epsilon_j\hspace{-0.05cm}+\hspace{-0.05cm}\epsilon_{j^{\prime}})}{(\epsilon_j\hspace{-0.05cm}+\hspace{-0.05cm}\epsilon_{j^{\prime}})^2 - \Omega^2} (u_j^* u_{j^{\prime}} \hspace{-0.05cm}\pm v_j^* v_{j^{\prime}})(u_j^{\prime} u_{j^{\prime}}^{\prime *}\hspace{-0.05cm}\pm v_j^{\prime} v_{j^{\prime}}^{\prime *}).
\label{Mplusminus}
\end{equation}
Here, $\Omega\in \mathbb{C}$, the prime on the summation stands for $\epsilon_j >h$, and we have used the notation $(u,v)\equiv (u(x),v(x))$ and $(u^{\prime},v^{\prime}) \equiv (u(x^{\prime}),v(x^{\prime}))$. The collective modes represent non-trivial solutions to Eq. \eqref{integraleqn}.

Periodicity of the soliton train leads to a Brillouin zone structure for the collective modes, i.e., one can write $\delta_{p,a}(x) = e^{i q x}\sum_n C^{\pm}_n e^{i n k_0 x}$, where $-k_0/2<q \leq k_0/2$ and $n \in \mathbb{Z}$. However, the stationary solution has an additional symmetry, $\Delta_0(x+\pi/k_0) = -\Delta_0(x)$, which causes the even and odd Fourier modes to decouple in Eq. \eqref{integraleqn}, effectively doubling the Brillouin zone \cite{edge2009signature}. Thus we consider only odd Fourier components, with $-k_0<q \leq k_0$. Substituting the Fourier expansion into Eq. \eqref{integraleqn} yields a matrix equation, $C^{\pm}_n = -g_{\mbox{\tiny{1D}}}\sum_m M^{\pm}_{nm}(q,\Omega)\hspace{0.05cm} C^{\pm}_m$, where
\begin{equation}
\hspace{-0.2cm}M^{\pm}_{nm} \hspace{-0.05cm}= \hspace{-0.03cm}\frac{k_0}{2\pi} \hspace{-0.03cm}\int_{-\pi\hspace{-0.03cm}/\hspace{-0.03cm}k_0}^{\pi\hspace{-0.03cm}/\hspace{-0.03cm}k_0} \hspace{-0.1cm}dx\hspace{-0.075cm} \int\hspace{-0.1cm} dx^{\prime} e^{-i(q+n k_0)x + i(q+m k_0)x^{\prime}} \hspace{-0.03cm}\mathcal{M}^{\pm}\hspace{-0.03cm}.
\label{Matrixplusminus}
\end{equation}
We find the collective-mode spectrum by solving $\det (I+g_{\mbox{\tiny{1D}}} M^{\pm}(q,\Omega))=0$. \hspace{-0.05cm}Note that $M^{\pm}(q,\Omega)$ has branch cuts on the real-$\Omega$ axis, which originate from particle-hole excitations. Thus while considering real frequencies ($\omega$), we set $\Omega \to \omega + i \hspace{0.05cm}0^+$. We find $\Omega$ is either real or imaginary for all collective modes.

The matrices $M^{\pm}$ are related to the pairing susceptibilities $\chi^{\pm}(q,\omega)$, which describe the linear response to a pairing field, as $\chi^{\pm} = -g_{\mbox{\tiny{1D}}}\hspace{0.01cm} \mbox{Tr}\hspace{0.01cm} \big[(I+g_{\mbox{\tiny{1D}}} M^{\pm})^{-1} M^{\pm}\hspace{-0.02cm}\big]$ (see Supplemental Material \cite{Note1} for a derivation). The spectral densities, $\mbox{Im} \hspace{0.05cm}\chi^{\pm}$, contain isolated poles corresponding to collective modes, and broad particle-hole continua.

The collective excitation spectrum is fully characterized by two dimensionless quantities: $n_s$, the number of unpaired fermions per soliton, and $k_1$, which describes the sharpness of the solitons. They are set by the parameters $k_0/k_{\mbox{\tiny{F}}}$, $k_{\mbox{\tiny{F}}} a_{\mbox{\tiny{1D}}}$, and $h/\epsilon_{\mbox{\tiny{F}}}$, $a_{\mbox{\tiny{1D}}}$ being the 1D scattering length ($a_{\mbox{\tiny{1D}}} = -2/g_{\mbox{\tiny{1D}}}$ \cite{olshanii1998atomic, *bergeman2003atom, *haller2010confinement, [{}] [{ and references therein.}]dutta2015dimensional}).\deleted[id=SD]{ In particular, stronger interactions (smaller $k_{\mbox{\tiny{F}}} a_{\mbox{\tiny{1D}}}$) have the same effect as a larger spacing between solitons (smaller $k_0/k_{\mbox{\tiny{F}}}$) \cite{Note1}.}\added[id=SA]{ To a good approximation, the dependence on $k_0/k_{\mbox{\tiny{F}}}$ and $k_{\mbox{\tiny{F}}} a_{\mbox{\tiny{1D}}}$ appears through the combination $w \equiv (k_0/k_{\mbox{\tiny{F}}}) \exp(\pi k_{\mbox{\tiny{F}}} a_{\mbox{\tiny{1D}}}/2)$, which measures the width of the Andreev bound states in units of the soliton spacing. For $h=0$ and $w \lesssim 2.5$, $k_1 \approx 1 - 8 \hspace{0.03cm} e^{-4\pi/w}$ \cite{Note1}.}
\begin{figure}[t]
\includegraphics[scale=1]{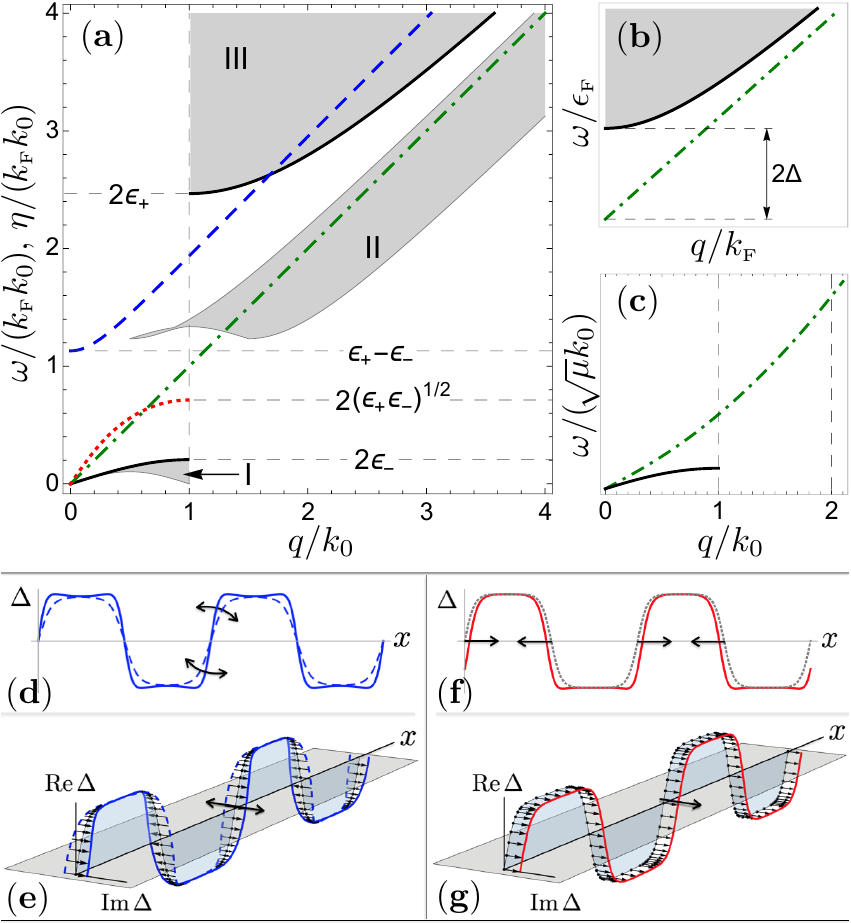}
\caption{\label{figure2}(Color online.) Collective-mode spectrum of ({\bf a}) a soliton train in a Fermi superfluid with no spin imbalance, for $k_0/k_{\mbox{\tiny{F}}} = 0.05$ and $k_{\mbox{\tiny{F}}} a_{\mbox{\tiny{1D}}} = 2.6$, ({\bf b}) a uniform Fermi superfluid, ({\bf c}) soliton train in a Bose-Einstein condensate, modeled by the Gross-Pitaevskii equation. There are two gapless Goldstone modes in ({\bf a}): a ``phonon'' mode (dot-dashed, green) and an ``elastic'' mode (solid, black) which describe phase twists and elastic deformations of the order parameter respectively. The ``phonon'' mode is the analog of the Anderson-Bogoliubov mode of a uniform superfluid in ({\bf b}). A second gapped branch of amplitude oscillations (solid, black) forms the remnant of the ``Higgs'' mode in ({\bf b}). Both ``elastic'' and ``Higgs'' modes in ({\bf a}) reside on an edge of the two-particle continua, shaded in gray, which originate from three types of particle-hole excitations, as shown in Fig.~\ref{figure1}(b). Additionally, we find novel, twofold degenerate gapped modes in ({\bf a}) (dashed, blue) which, for small $q$, describe width and grayness oscillations of each soliton, as illustrated in ({\bf d}) and ({\bf e}). The soliton train also has instabilities toward a uniform superfluid state, which show up as twofold degenerate unstable modes. The dotted (red) curve in ({\bf a}) gives the growth rate $\eta$ of these modes. The most unstable mode consists of pairs of solitons annihilating one another ({\bf f}) or the order parameter moving off in the complex plane ({\bf g}). In contrast, the spectrum of a bosonic soliton train in ({\bf c}) only contains two gapless Goldstone modes.}
\end{figure}

Figure~\ref{figure2}(a) shows the collective-mode spectrum for $n_s=0$, $k_0/k_{\mbox{\tiny{F}}}=0.05$, and $k_{\mbox{\tiny{F}}} a_{\mbox{\tiny{1D}}}=2.6$ in the extended-zone scheme. Its structure is representative of the $n_s \hspace{-0.05cm}=\hspace{-0.02cm} 0$ case. The two-particle continuum has three separate regions, corresponding to particle-hole excitation between different bands of the quasiparticle spectrum [Fig.~\ref{figure1}(b)].

We find two gapless Goldstone modes. The Goldstone phase mode is described by $\delta_p(x) \propto\Delta_0(x) e^{i q x}$ and $\omega = k_{\mbox{\tiny{F}}} q$. It is the analog of the Anderson-Bogoliubov phonon mode in a uniform Fermi superfluid \cite{bogoljubov1958new, *anderson1958random, *nambu1960quasi}. The Goldstone amplitude mode represents elastic deformations of $\Delta$ and has a second gapped branch extending to large wave-vectors, which forms the analog of the ``Higgs'' mode in a uniform Fermi superfluid \cite{pekker2014amplitude, *littlewood1982amplitude, *matsunaga2013higgs, anderson2015superconductivity, *matsunaga2014light}. Both branches are expressed by $\delta_a(x) \propto u_{\frac{q}{2}}(x) \hspace{0.05cm} v_{\frac{q}{2}}(x)$ and $\omega = 2\hspace{0.05cm} \epsilon \hspace{0.02cm}(\frac{q}{2})$, where $\epsilon(k)$ is the single-particle dispersion. Like the ``Higgs'' mode, both branches sit on the threshold for particle-hole excitations and will therefore be damped \cite{pekker2014amplitude, *littlewood1982amplitude, *matsunaga2013higgs, [{}] [{ [\href{http://www.jetp.ac.ru/cgi-bin/e/index/e/38/5/p1018?a=list}{Sov. Phys. JETP {\bf 38}, 1018 (1974)}]; }]volkov1974collisionless, *barankov2006synchronization, *barankov2004collective, *cea2015nonrelativistic, *podolsky2011visibility, *cea2014nature, *han2016observability, *sherman2015higgs}. In contrast, the excitation spectrum of a soliton train in a Bose superfluid, modeled by the Gross-Pitaevskii equation, is comprised only of two undamped gapless modes [Fig.~\ref{figure2}(c)]. They have a similar dispersion to the fermionic case for small $q$, but each mode contains both amplitude and phase variations \cite{Note1}.

In Fig.~\ref{figure2}(a), we also show a gapped mode that is not present in either a Bose superfluid or a uniform Fermi superfluid (dashed, blue curve). This mode is twofold degenerate, with a phase- and an amplitude sector. For small $q$, they describe oscillations in the grayness and width of each soliton [Fig.~\ref{figure2}(d)-(e)]. In particular, at $q=0$, these sectors are expressed by $\delta_p(x) \propto \mbox{cn} (\Delta_1 x/k_{\mbox{\tiny{F}}}, k_1)$, $\delta_a(x) \propto \mbox{sn} (\Delta_1 x/k_{\mbox{\tiny{F}}}, k_1) \hspace{0.05cm}\mbox{dn} (\Delta_1 x/k_{\mbox{\tiny{F}}}, k_1)$, and have an energy $\omega = \epsilon_+ - \epsilon_-$. Surprisingly, we find $\delta_a(x) \hspace{-0.05cm}\propto \delta_p^{\prime}(x) \hspace{0.1cm}\forall \hspace{0.05cm} q$. Being outside the continua, these ``core'' modes should be long-lived and hence suitable for experimental detection. One can excite the amplitude ``core'' mode by a fast ramp to a different interaction strength\deleted[id=SD]{, so that the new stationary solution has a different soliton width} [\added[id=SA]{see }Fig.~\ref{figure1}(a)].
\begin{figure}[b]
\includegraphics[scale=1]{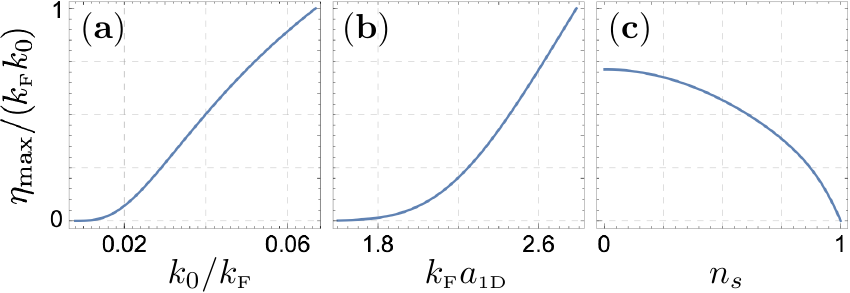}
\caption{\label{figure3}Maximum instability rate vs ({\bf a}) inverse soliton separation, with $n_s = 0$, $k_{\mbox{\tiny{F}}} a_{\mbox{\tiny{1D}}} = 2.6$, ({\bf b}) interaction strength, with $n_s = 0$, $k_0/k_{\mbox{\tiny{F}}} = 0.05$, and ({\bf c}) spin imbalance, with $k_0/k_{\mbox{\tiny{F}}} = 0.05$, $k_{\mbox{\tiny{F}}} a_{\mbox{\tiny{1D}}} = 2.6$. By making the rate sufficiently small, one can investigate the stable collective modes.}
\end{figure}

The balanced soliton train ($n_s\hspace{-0.02cm}=0$) has dynamical instabilities toward a uniform superfluid state, which show up as two degenerate solutions to Eq. \eqref{integraleqn} with an imaginary frequency. The unstable amplitude mode is associated with pairs of solitons approaching one another and annihilating, whereas the unstable phase mode involves the order parameter moving off into the complex plane [Fig.~\ref{figure2}(f)-(g)]. The maximum instability occurs at $q=k_0$, where $\delta_a(x) \propto \mbox{dn}^2 (\Delta_1 x/k_{\mbox{\tiny{F}}}, k_1)$, $\delta_p(x) = \text{constant}$, and the fluctuations grow at a rate $\eta_{\mathrm{max}} = 2 (\epsilon_+ \epsilon_-)^{1/2}$. For a given soliton spacing, $\eta_{\mathrm{max}}$ is highest at weak interactions, approaching $k_{\mbox{\tiny{F}}} k_0$. One can lower $\eta_{\mathrm{max}}$ by creating solitons farther apart or increasing the interaction strength [Fig.~\ref{figure3}(a)-(b)]. We have verified the instability by direct simulations of the BdG equations without the Andreev approximation. We find a lower bound on the soliton lifetime $\tau_{\mathrm{min}} \sim 8/k_{\mbox{\tiny{F}}} k_0$\added[id=SA]{, which}\deleted[id=SD]{. This bound} is saturated at weak interactions. For $^{6}$Li atoms with $\epsilon_{\mbox{\tiny{F}}} = 1.2\;\mu$K (as in \cite{liao2010spin, *revelle20161d}) and $k_0/k_{\mbox{\tiny{F}}}=0.05$, $\tau_{\mathrm{min}}\approx 0.5$ ms. The instability becomes unnoticeable for $k_{\mbox{\tiny{F}}} a_{\mbox{\tiny{1D}}} \lesssim 2$, where adjacent solitons collide elastically, in agreement with previous findings on two-soliton collisions \cite{bulgac2012quantum, scott2012decay}. We present the simulations in the Supplemental Material \cite{Note1}, along with collective-mode spectra at different interactions.

An alternate way to stabilize the soliton train is by filling solitons with unpaired fermions \cite{reichl2017core}. As we increase $n_s$ from 0, the instability rate falls, becoming zero at $n_s = 1$ for the C-FFLO phase [Fig.~\ref{figure3}(c)]. The stability of the C-FFLO phase originates from the absence of zero-energy particle-hole excitations, as the chemical potentials lie within gaps in the single-particle spectrum. For \smash{$n_s > 1$}, one again has instabilities (see Supplemental Material \cite{Note1} for more details).
\begin{figure}[t]
\includegraphics[width=\columnwidth]{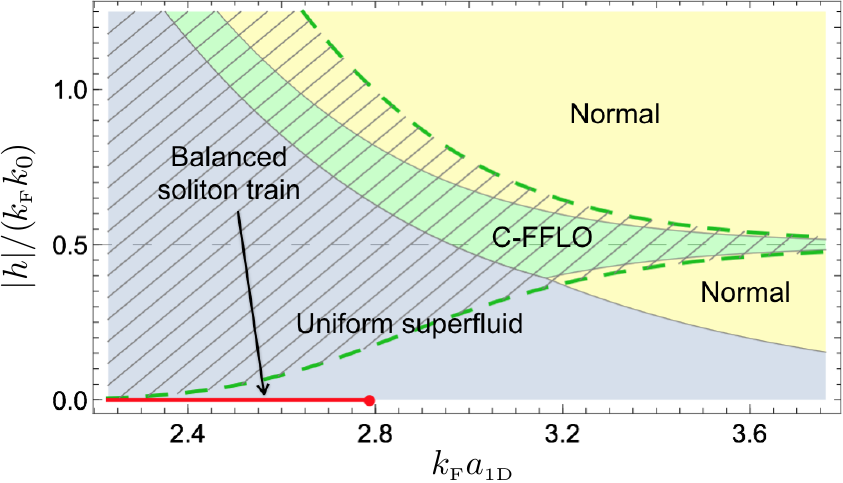}
\caption{\label{figure4}(Color online.) Phase diagram obtained by comparing mean-field energies of homogeneous phases, and soliton train states with $k_0/k_{\mbox{\tiny{F}}} = 0.05$. Solid regions show the lowest-energy states. The C-FFLO phase exists and is dynamically stable throughout the hatched region. The balanced soliton train exists above a minimum interaction strength\deleted[id=SD]{ (shown as a solid red line on the horizontal axis)}\added[id=SA]{ ($w \lesssim 4$)}. To see where the other soliton train solutions exist, see\deleted[id=SD]{ Supplemental Material} \cite{Note1}.}
\end{figure}

Past studies on FFLO have focused on the phase that minimizes the free energy, which occurs at specific values of $k_0$ within a limited region of the phase diagram \cite{[{}] [{ and references therein.}]dutta2015dimensional, liu2007fulde, *liu2008finite, *parish2007quasi, *mizushima2005direct, *sun2012oscillatory, *radzihovsky2010imbalanced, *baksmaty2011bogoliubov}. Low-energy collective excitations of energetically stable FFLO states have been explored using different theoretical techniques \cite{edge2009signature, edge2010collective, heikkinen2011collective, feiguin2009spectral, *samokhin2010goldstone, *samokhin2011spectrum, *radzihovsky2009quantum, *radzihovsky2011fluctuations}, and methods for detecting such states have been proposed \cite{lutchyn2011spectroscopy}. However, we find that a C-FFLO phase is always dynamically stable, even when there are lower-energy states available. To see this, we compare the energies of competing states \cite{Note1} to arrive at a phase diagram, shown in Fig.~\ref{figure4}. Despite its large region of stability, the C-FFLO phase has the lowest energy in a relatively small region. Moreover, the optimal value of $k_0/k_{\mbox{\tiny{F}}}$ varies continuously with $h$, a feature not apparent in Fig.~\ref{figure4} which is concerned with a fixed value of $k_0/k_{\mbox{\tiny{F}}}$. The metastability in this system implies that energetic considerations are of less importance than how the cloud is prepared. In particular, one can engineer\deleted[id=SD]{ stable C-FFLO states (as well as long-lived FFLO with $n_s<1$ at strong enough interactions)}\added[id=SA]{ long-lived FFLO states} by phase imprinting. In Ref. \cite{[{}] [{ (2017).}]dutta2017protocol} we propose a simple protocol for this engineering, briefly outlined in the Supplemental Material \cite{Note1}.

Our results carry over to other physical systems where solitons arise in a BdG formalism. This includes quasi-1D superconductors in a magnetic field \cite{[{}] [{ [\href{http://www.jetp.ac.ru/cgi-bin/e/index/e/58/2/p428?a=list}{Sov. Phys. JETP {\bf 58}, 428 (1983)}]; }]buzdin1983phase, *[{}] [{ [\href{http://www.jetp.ac.ru/cgi-bin/e/index/e/66/2/p422?a=list}{Sov. Phys. JETP {\bf 66}, 422 (1987)}]. }]buzdin1987nonuniform, machida1984superconductivity}, electron-phonon model of conducting polymers \cite{takayama1980continuum, [{}] [{ [\href{http://www.jetpletters.ac.ru/ps/1354/article_20458.shtml}{JETP Lett. {\bf 31}, 456 (1980)}]; }]brazovskii1980exact, *horovitz1981soliton, *mertsching1981incommensurate, *[{}] [{ [\href{http://www.jetp.ac.ru/cgi-bin/e/index/e/59/2/p434?a=list}{Sov. Phys. JETP {\bf 59}, 434 (1984)}].}]brazovskii1984peierls}, and Gross-Neveu models in quantum field theory \cite{thies2006relativistic, *thies2004analytical, *campbell1982soliton}. The gapped modes describing width- and grayness oscillations of solitons could be more generic features associated with mesoscale structures; e.g., we find such modes in soliton trains described by a nonlinear Klein-Gordon equation, which also have unstable modes \cite{Note1}. 
Although defined by pairing oscillations, these modes should be visible in many different spectroscopic channels. For example, the techniques demonstrated in \cite{yefsah2013heavy, ku2014motion} for observing the oscillation of a single soliton are well-suited for probing the ``elastic'' modes. The instabilities can be studied using techniques from \cite{ku2016cascade, aycock2017brownian}. The ``core'' modes may be accessible through radio-frequency or modulation spectroscopy \cite{lutchyn2011spectroscopy, edge2010collective, torma2016physics, *endres2012higgs}. The dynamical stability of the C-FFLO phase should pave the way to its realization via phase imprinting \cite{Note1}. Other techniques might also be feasible; e.g., in Bose-Einstein condensates, soliton trains spontaneously form in rapid quenches of interaction strength \cite{khaykovich2002formation, strecker2002formation, nguyen2017formation, [{}] [{ (2017).}]everitt2017observation} or temperature \cite{lamporesi2013spontaneous}, or when two condensates collide \cite{weller2008experimental, shomroni2009evidence}. These processes could have analogs in Fermi superfluids. There exist theoretical methods complementary to BdG such as effective field theories \cite{klimin2014finite, *lombardi2016soliton, simonucci2015vortex, *klimin2015finite} and density-functional theories \cite{bulgac2012quantum, bulgac2014quantized} which could be extended to study soliton trains at strong interactions and finite temperatures. Our analysis provides a useful benchmark for such future investigations.\\

We thank Matthew Reichl for useful discussions. This material is based upon work supported by the National Science Foundation under Grant No. PHY-1508300 and the ARO-MURI Non-equilibrium Many-body Dynamics Grant No. W9111NF-14-1-0003.

\begingroup
\renewcommand{\addcontentsline}[3]{}
\renewcommand{\section}[2]{}


%

\endgroup


\onecolumngrid
\clearpage

\begin{center}
\textbf{\large Supplemental Material for\\ ``Collective Modes of a Soliton Train in a Fermi Superfluid"}
\end{center}

\setcounter{equation}{0}
\setcounter{figure}{0}
\setcounter{table}{0}
\setcounter{page}{1}
\makeatletter

\renewcommand{\thefigure}{S\arabic{figure}}
\renewcommand{\theequation}{S\arabic{equation}}
\renewcommand{\bibnumfmt}[1]{[S#1]}
\renewcommand{\citenumfont}[1]{S#1}






\tableofcontents

\section{\label{conventions}Conventions for Bogoliubov operators}
There are at least two competing conventions in the literature for defining the Bogoliubov operators for spin-1/2 fermions: in one, the quasiparticle spectrum is symmetric for positive and negative energies, and there is only one type of Bogoliubov mode $\hat{\gamma}_j$. In the other, the spectrum has only positive energies, but there are two types of Bogoliubov modes,  $\hat{\gamma}_j$ and \smash{$\hat{\zeta}_j$}. We follow the former convention in the paper, though the latter is perhaps more commonly used. To avoid any confusion, here we summarize both conventions, and discuss how they relate to one another.

We consider a system of spin-1/2 fermions with attractive interactions, described by the Hamiltonian
\begin{equation}
\hat{H} = \int\hspace{-0.05cm} dx \left[\hspace{0.1cm}\sum\nolimits_{\sigma=\uparrow,\downarrow}\hspace{-0.1cm} \hat{\Psi}_{\sigma}^{\dagger}(x) (\hat{H}_0 - \mu_{\sigma}) \hat{\Psi}_{\sigma}(x) + g_{\mbox{\tiny{1D}}} \hat{\Psi}_{\uparrow}^{\dagger}(x)\hat{\Psi}_{\downarrow}^{\dagger}(x) \hat{\Psi}_{\downarrow}(x) \hat{\Psi}_{\uparrow} (x)\right],
\label{Hspinhalf}
\end{equation}
where $\hat{H}_0$ is the single-particle Hamiltonian, and $\mu_{\uparrow,\downarrow} \equiv \mu \mp h$ are the chemical potentials of the two spins. In terms of the order parameter $\Delta_0(x) \equiv g_{\mbox{\tiny{1D}}}\langle \hat{\Psi}_{\downarrow} (x) \hat{\Psi}_{\uparrow} (x)\rangle$, the mean-field Hamiltonian is given by
\begin{equation}
\hat{H}_{\mbox{\tiny{MF}}} \hspace{-0.03cm}=\hspace{-0.1cm} \int \hspace{-0.1cm} d x\hspace{0.05cm} \bigg[
\big(\hat{\Psi}^{\dagger}_{\uparrow}(x) \;\;\; \hat{\Psi}_{\downarrow}(x)\big)
\begin{pmatrix} \hat{H}_0 - \mu + h & \Delta_0(x) \\ \Delta_0^* (x) & -\hat{H}_0 + \mu + h \end{pmatrix} 
\bigg(\hspace{-0.1cm}\begin{array}{c} \hat{\Psi}_{\uparrow}(x) \\ \hat{\Psi}^{\dagger}_{\downarrow}(x) \end{array}\hspace{-0.1cm}\bigg) - g_{\mbox{\tiny{1D}}}^{-1} \big|\Delta_0(x)\big|^2\bigg] + \text{Tr}\hspace{0.03cm}\big(\hat{H}_0 - \mu - h\big) \hspace{0.05cm}.
\label{HMFspinhalf}
\end{equation}
This Hamiltonian can be diagonalized by solving the BdG equations
\begin{equation}
\begin{pmatrix} \hat{H}_0 - \mu & \Delta_0(x) \\ \Delta_0^* (x) & -\hat{H}_0 + \mu \end{pmatrix} 
\bigg(\hspace{-0.1cm}\begin{array}{c} u_j (x) \\ v_j (x) \end{array}\hspace{-0.1cm}\bigg) = \epsilon_j \bigg(\hspace{-0.1cm}\begin{array}{c} u_j (x) \\ v_j (x) \end{array}\hspace{-0.1cm}\bigg) \hspace{0.02cm},
\label{BdGspinhalf}
\end{equation}
which has both positive and negative eigenvalues $\epsilon_j$. In fact, the spectrum is symmetric: if $(u_j(x)\;\;v_j(x))^{T}$ is an eigenvector with eigenvalue $\epsilon_j$, then $(-v^*_j(x)\;\;u^*_j(x))^T$ is an eigenvector with eigenvalue $-\epsilon_j$. The eigenvectors form an orthonormal set, i.e., \smash{$\int dx \hspace{0.03cm}(u_j^* (x) u_{j^{\prime}}(x) + v_j^* (x) v_{j^{\prime}}(x)) = \delta_{j j^{\prime}}$}.

In our convention we define the Bogoliubov operators $\hat{\gamma}_j$ as
\begin{equation}
\bigg(\hspace{-0.1cm}\begin{array}{c} \hat{\Psi}_{\uparrow}(x) \\ \hat{\Psi}^{\dagger}_{\downarrow}(x) \end{array}\hspace{-0.1cm}\bigg) = \sum_j \bigg(\hspace{-0.1cm}\begin{array}{c} u_j (x) \\ v_j (x) \end{array}\hspace{-0.1cm}\bigg) \hat{\gamma}_j \hspace{0.05cm},
\label{ourdefinition}
\end{equation}
where the sum is over both positive and negative energies. The orthonormality of the eigenvectors ensures that the modes $\hat{\gamma}_j$ are fermionic, i.e., $\{\hat{\gamma}_j,\hat{\gamma}^{\dagger}_{j^{\prime}}\} = \delta_{j j^{\prime}}$. Substituting Eq.~\eqref{ourdefinition} into Eq.~\eqref{HMFspinhalf}, we find
\begin{equation}
\hat{H}_{\mbox{\tiny{MF}}} \hspace{-0.03cm}= \sum_j (\epsilon_j + h) \hat{\gamma}^{\dagger}_j \hat{\gamma}_j + \text{Tr}\hspace{0.03cm}\big(\hat{H}_0 - \mu - h\big) - g_{\mbox{\tiny{1D}}}^{-1} \int dx \big|\Delta_0(x)\big|^2 \hspace{0.05cm}.
\label{HMFourconvention}
\end{equation}
The occupation of the modes is given by $\langle \hat{\gamma}^{\dagger}_j \hat{\gamma}_j \rangle = n_{\mbox{\tiny{F}}}(\epsilon_j + h)$ where $n_{\mbox{\tiny{F}}}$ is the Fermi function. Thus at zero temperature, all quasiparticle modes with energy $\epsilon_j < -h$ are occupied, and all other modes are empty. In particular, for $h=0$ (no imbalance), all negative energy modes are occupied and positive energy modes are empty. When $\mu_{\downarrow} > \mu_{\uparrow}$ ($h>0$), one has to remove quasiparticles from the modes with energy between 0 and $-h$, resulting in a net excess of $\downarrow$-spins. Similarly, if $\mu_{\uparrow} > \mu_{\downarrow}$, one populates the modes between 0 and $|h|$, resulting in a net excess of $\uparrow$-spins.


One arrives at the other convention by noting that Eq.~\eqref{ourdefinition} can be written as
\begin{equation}
\bigg(\hspace{-0.1cm}\begin{array}{c} \hat{\Psi}_{\uparrow} \\ \hat{\Psi}^{\dagger}_{\downarrow} \end{array}\hspace{-0.1cm}\bigg) = \sum_{\epsilon_j>0} \bigg(\hspace{-0.1cm}\begin{array}{c} u_j  \\ v_j  \end{array}\hspace{-0.1cm}\bigg) \hat{\gamma}_j + \sum_{\epsilon_j<0} \bigg(\hspace{-0.1cm}\begin{array}{c} u_j  \\ v_j  \end{array}\hspace{-0.1cm}\bigg) \hat{\gamma}_j = \sum_{\epsilon_j>0} \bigg(\hspace{-0.1cm}\begin{array}{c} u_j  \\ v_j  \end{array}\hspace{-0.1cm}\bigg) \hat{\gamma}_j + \sum_{\epsilon_j>0} \bigg(\hspace{-0.1cm}\begin{array}{c} -v^*_j  \\ u^*_j \end{array}\hspace{-0.1cm}\bigg) \hat{\zeta}_j^{\dagger} =  
\sum_{\epsilon_j>0} \begin{pmatrix} u_j  & -v^*_j  \\ v_j & u^*_j  \end{pmatrix}
 \bigg(\hspace{-0.1cm}\begin{array}{c} \hat{\gamma}_j \\ \hat{\zeta}_j^{\dagger} \end{array}\hspace{-0.1cm}\bigg) \hspace{0.05cm},
\label{otherdefinition}
\end{equation}
where we have used the fact that for each state $(u_j\;\;v_j)^T$ with energy $\epsilon_j$, there is a state $(-v_j^*\;\;u_j^*)^T$ with energy $-\epsilon_j$, and defined new fermionic operators \smash{$\hat{\zeta}_j \equiv \hat{\gamma}_j^{\dagger}$} for $\epsilon_j <0$. The operators $\hat{\gamma}_j$ and \smash{$\hat{\zeta}_j$} in Eq.~\eqref{otherdefinition} represent the Bogoliubov modes in this other convention. Substituting Eq.~\eqref{otherdefinition} into Eq.~\eqref{HMFspinhalf}, we obtain
\begin{equation}
\hat{H}_{\mbox{\tiny{MF}}} \hspace{-0.03cm}= \sum_{\epsilon_j>0} \big[(\epsilon_j + h) \hat{\gamma}^{\dagger}_j \hat{\gamma}_j + (\epsilon_j - h)\hspace{0.03cm} \hat{\zeta}^{\dagger}_j\hat{\zeta}_j - (\epsilon_j - h) \big]  + \text{Tr}\hspace{0.03cm}\big(\hat{H}_0 - \mu - h\big) - g_{\mbox{\tiny{1D}}}^{-1} \int dx \big|\Delta_0(x)\big|^2 \hspace{0.05cm}.
\label{HMFotherconvention}
\end{equation}
The occupations of the modes are given by $\langle \hat{\gamma}^{\dagger}_j \hat{\gamma}_j \rangle = n_{\mbox{\tiny{F}}}(\epsilon_j + h)$ and \smash{$\langle \hat{\zeta}^{\dagger}_j \hat{\zeta}_j \rangle = n_{\mbox{\tiny{F}}}(\epsilon_j - h)$}. At zero temperature, only the $\hat{\gamma}$ modes with $\epsilon_j < -h$ and the \smash{$\hat{\zeta}$} modes with $\epsilon_j < h$ are occupied. However $\epsilon_j >0$, so there are no negative energy modes, which means in the balanced case ($h=0$), all Bogoliubov modes are empty. Excess $\downarrow$-spins ($h>0$) are incorporated by filling up only the \smash{$\hat{\zeta}$} modes with $0<\epsilon_j < h$, whereas excess $\uparrow$-spins ($h<0$) are incorporated by filling up only the \smash{$\hat{\gamma}$} modes with $0<\epsilon_j < |h|$.


Although the two conventions yield different descriptions of a state, they are formally equivalent. This can be checked, e.g., by calculating the energy $E = \langle \hat{H}_{\mbox{\tiny{MF}}} \rangle$ of a state. In the second convention, the energy is given by
\begin{align}
\nonumber \langle\hat{H}_{\mbox{\tiny{MF}}}\rangle \hspace{-0.03cm}&= \sum_{\epsilon_j>0} \big[(\epsilon_j + h)\hspace{0.05cm} n_{\mbox{\tiny{F}}}(\epsilon_j + h) + (\epsilon_j - h)\hspace{0.03cm}\hspace{0.05cm} n_{\mbox{\tiny{F}}}(\epsilon_j - h) - (\epsilon_j - h) \big]  + \text{Tr}\hspace{0.03cm}\big(\hat{H}_0 - \mu - h\big) - g_{\mbox{\tiny{1D}}}^{-1} \int dx \big|\Delta_0(x)\big|^2\\
\nonumber &=\sum_{\epsilon_j>0} \big[(\epsilon_j + h) \hspace{0.05cm} n_{\mbox{\tiny{F}}}(\epsilon_j + h)  + (\epsilon_j - h)\hspace{0.03cm}\hspace{0.05cm} n_{\mbox{\tiny{F}}}(-\epsilon_j + h)\big]  + \text{Tr}\hspace{0.03cm}\big(\hat{H}_0 - \mu - h\big) - g_{\mbox{\tiny{1D}}}^{-1} \int dx \big|\Delta_0(x)\big|^2\\
& = \sum_{j} (\epsilon_j + h) \hspace{0.05cm} n_{\mbox{\tiny{F}}}(\epsilon_j + h)  + \text{Tr}\hspace{0.03cm}\big(\hat{H}_0 - \mu - h\big) - g_{\mbox{\tiny{1D}}}^{-1} \int dx \big|\Delta_0(x)\big|^2 \hspace{0.05cm},
\label{Eotherconvention}
\end{align}
which is the same as the energy in the first convention.

\section{\label{experimental scheme}Experimental protocol for creating soliton train states}

\subsection{\label{creating balanced soliton train}Protocol for creating balanced soliton trains}
\begin{figure}[h]
\includegraphics[width=0.43\textwidth]{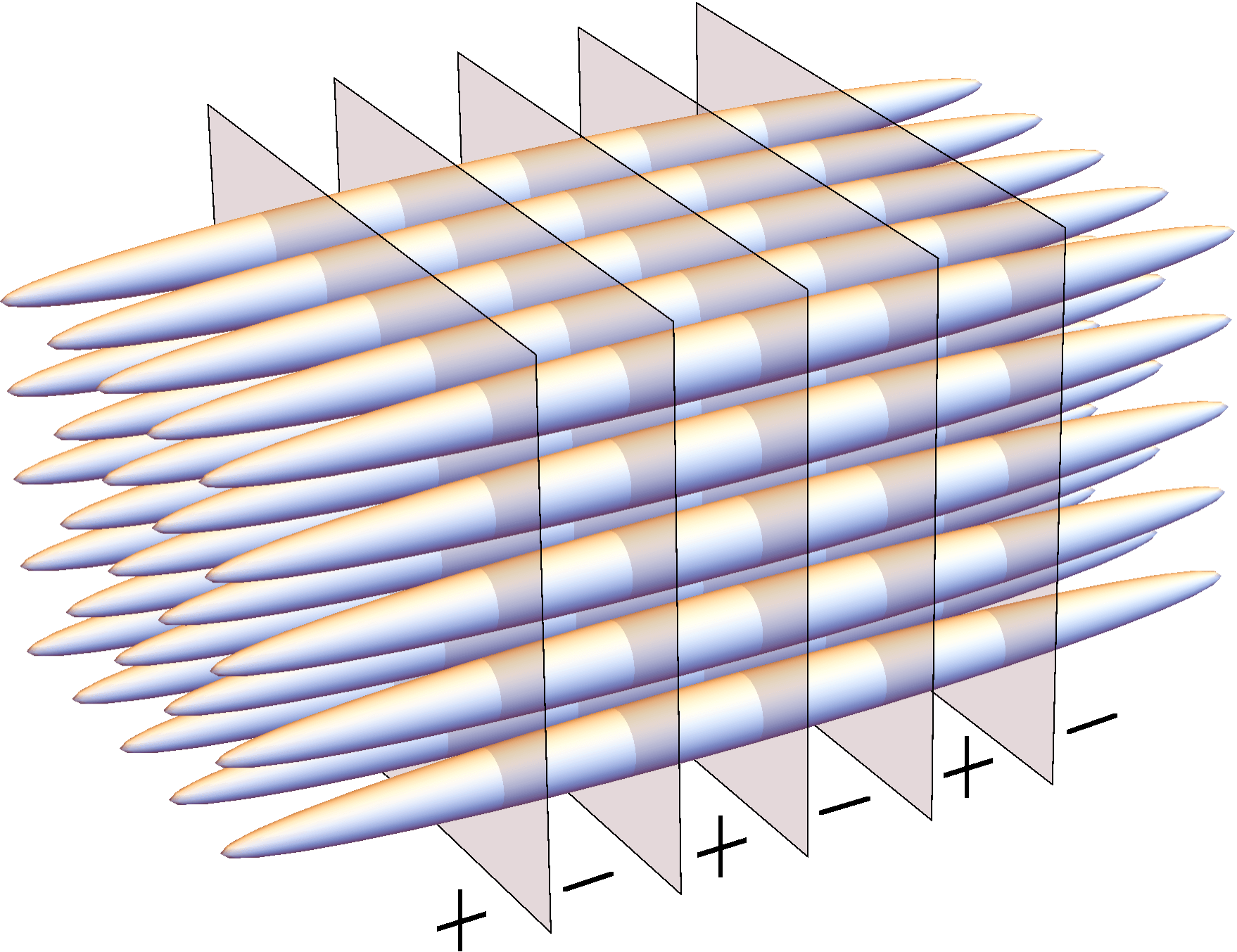}
\caption{\label{balancedprotocolfig}Schematic experimental set-up for producing balanced soliton trains in an array of weakly-coupled tubes. First uniform superfluids are prepared in each tube by cooling attractively interacting fermions near a Feshbach resonance. Then solitons are imprinted by shining off-resonant lasers in alternate regions labeled `$-$' to reverse the sign of the local order parameter.}
\end{figure}

To produce a balanced soliton train, one traps equal mixtures of $\uparrow$- and $\downarrow$-fermions (e.g., two hyperfine states of $^{6}$Li or $^{40}$K atoms) in an array of weakly-coupled 1D tubes (Fig.~\ref{balancedprotocolfig}). As demonstrated experimentally in \cite{sliao2010spin, srevelle20161d}, a superfluid is formed when the atoms are cooled near a Feshbach resonance. Following the strategy used in 3D gases \cite{syefsah2013heavy, *sku2014motion, *sku2016cascade, *ssacha2014proper, *skarpiuk2002solitons}, one can create solitons in these superfluids by phase imprinting, whereby one shines an off-resonant laser on selected portions of a superfluid for a short duration to rotate the phase of the local order parameter by a given amount. To generate soliton trains, one can imprint a $\pi$ phase in alternate regions of each tube, as shown in Fig.~\ref{balancedprotocolfig}.

\subsection{\label{creating C-FFLO}Protocol for creating C-FFLO states}
To produce a C-FFLO phase, we advocate starting with balanced soliton trains in an array of weakly-coupled tubes. As we detail in \cite{[{}] [{ (2017).}]sdutta2017protocol}, one can then use radio waves to selectively break up pairs in the soliton cores, transferring spin-$\uparrow$ atoms at those locations to a third spin state $|\phi\rangle$, which does not interact with the $\uparrow$- and $\downarrow$-spin states, thus leaving behind a C-FFLO state with unpaired $\downarrow$-spins. For example, in $^{40}$K one could use $|\hspace{-0.12cm}\uparrow\rangle = |9/2, -7/2\rangle$, $|\hspace{-0.12cm}\downarrow\rangle = |9/2, -9/2\rangle$, and $|\phi\rangle = |9/2,-5/2\rangle$, where the two numbers denote the total atomic spin $F$ and its projection $m_{\text{F}}$ \cite{ssagi2015breakdown}. Unwanted bulk excitations can be eliminated by Pauli blocking if one starts with an appropriate density of $|\phi\rangle$-atoms. Even without Pauli blocking, our approach gives relatively few bulk excitations at strong enough interactions.

Within our convention for Bogoliubov operators (detailed in Sec.~\ref{conventions}), the quasiparticle spectrum of a soliton train is symmetric for positive and negative energies, with delocalized bulk modes for $|\epsilon|>\epsilon_+$, and midgap modes bound to soliton cores for $|\epsilon|<\epsilon_-$ [Fig.~1(b) in the main text]. All negative energy modes are occupied in a balanced soliton train. The C-FFLO state with excess $\downarrow$-spins is formed by removing all quasiparticles from the midgap modes. Our key idea is to use a Rapid Adiabatic Passage protocol which vacates the midgap modes by a radio-frequency sweep, while a preformed Fermi sea of $|\phi\rangle$-atoms prevents any bulk excitation. Adiabaticity requires that the sweep is sufficiently slow. However, the finite lifetime of the soliton train sets an upper bound on the sweep rate. Fortunately, when the interactions are sufficiently strong there is a separation of scales.

\section{\label{stationary solution}Stationary soliton train solution}
Here we summarize the properties of the stationary soliton train solution that are relevant for examining the collective modes. For further details of derivation, we refer the reader to Refs. \cite{[{}] [{ [\href{http://www.jetpletters.ac.ru/ps/1354/article_20458.shtml}{JETP Lett. {\bf 31}, 456 (1980)}].}]sbrazovskii1980exact, shorovitz1981soliton, smertsching1981incommensurate, [{}] [{ [\href{http://www.jetp.ac.ru/cgi-bin/e/index/e/59/2/p434?a=list}{Sov. Phys. JETP {\bf 59}, 434 (1984)}].}]sbrazovskii1984peierls, [{}] [{ [\href{http://www.jetp.ac.ru/cgi-bin/e/index/e/58/2/p428?a=list}{Sov. Phys. JETP {\bf 58}, 428 (1983)}].}]sbuzdin1983phase, [{}] [{ [\href{http://www.jetp.ac.ru/cgi-bin/e/index/e/66/2/p422?a=list}{Sov. Phys. JETP {\bf 66}, 422 (1987)}].}]sbuzdin1987nonuniform, smachida1984superconductivity, swhittaker1996course}.

We first derive the stationary BdG equations, starting from the many-body Hamiltonian in Heisenberg picture,
\begin{equation}
\hat{H} = \int\hspace{-0.05cm} dx \left[\hspace{0.1cm}\sum\nolimits_{\sigma=\uparrow,\downarrow}\hspace{-0.1cm} \hat{\Psi}_{\sigma}^{\dagger}(x,t) \hat{H}_{\sigma}^{(0)} \hat{\Psi}_{\sigma}(x,t) + g_{\mbox{\tiny{1D}}} \hat{\Psi}_{\uparrow}^{\dagger}(x,t)\hat{\Psi}_{\downarrow}^{\dagger}(x,t) \hat{\Psi}_{\downarrow}(x,t) \hat{\Psi}_{\uparrow} (x,t)\right],
\label{manybodyH}
\end{equation}
where $\hat{H}_{\uparrow,\downarrow}^{(0)} \equiv -\partial_x^2/2-\epsilon_{\text{F}} \pm h$. We encode superfluid pairing in the order parameter $\Delta(x,t) = g_{\mbox{\tiny{1D}}} \langle \hat{\Psi}_{\downarrow}(x,t) \hat{\Psi}_{\uparrow}(x,t)\rangle$, and ignore quadratic fluctuations about $\Delta$, yielding the mean-field Hamiltonian
\begin{equation}
\hspace{-0.2cm}\hat{H}_{\mbox{\tiny{MF}}} \hspace{-0.05cm}=\hspace{-0.05cm} \int\hspace{-0.05cm} dx \left[\hspace{0.1cm}\sum\nolimits_{\sigma=\uparrow,\downarrow}\hspace{-0.1cm} \hat{\Psi}_{\sigma}^{\dagger}(x,t) \hat{H}_{\sigma}^{(0)} \hat{\Psi}_{\sigma}(x,t) + \Delta(x,t) \hat{\Psi}_{\uparrow}^{\dagger}(x,t) \hat{\Psi}_{\downarrow}^{\dagger}(x,t) + \Delta^*(x,t) \hat{\Psi}_{\downarrow}(x,t) \hat{\Psi}_{\uparrow} (x,t) - g_{\mbox{\tiny{1D}}}^{-1}|\Delta(x,t)|^2\right].\hspace{-0.1cm}
\label{HMF}
\end{equation}
The Heisenberg equations of motion for the field operators, $i \partial_t \hat{\Psi}_{\sigma} = [\hat{H}_{\mbox{\tiny{MF}}},\hat{\Psi}_{\sigma}]$, can be expressed as
\begin{equation}
i \partial_t \hat{\Psi}(x,t) = 
\begin{pmatrix}  -\partial_x^2/2 - \epsilon_{\mbox{\tiny{F}}} + h & \Delta(x,t) \\ \Delta^*(x,t) & \partial_x^2/2 + \epsilon_{\mbox{\tiny{F}}} + h\end{pmatrix} \hat{\Psi}(x,t) \hspace{0.05cm},
\label{heisenbergeom}
\end{equation}
where $\hat{\Psi} \equiv (\hat{\Psi}_{\uparrow} \;\; \hat{\Psi}_{\downarrow}^{\dagger})^{T}$. In the Andreev approximation, we write $\hat{\Psi}(x,t)$ as a sum over right-moving and left-moving fermionic quasiparticle modes, $\smash{\hat{\Psi}(x,t) = \sum_{s=\pm,j} e^{i s k_{\text{F}} x} (U^{s}_j(x,t)\;\;V^{s}_j(x,t))^{T} \hat{\gamma}^{s}_j}$ where
\begin{equation}
\big(\partial_x^2/2 - \epsilon_{\text{F}}\big)\bigg[\bigg(\hspace{-0.1cm}\begin{array}{c} U_j^{\pm}(x,t) \\ V_j^{\pm}(x,t) \end{array}\hspace{-0.1cm}\bigg) e^{\pm i k_{\text{F}} x}\bigg] \approx \bigg[\hspace{-0.05cm}\mp i k_{\text{F}} \partial_x \bigg(\hspace{-0.1cm}\begin{array}{c} U_j^{\pm}(x,t) \\ V_j^{\pm}(x,t) \end{array}\hspace{-0.1cm}\bigg)\bigg] e^{\pm i k_{\text{F}} x}\hspace{0.05cm},\quad\text{and}\quad \langle \hat{\gamma}_{j}^{s \dagger} \hat{\gamma}_{j^{\prime}}^{s^{\prime}} \rangle = \delta_{s s^{}\prime}\delta_{j j^{\prime}} \langle \hat{\gamma}_j^{s \dagger} \hat{\gamma}_j^s \rangle\hspace{0.05cm}.
\label{andreevapprox}
\end{equation}
Substituting this expansion into Eq.~\eqref{heisenbergeom} and in the definition of the order parameter, we find the BdG equations
\begin{equation}
\hspace{-0.1cm}i \partial_t \left(\hspace{-0.1cm}\begin{array}{c} U^{\pm}_j(x,t) \\ V^{\pm}_j(x,t) \end{array}\hspace{-0.1cm}\right) \hspace{-0.05cm}=  \hspace{-0.05cm}
\begin{pmatrix}  \mp i k_{\mbox{\tiny{F}}} \partial_x + h & \Delta(x,t) \\ \Delta^*(x,t) & \pm i k_{\mbox{\tiny{F}}} \partial_x + h\end{pmatrix} 
\left(\hspace{-0.1cm}\begin{array}{c} U^{\pm}_j(x,t) \\ V^{\pm}_j(x,t) \end{array}\hspace{-0.1cm}\right) ,\quad\hspace{-0.05cm}\text{with}\quad\hspace{-0.05cm}\Delta(x,t) = g_{\mbox{\tiny{1D}}} \sum_{s=\pm,j} \langle \hat{\gamma}^{s \dagger}_j \hat{\gamma}^{s}_j \rangle\hspace{0.05cm} U^{s}_j(x,t) V^{s *}_j(x,t)\hspace{0.05cm}.
\label{FullBdG}
\end{equation}
For a stationary solution $\Delta(x,t)\hspace{-0.05cm}=\hspace{-0.05cm}\Delta_0(x)$ with quasiparticle energies $\epsilon_j^{\pm}$, $\langle \hat{\gamma}^{\pm \dagger}_j \hat{\gamma}^{\pm}_j \rangle = n_{\mbox{\tiny{F}}} (\epsilon_j^{\pm}+h)$ where $n_{\mbox{\tiny{F}}}$ is the Fermi distribution, and $\smash{(U_j^{\pm}(x,t)\hspace{-0.03cm},\hspace{-0.03cm} V_j^{\pm}(x,t)) \hspace{-0.05cm}=\hspace{-0.05cm} (u_j^{\pm}(x),v^{\pm}_j(x))\hspace{0.02cm} e^{-i(\epsilon_j^{\pm} + h)t}}$. Using these expressions in Eq.~\eqref{FullBdG}, we find
\begin{equation}
\begin{pmatrix} \mp i k_{\text{F}} \partial_x & \Delta_0(x) \\ \Delta_0^*(x) & \pm i k_{\text{F}}\partial_x \end{pmatrix} 
\left(\hspace{-0.1cm}\begin{array}{c} u^{\pm}_j(x) \\ v^{\pm}_j(x) \end{array}\hspace{-0.1cm}\right) = \epsilon^{\pm}_j \left(\hspace{-0.1cm}\begin{array}{c} u^{\pm}_j(x) \\ v^{\pm}_j(x) \end{array}\hspace{-0.1cm}\right), \quad
\text{with} \quad
\Delta_0(x) = g_{\mbox{\tiny{1D}}} \sum_{s=\pm,j} n_{\mbox{\tiny{F}}} (\epsilon^s_j + h)\hspace{0.05cm} u^{s}_j(x) v^{s *}_j(x) \hspace{0.05cm}.
\label{stationaryBdGAndreev}
\end{equation}
For real $\Delta_0 (x)$, the right- and left-moving branches are related by a complex conjugation: $(u^-,v^-) = (u^+,v^+)^*$ and $\epsilon^- = \epsilon^+$. Thus we can only consider the right-moving branch, drop the superscript `+', and write
\begin{equation}
\begin{pmatrix} - i k_{\text{F}} \partial_x & \Delta_0(x) \\ \Delta_0(x) & + i k_{\text{F}} \partial_x \end{pmatrix} 
\left(\hspace{-0.1cm}\begin{array}{c} u_j(x) \\ v_j(x) \end{array}\hspace{-0.1cm}\right) = \epsilon_j \left(\hspace{-0.1cm}\begin{array}{c} u_j(x) \\ v_j(x) \end{array}\hspace{-0.1cm}\right), \quad
\text{with} \quad
\Delta_0(x) = 2g_{\mbox{\tiny{1D}}} \sum\nolimits_j n_{\mbox{\tiny{F}}} (\epsilon_j + h)\hspace{0.05cm} \text{Re}\hspace{-0.05cm}\left[ u_j(x) v^*_j(x)\right] \hspace{0.05cm}.
\label{stationaryBdGAndreevright}
\end{equation}
Past studies have shown that a periodic solution to Eq.~\eqref{stationaryBdGAndreevright} has the soliton train profile $\Delta_0 (x) = \Delta_1 k_1 \text{sn}(\Delta_1 x/k_{\text{F}}, k_1)$, where $\Delta_1 = 2 k_{\text{F}} k_0 K(k_1)/\pi$. Here $2\pi/k_0$ denotes the period, $K$ denotes the complete elliptic integral of the first kind, and $k_1 \in (0,1)$ parametrizes the sharpness of the solitons.

Since $\Delta_0 (x)$ is periodic, each quasiparticle wavefunction $(u(x),v(x))$ in Eq.~\eqref{stationaryBdGAndreevright} can be labeled by a quasimomentum $k$, with $-k_0/2<k<k_0/2$ representing the first Brillouin zone. In addition, the solutions have the following properties: (i) $(-v_k, u_k)$ is a wavefunction with energy $-\epsilon_k$, and (ii) $(v_k^*, u_k^*)$ is another wavefunction with energy $\epsilon_k$, i.e., $(u_{-k}, v_{-k}) = (v_k^*, u_k^*)$ and $\epsilon_{-k} = \epsilon_k$. Therefore, the quasiparticle spectrum $\epsilon(k)$ is symmetric about both $\epsilon$ and $k$ axes. For $k\geq 0$ and $\epsilon \geq 0$, it is given by (in the extended zone representation)
\begin{equation}
\frac{k}{k_0} = \frac{1}{\pi} \frac{\epsilon}{\epsilon_+} \text{Re} \hspace{-0.05cm} \left[ \sqrt{\frac{\epsilon_-^2 - \epsilon^2}{\epsilon_+^2 - \epsilon^2}} \hspace{0.1cm}\Pi \hspace{-0.05cm} \left(\frac{\epsilon_+^2 - \epsilon_-^2}{\epsilon_+^2 - \epsilon^2}, \sqrt{1-\frac{\epsilon_-^2}{\epsilon_+^2}}\right) \right],
\label{quasiparticlespectrum}
\end{equation}
where $\epsilon_{\pm} \equiv \frac{1}{2}(1 \pm k_1) \Delta_1$, and $\Pi$ denotes the complete elliptic integral of the third kind. Figure 1(b) of the main article shows the spectrum for $k_1=0.65$. It has a band of Andreev bound states with $|\epsilon | \leq \epsilon_-$, and continua of free states with $|\epsilon | \geq \epsilon_+$. The dispersion is linear as $k\to 0$ and $k\to\infty$, with $\epsilon \approx k_{\text{F}} k$ for $k \gg k_0$. Interestingly, there is no gap in the spectrum at $k=n(k_0/2)$ with $n=\pm 2, \pm 3,\dots$. This is because $\Delta_0(x)$ presents a reflectionless potential (a 1-gap Lam\'{e} potential) to the Bogoliubov quasiparticles (see \cite{smertsching1981incommensurate, smachida1984superconductivity, swhittaker1996course, sbelokolos2001exact, scooper1995supersymmetry, snovikov1984theory} for more details).

Hereafter we'll use `tilde' (\textasciitilde) to denote nondimensionalized quantities, with energies rescaled by $k_{\text{F}} k_0$, and momenta rescaled by $k_0$, e.g., $\tilde{\epsilon}_{\pm} \equiv \epsilon_{\pm} / (k_{\text{F}} k_0)$, $\tilde{k} \equiv k/k_0$. The density of states is given by
\begin{equation}
\tilde{\rho}(\tilde{\epsilon}) \equiv \frac{1}{\pi} \frac{d\tilde{k}}{d\tilde{\epsilon}} = \frac{1}{\pi}\hspace{0.05cm}\text{Re}\bigg[\frac{|\tilde{\epsilon}^2 - \tilde{\epsilon}_g^2|}{((\tilde{\epsilon}^2 - \tilde{\epsilon}_-^2)(\tilde{\epsilon}^2 - \tilde{\epsilon}_+^2))^{1/2}}\bigg]\hspace{0.05cm},\quad\text{where}\quad
\epsilon_g^2 \equiv \epsilon_+^2 \hspace{0.1cm} E\left(\sqrt{1-\frac{\epsilon_-^2}{\epsilon_+^2}}\right)\bigg/K\left(\sqrt{1-\frac{\epsilon_-^2}{\epsilon_+^2}}\right).
\label{DOS}
\end{equation}
Here $E$ denotes the complete elliptic integral of the second kind. Note that the density of states diverges as $\epsilon \to\epsilon_{\pm}$, as expected for band edges in 1D. The energy scale $\epsilon_g$ satisfies the inequality $\epsilon_- < \epsilon_g < \epsilon_+$.

The quasiparticle wavefunctions can be expressed in terms of a `spectral parameter' $a_k \in [-\alpha,\hspace{0.05cm}\alpha]$ where $\alpha \equiv \tilde{\epsilon}_+^{-1} K(\epsilon_-/\epsilon_+)$. The continuum of free states with $\epsilon \geq \epsilon_+$ and $k \geq k_0/2$ is described by
\begin{equation}
\left(\hspace{-0.1cm}\begin{array}{c} u_k(x) \vspace{0.1cm}\\ v_k(x) \end{array}\hspace{-0.1cm}\right) = \frac{e^{i k x}}{2\sqrt{L (\tilde{\epsilon}_k^2 - \tilde{\epsilon}_g^2)}} \left(\hspace{-0.1cm}\begin{array}{c} \sum_{n \text{ even}}\vspace{0.15cm} \\ -i\sum_{n \text{ odd}} \end{array}\hspace{-0.1cm}\right) \frac{e^{i n k_0 x}}{\sinh (n \alpha + a_k/2)} \hspace{0.05cm},
\label{freestatesuv}
\end{equation}
where the momentum $k$ and energy $\epsilon_k$ are parametrized as
\begin{equation}
\tilde{k} = (i/\pi)\big[\pi \hspace{0.05cm}\zeta(i a_{\tilde{k}} |\pi, i\alpha) - i a_{\tilde{k}} \hspace{0.05cm} \zeta (\pi |\pi, i\alpha)\big]\hspace{0.05cm}, \quad \text{and}\quad \quad \tilde{\epsilon}_{\tilde{k}} = \sqrt{(\tilde{\epsilon}_+^2 + \tilde{\epsilon}_-^2)/3 - \wp (i a_{\tilde{k}} | \pi,i\alpha)}\hspace{0.05cm}.
\label{freestatesparam}
\end{equation}
Here $\zeta$ and $\wp$ denote Weierstrass elliptic functions with half-periods $\pi$ and $i\alpha$, and $L$ in Eq.~\eqref{freestatesuv} denotes the length of the system. As $k$ varies from $k_0/2$ to $\infty$, $a_k$ decreases monotonically from $\alpha$ to 0, and $\epsilon_k$ grows from $\epsilon_+$ to $\infty$. Similarly, the bound states with $0\leq k \leq k_0/2$ and $0\leq \epsilon \leq \epsilon_-$ are described by
\begin{equation}
\left(\hspace{-0.1cm}\begin{array}{c} u_k(x) \vspace{0.1cm}\\ v_k(x) \end{array}\hspace{-0.1cm}\right) = \frac{e^{i k x}}{2\sqrt{L (\tilde{\epsilon}_g^2 - \tilde{\epsilon}_k^2)}} \left(\hspace{-0.1cm}\begin{array}{c} \sum_{n \text{ even}}\vspace{0.15cm} \\ -i\sum_{n \text{ odd}} \end{array}\hspace{-0.1cm}\right) \frac{e^{i n k_0 x}}{\cosh (n \alpha + a_k/2)} \hspace{0.05cm},
\label{boundstatesuv}
\end{equation}
\begin{equation}
\text{with} \quad \tilde{k} = (i/\pi)\big[\pi \hspace{0.05cm}\zeta(\pi + i a_{\tilde{k}} |\pi, i\alpha) - (\pi + i a_{\tilde{k}}) \hspace{0.05cm} \zeta (\pi |\pi, i\alpha)\big]\hspace{0.05cm}, \quad \text{and}\quad \quad \tilde{\epsilon}_{\tilde{k}} = \sqrt{(\tilde{\epsilon}_+^2 + \tilde{\epsilon}_-^2)/3 - \wp (\pi + i a_{\tilde{k}} | \pi,i\alpha)}\hspace{0.05cm}.
\label{boundstatesparam}
\end{equation}
As $k$ is varied from $0$ to $k_0/2$, $a_k$ increases monotonically from 0 to $\alpha$, and $\epsilon_k $ grows from 0 to $\epsilon_-$.

Note that the spectrum and the wavefunctions are completely specified (in rescaled coordinates) by the sharpness parameter $k_1$. This parameter is in turn set by $k_0$, $k_{\text{F}}$, $h$, and $a_{\mbox{\tiny{1D}}}$ at zero temperature through the self-consistency condition in Eq.~\eqref{stationaryBdGAndreevright}. To see this, we use $n_{\mbox{\tiny{F}}}(\epsilon) = \Theta (-\epsilon)$ at zero temperature, $\Theta$ being the unit-step function, and write the self-consistency condition in terms of the quasiparticle states with $\epsilon, k \geq 0$ as
\begin{equation}
\Delta_0 (x) = -4 g_{\mbox{\tiny{1D}}} \sum_{k\geq 0} \Theta(\epsilon_k - h) \hspace{0.05cm}\text{Re}\left[ u_k(x) v^*_k(x)\right] \hspace{0.05cm}.
\label{selfconsistency}
\end{equation}
Using properties of elliptic functions, one can show that $\text{Re}\left[u_k(x) v^*_k(x)\right] = \Delta_0(x) \epsilon_k/\big(2L (\epsilon_k^2 - \epsilon_g^2)\big)\hspace{0.1cm}\forall \hspace{0.05cm} k \geq 0$ \cite{smachida1984superconductivity}. Substituting this result and the relation $g_{\mbox{\tiny{1D}}} = -2/a_{\mbox{\tiny{1D}}}$ \cite{solshanii1998atomic, sbergeman2003atom} into Eq.~\eqref{selfconsistency}, we get, in the limit $L \to \infty$,
\begin{equation}
\frac{2}{\pi a_{\mbox{\tiny{1D}}}} \int_0^{k_c} d k \hspace{0.05cm} \frac{\Theta \big(\epsilon_k - h \big) \hspace{0.05cm}\epsilon_k}{\epsilon_k^2 - \epsilon_g^2} = 1.
\label{selfconsistencycontinuum}
\end{equation}
Here we have introduced an ultraviolet cutoff $k_c$ because the integral has a logarithmic divergence at high energies, as the dispersion is linear at large $k$. This is an artifact of the Andreev approximation, and not present in the full model. We choose the cutoff by requiring that a uniform (BCS-type) solution to Eq.~\eqref{stationaryBdGAndreevright} match the known solution in the full model, as was done in Refs. \cite{smertsching1981incommensurate, smachida1984superconductivity}. This procedure yields $k_c/k_{\text{F}}$ as a function of $k_{\text{F}} a_{\mbox{\tiny{1D}}}$ (see Sec.~\ref{cutoff}). We find $k_c \approx 2 k_{\text{F}}$ throughout the weakly-interacting regime ($k_{\text{F}} a_{\mbox{\tiny{1D}}} \gtrsim 1$). We have verified that the soliton train profiles obtained using this cutoff closely match the numerically obtained profiles in the full model. Further, barring the weak dependence of $k_1$ on $k_c$, the collective modes are insensitive to the choice of the cutoff. We can rewrite Eq.~\eqref{selfconsistencycontinuum} as
\begin{equation}
\int_0^{\tilde{k}_c} d \tilde{k}\hspace{0.05cm} \frac{\Theta(\tilde{\epsilon}_{\tilde{k}} - \tilde{h}) \hspace{0.05cm}\tilde{\epsilon}_{\tilde{k}}}{\tilde{\epsilon}_{\tilde{k}}^2 - \tilde{\epsilon}_g^2} = \frac{\pi}{2} k_{\text{F}} a_{\mbox{\tiny{1D}}} \hspace{0.05cm},\quad\text{or}\quad
\int_{\tilde{h}}^{\tilde{\epsilon}_c} \frac{\tilde{\epsilon}\hspace{0.05cm} \tilde{\rho} (\tilde{\epsilon}) \hspace{0.05cm} d\tilde{\epsilon}}{\tilde{\epsilon}^2 - \tilde{\epsilon}_m^2} = \frac{1}{2} k_{\text{F}} a_{\mbox{\tiny{1D}}}\hspace{0.05cm},
\label{selfconsistencyenergy}
\end{equation}
where $\tilde{\epsilon}_c \equiv \tilde{\epsilon}_{\tilde{k}_c}$. Substituting the expression for $\tilde{\rho} (\tilde{\epsilon})$ from Eq.~\eqref{DOS}, and evaluating the integral, we get
\begin{equation}
\text{Re}\left[\ln \left(\frac{(\tilde{\epsilon}_c^2 - \tilde{\epsilon}_-^2)^{1/2} + (\tilde{\epsilon}_c^2 - \tilde{\epsilon}_+^2)^{1/2}}{(\tilde{\epsilon}_-^2 - \tilde{h}^2)^{1/2} + (\tilde{\epsilon}_+^2 - \tilde{h}^2)^{1/2}}\right)\right] = \frac{\pi}{2} k_{\text{F}} a_{\mbox{\tiny{1D}}} \hspace{0.05cm}.
\label{selfconsistencyfinal}
\end{equation}
Equation~\eqref{selfconsistencyfinal} determines $k_1$ for given values of $k_0/k_{\text{F}}$, $k_{\text{F}} a_{\mbox{\tiny{1D}}}$, and $|h|/(k_{\text{F}} k_0)$ at zero temperature.

The collective modes are characterized by $k_1$ and $n_s$, where $n_s$ denotes the number of unpaired fermions per soliton. To see how $n_s$ depends on $k_0$, $k_{\text{F}}$, $a_{\mbox{\tiny{1D}}}$, and $h$, we first write the expressions for the densities of up- and down-spins:
\begin{align}
n_{\uparrow}(x) &= \langle \hat{\Psi}_{\uparrow}^{\dagger}(x) \hat{\Psi}_{\uparrow}(x) \rangle = 2\sum\nolimits_j n_{\mbox{\tiny{F}}} (\epsilon_j + h) |u_j (x)|^2 = \sum\nolimits_j n_{\mbox{\tiny{F}}} (\epsilon_j + h) \left(|u_j (x)|^2 + |v_j (x)|^2\right) \hspace{0.05cm},\\
n_{\downarrow}(x) &= \langle \hat{\Psi}_{\downarrow}^{\dagger}(x) \hat{\Psi}_{\downarrow}(x) \rangle = 2\sum\nolimits_j n_{\mbox{\tiny{F}}} (-\epsilon_j - h) |v_j (x)|^2 = \sum\nolimits_j n_{\mbox{\tiny{F}}} (-\epsilon_j - h) \left(|u_j (x)|^2 + |v_j (x)|^2\right) \hspace{0.05cm}.
\label{densities}
\end{align}
In the last step of the above equations, we have made use of the symmetry $(u_{-k}, v_{-k}) = (v_k^*, u_k^*)$ and $\epsilon_{-k} = \epsilon_k$. Thus, at zero temperature, the density of unpaired fermions is given by
\begin{equation}
\delta n (x) = \sum_{-|h|<\epsilon_j<|h|} |u_j (x)|^2 + |v_j (x)|^2 = \sum_{0\leq \epsilon_k<|h|} 2\left(|u_k (x)|^2 + |v_k (x)|^2\right)\hspace{0.05cm}.
\label{unpaireddensity}
\end{equation}
Here we have used $(-v_k (x), u_k(x))$ and $(u_k^* (x), -v_k^*(x))$ for the two states with energy $-\epsilon_k$. From Eqs.~\eqref{freestatesuv} and \eqref{boundstatesuv}, we see that $\delta n(x+\pi/k_0) = \delta n (x)$. Thus one can find $n_s$ by simply integrating $\delta n(x)$ over all $x$, then dividing by the number of solitons $N_s = L/(\pi/k_0) = k_0 L/\pi$. However, $\int \hspace{-0.05cm}dx \left(|u_k (x)|^2 + |v_k (x)|^2\right) = 1$ from normalization. Hence,
\begin{equation}
n_s = \frac{2\pi}{k_0 L} \sum_{0\leq \epsilon_k<|h|} 1 \hspace{0.05cm}\xrightarrow{\: L \to \infty \: } \left.2\hspace{0.05cm}\frac{k_{\text{h}}}{k_0} \right\vert_{\epsilon_{k_{\text{h}}}=h} = \frac{2}{\pi} \frac{\tilde{h}}{\tilde{\epsilon}_+} \text{Re} \hspace{-0.05cm} \left[ \sqrt{\frac{\tilde{\epsilon}_-^2 - \tilde{h}^2}{\tilde{\epsilon}_+^2 - \tilde{h}^2}} \hspace{0.1cm}\Pi \hspace{-0.05cm} \left(\frac{\tilde{\epsilon}_+^2 - \tilde{\epsilon}_-^2}{\tilde{\epsilon}_+^2 - \tilde{h}^2}, \sqrt{1-\frac{\tilde{\epsilon}_-^2}{\tilde{\epsilon}_+^2}}\right) \right]\hspace{0.05cm},
\label{ns}
\end{equation}
where we have used Eq.~\eqref{quasiparticlespectrum} for the dispersion. Note that $\tilde{\epsilon}_{\pm} = (1\pm k_1) K(k_1)/\pi$. Thus Eq.~\eqref{ns} yields $n_s$ for given values of $|h|/(k_{\text{F}} k_0)$ and $k_1$ at zero temperature. When $h=0$, $n_s = 0$, and we get a balanced soliton train, whereas for $\tilde{\epsilon}_- < |h|/(k_{\text{F}} k_0) < \tilde{\epsilon}_+$, $n_s=1$, and we get a commensurate FFLO (C-FFLO) state.

\begin{figure}[h]
\includegraphics[scale=1]{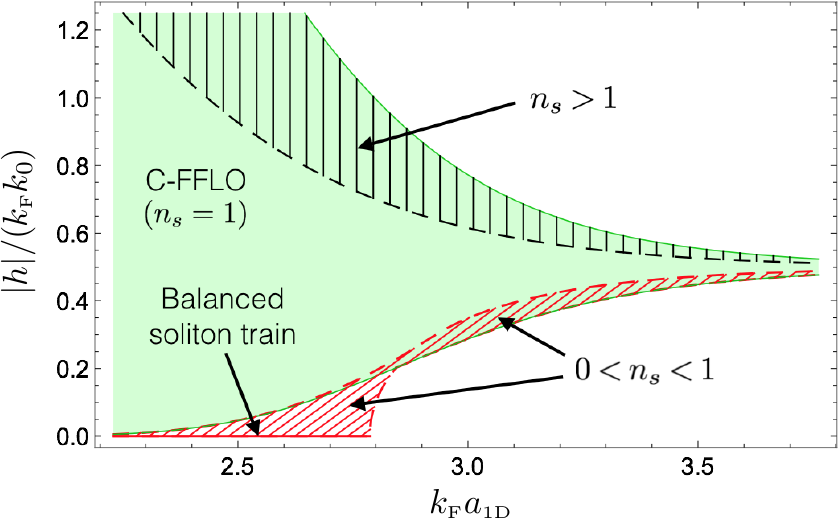}
\caption{\label{stationarysolutions}(Color online.) Stationary soliton train solutions in different regions of the $k_{\text{F}} a_{\mbox{\tiny{1D}}}\hspace{-0.05cm}$ -- $\hspace{-0.05cm}|h|/(k_{\text{F}} k_0)$ plane for $k_0/k_{\text{F}}=0.05$. The solutions are obtained by imposing the self-consistency condition in Eq.~\eqref{selfconsistencyfinal}. The states are classified in terms of $n_s$, the number of unpaired fermions per soliton [Eq.~\eqref{ns}]. Solid (green), vertically hatched (black), and oblique hatched (red) regions contain, respectively, a solution with $n_s=1$, $n_s>1$, and $0<n_s<1$. Overlapping regions contain multiple solutions. Figure 4 in the main article is similar, but shows only stable solutions for $|h|>0$.}
\end{figure}

Equations~\eqref{selfconsistencyfinal} and \eqref{ns} determine $k_1$ and $n_s$ for given values of $k_0/k_{\text{F}}$, $k_{\text{F}} a_{\mbox{\tiny{1D}}}$, and $|h|/(k_{\text{F}} k_0)$. In general, there can be zero, one, or multiple solutions, as the left-hand side of Eq.~\eqref{selfconsistencyfinal} is a non-monotonic function of $k_1$. Figure~\ref{stationarysolutions} shows regions in the $k_{\text{F}} a_{\mbox{\tiny{1D}}}\hspace{-0.05cm}$ -- $\hspace{-0.05cm}|h|/(k_{\text{F}} k_0)$ plane where different types of solutions exist for $k_0/k_{\text{F}}=0.05$. As $k_0/k_{\text{F}}$ is decreased (increased), the regions remain structurally similar, but translate to weaker (stronger) interactions. In particular, the balanced soliton train with a given period exists only above a minimum interaction strength, given by $k_0/k_{\text{F}} \lesssim 4 e^{-\frac{\pi}{2} k_{\text{F}} a_{\mbox{\tiny{1D}}}}$. Conversely, the C-FFLO phase exists for all values of $k_0/k_{\text{F}}$ and $k_{\text{F}} a_{\mbox{\tiny{1D}}}$, although for very weak interactions, it is confined to a small interval of $|h|/(k_{\text{F}} k_0)$ near $1/2$. We also note that the C-FFLO state for a given $k_0/k_{\text{F}}$ and $k_{\text{F}} a_{\mbox{\tiny{1D}}}$ does not vary with $h$, as Eqs.~\eqref{selfconsistencyfinal} and \eqref{ns} become independent of $h$ for $\tilde{\epsilon}_- < |h|/(k_{\text{F}} k_0) < \tilde{\epsilon}_+$.

\section{\label{cutoff}High-energy cutoff in the Andreev approximation}
In this section we find an expression for the cutoff $k_c$ in Eq.~\eqref{selfconsistencycontinuum} by requiring that a uniform superfluid solution to Eq.~\eqref{stationaryBdGAndreevright}, $\Delta_0(x) = \Delta_{\text{BCS}}$, match the corresponding solution in the full model (at $h=0$).

At zero temperature, the stationary BdG equations in the full model, with $\Delta_0 (x) = \Delta_{\text{BCS}} \in \mathbb{R}$, are given by
\begin{equation}
\begin{pmatrix} - \partial_x^2/2 - \epsilon_{\mbox{\tiny{F}}} & \Delta_{\text{BCS}} \\ \Delta_{\text{BCS}} & \partial_x^2/2 + \epsilon_{\mbox{\tiny{F}}} \end{pmatrix} 
\left(\hspace{-0.1cm}\begin{array}{c} u_j(x) \\ v_j(x) \end{array}\hspace{-0.1cm}\right) = \epsilon_j \left(\hspace{-0.1cm}\begin{array}{c} u_j(x) \\ v_j(x) \end{array}\hspace{-0.1cm}\right), \quad
\text{with} \quad
\Delta_0(x) = g_{\mbox{\tiny{1D}}} \sum\nolimits_j \Theta (-\epsilon_j)\hspace{0.05cm} u_j(x) v^*_j(x) \hspace{0.05cm}.
\label{stationaryBdGfullmodel}
\end{equation}
The Hamiltonian has plane wave eigenfunctions
\begin{equation}
\big(u_k^{\pm}(x), v_k^{\pm}(x)\big) = \frac{e^{i k x}}{\sqrt{2 L}}\bigg(\Big(1+ \frac{\xi_k}{\epsilon_k^{\pm}}\Big)^{1/2}, \pm\Big(1- \frac{\xi_k}{\epsilon_k^{\pm}}\Big)^{1/2}\bigg)\hspace{0.05cm},\quad\text{with energies}\quad
\epsilon_k^{\pm} = \pm \sqrt{\xi_k^2 + \Delta_{\text{BCS}}^2}\hspace{0.05cm},
\label{uniformfull}
\end{equation}
where $\xi_k \equiv k^2/2 - \epsilon_{\mbox{\tiny{F}}}$. Using these expressions in the self-consistency condition in Eq.~\eqref{stationaryBdGfullmodel}, we obtain
\begin{equation}
\Delta_{\text{BCS}} = -\frac{g_{\mbox{\tiny{1D}}}}{2 L} \sum_k \frac{\Delta_{\text{BCS}}}{\big(\xi_k^2 + \Delta_{\text{BCS}}^2\big)^{1/2}}\hspace{0.05cm},\quad \text{or} \quad \int_0^{\infty} \frac{d \bar{k}}{\left((\bar{k}^2-1)^2 + (\Delta_{\text{BCS}}/\epsilon_{\mbox{\tiny{F}}})^2\right)^{1/2}} = \frac{\pi}{2} k_{\text{F}} a_{\mbox{\tiny{1D}}}\hspace{0.05cm},
\label{selfconsistencyuniformfull}
\end{equation}
where $\bar{k} \equiv k/k_{\text{F}}$, and we have taken the limit $L\to\infty$. Evaluating the integral in Eq.~\eqref{selfconsistencyuniformfull} yields
\begin{equation}
\frac{1}{(1+(\Delta_{\text{BCS}}/\epsilon_{\mbox{\tiny{F}}})^2)^{1/4}}\hspace{0.05cm} F \left(\pi\bigg| \frac{1}{2}+\frac{1}{2\hspace{0.05cm}(1+(\Delta_{\text{BCS}}/\epsilon_{\mbox{\tiny{F}}})^2)^{1/2}}\right) = \pi k_{\text{F}} a_{\mbox{\tiny{1D}}}\hspace{0.05cm},
\label{DeltaBCS}
\end{equation}
where $F$ denotes the incomplete elliptic integral of the first kind. Inverting this equation gives $\Delta_{\text{BCS}}/\epsilon_{\mbox{\tiny{F}}}$ as a function of $k_{\text{F}} a_{\mbox{\tiny{1D}}}$. In the weakly-interacting regime ($k_{\text{F}} a_{\mbox{\tiny{1D}}} \gtrsim 1$), $\Delta_{\text{BCS}}$ decays exponentially with $k_{\text{F}} a_{\mbox{\tiny{1D}}}$ as $\Delta_{\text{BCS}} \approx 8 \hspace{0.05cm}\epsilon_{\mbox{\tiny{F}}}\hspace{0.05cm} e^{-\frac{\pi}{2} k_{\text{F}} a_{\mbox{\tiny{1D}}}}$.

Next we solve Eq.~\eqref{stationaryBdGAndreevright} with $\Delta_0(x) = \Delta_{\text{BCS}}$ and $h=0$ at zero temperature to determine the cutoff $k_c$:
\begin{equation}
\begin{pmatrix} - i k_{\text{F}} \partial_x & \Delta_{\text{BCS}} \\ \Delta_{\text{BCS}} & + i k_{\text{F}} \partial_x \end{pmatrix} 
\left(\hspace{-0.1cm}\begin{array}{c} u_j(x) \\ v_j(x) \end{array}\hspace{-0.1cm}\right) = \epsilon_j \left(\hspace{-0.1cm}\begin{array}{c} u_j(x) \\ v_j(x) \end{array}\hspace{-0.1cm}\right), \quad
\text{with} \quad
\Delta_0(x) = 2g_{\mbox{\tiny{1D}}} \sum\nolimits_j^{\prime} \Theta (-\epsilon_j) \hspace{0.05cm} \text{Re}\hspace{-0.05cm}\left[ u_j(x) v^*_j(x)\right] \hspace{0.05cm},
\label{stationaryBdGuniformAndreev}
\end{equation}
where the prime on the sum stands for the cutoff. We can again solve the system in terms of plane waves
\begin{equation}
\big(u_k^{\pm}(x), v_k^{\pm}(x)\big) = \frac{e^{i k x}}{\sqrt{2 L}}\bigg(\Big(1+ \frac{k_{\text{F}} k}{\epsilon_k^{\pm}}\Big)^{1/2}, \pm\Big(1- \frac{k_{\text{F}} k}{\epsilon_k^{\pm}}\Big)^{1/2}\bigg)\hspace{0.05cm},\quad\text{where}\quad
\epsilon_k^{\pm} = \pm \sqrt{(k_{\text{F}} k)^2 + \Delta_{\text{BCS}}^2}\hspace{0.05cm}.
\label{uniformAndreev}
\end{equation}
Substituting these expressions in the self-consistency condition in Eq.~\eqref{stationaryBdGuniformAndreev}, we get
\begin{equation}
\Delta_{\text{BCS}} = -\frac{g_{\mbox{\tiny{1D}}}}{L} \sideset{}{'}\sum_k \frac{\Delta_{\text{BCS}}}{\big((k_{\text{F}} k)^2 + \Delta_{\text{BCS}}^2\big)^{1/2}}\hspace{0.05cm},\quad \text{or} \quad \int_0^{k_c/k_{\text{F}}} \frac{d \bar{k}}{\left(\bar{k}^2 + (\Delta_{\text{BCS}}/2\epsilon_{\mbox{\tiny{F}}})^2\right)^{1/2}} = \frac{\pi}{2} k_{\text{F}} a_{\mbox{\tiny{1D}}}\hspace{0.05cm},
\label{selfconsistencyuniformAndreev}
\end{equation}
in the limit $L\to\infty$. Evaluating the integral yields
\begin{equation}
k_c = (\Delta_{\text{BCS}}/k_{\text{F}}) \sinh(\pi k_{\text{F}} a_{\mbox{\tiny{1D}}}/2)\hspace{0.05cm}.
\label{kc}
\end{equation}
Combining Eqs.~\eqref{DeltaBCS} and \eqref{kc}, we obtain $k_c/k_{\text{F}}$ as a function of $k_{\text{F}} a_{\mbox{\tiny{1D}}}$. For $k_{\text{F}} a_{\mbox{\tiny{1D}}} \gtrsim 1$, $k_c \approx 2 k_{\text{F}}$.

\section{\label{energy}Energy of stationary states and phase diagram}
Here we calculate the mean-field energy of stationary states in the Andreev approximation, which will let us compare the energies of the phases in Fig.~\ref{stationarysolutions}, as well as uniform states with $\Delta_0(x) = $ constant, to arrive at a phase diagram.

We rewrite the mean-field Hamiltonian in the Eq.~\eqref{HMF} for a stationary state $\Delta(x,t) = \Delta_0(x)\in\mathbb{R}$ as
\begin{align}
\nonumber \hat{H}_{\mbox{\tiny{MF}}} \hspace{-0.03cm}=\hspace{-0.1cm} \int \hspace{-0.1cm} d x\hspace{0.05cm} &\bigg[
\big(\hat{\Psi}^{\dagger}_{\uparrow}(x,t) \;\;\; \hat{\Psi}_{\downarrow}(x,t)\big)
\begin{pmatrix} -\partial_x^2/2 - \epsilon_{\text{F}} + h & \Delta_0(x) \\ \Delta_0(x) & \partial_x^2/2 + \epsilon_{\text{F}} + h \end{pmatrix} 
\bigg(\hspace{-0.1cm}\begin{array}{c} \hat{\Psi}_{\uparrow}(x,t) \\ \hat{\Psi}^{\dagger}_{\downarrow}(x,t) \end{array}\hspace{-0.1cm}\bigg) \\
&+ \hat{\Psi}_{\downarrow}(x,t) \hspace{0.05cm}\big(\hspace{-0.1cm}-\partial_x^2/2 - \epsilon_{\text{F}} - h\big) \hat{\Psi}^{\dagger}_{\downarrow} (x,t) + \hat{\Psi}^{\dagger}_{\downarrow}(x,t) \hspace{0.05cm}\big(\hspace{-0.1cm}-\partial_x^2/2 - \epsilon_{\text{F}} - h\big) \hat{\Psi}_{\downarrow} (x,t) - g_{\mbox{\tiny{1D}}}^{-1} \big(\Delta_0(x)\big)^2\bigg].
\label{mfhamiltonian}
\end{align}
In the Andreev approximation, we diagonalize the Hamiltonian by the Bogoliubov transformation (see Sec.~\ref{stationary solution})
\begin{equation}
\bigg(\hspace{-0.1cm}\begin{array}{c} \hat{\Psi}_{\uparrow}(x,t) \\ \hat{\Psi}^{\dagger}_{\downarrow}(x,t) \end{array}\hspace{-0.1cm}\bigg) = \sum\nolimits_j^{\prime} \bigg[\bigg(\hspace{-0.1cm}\begin{array}{c} u_j(x) \\ v_j(x) \end{array}\hspace{-0.1cm}\bigg) e^{i k_{\text{F}} x - i \epsilon_j t} \hspace{0.05cm}\hat{\gamma}^+_j + \bigg(\hspace{-0.1cm}\begin{array}{c} u^*_j(x) \\ v^*_j(x) \end{array}\hspace{-0.1cm}\bigg) e^{-i k_{\text{F}} x - i \epsilon_j t} \hspace{0.05cm}\hat{\gamma}^-_j\bigg]\hspace{0.05cm},
\label{bogoliubovtransformation}
\end{equation}
where the wavefunctions $(u_j(x)\;\;v_j(x))^T$ satisfy Eq.~\eqref{stationaryBdGAndreevright}, and the prime on the sum indicates we only include modes with energies below the high-energy cutoff. Next we use $\smash{\langle \hat{\gamma}^{s \dagger}_{j} \hat{\gamma}^{s^{\prime}}_{j^{\prime}} \rangle = n_{\mbox{\tiny{F}}} (\epsilon_j + h) \hspace{0.05cm}\delta_{j j^{\prime}}\delta_{s s^{\prime}}}$ and
\begin{equation}
\big(\partial_x^2/2 - \epsilon_{\text{F}}\big)\bigg[\bigg(\hspace{-0.1cm}\begin{array}{c} u_j(x) \\ v_j(x) \end{array}\hspace{-0.1cm}\bigg) e^{i k_{\text{F}} x}\bigg] \approx \bigg[\hspace{-0.05cm}- i k_{\text{F}} \partial_x \bigg(\hspace{-0.1cm}\begin{array}{c} u_j(x) \\ v_j(x) \end{array}\hspace{-0.1cm}\bigg)\bigg] e^{i k_{\text{F}} x}
\label{approx}
\end{equation}
in Eq.~\eqref{mfhamiltonian} to obtain the mean-field energy $E \equiv \langle \hat{H}_{\mbox{\tiny{MF}}} \rangle$ as
\begin{equation}
E = \sum\nolimits_j^{\prime}\bigg[2 (\epsilon_j + h)\hspace{0.05cm} n_{\mbox{\tiny{F}}} (\epsilon_j + h) +\hspace{-0.05cm} \int \hspace{-0.1cm} dx \hspace{0.05cm}\Big(\hspace{-0.05cm} i k_{\text{F}}  \hspace{0.05cm}v_j (x) \hspace{0.05cm} \partial_x v_j^* (x) - i k_{\text{F}}\hspace{0.05cm} v_j^* (x) \hspace{0.05cm}\partial_x v_j (x) - 2 h |v_j (x)|^2 \Big)\bigg] - g_{\mbox{\tiny{1D}}}^{-1} \int \hspace{-0.1cm} dx \hspace{0.05cm} \big(\Delta_0(x)\big)^2\hspace{0.05cm}.
\label{Emf1}
\end{equation}
The above expression can be simplified by noting that corresponding to a state $(u_j(x)\;\;v_j(x))^T$ with energy $\epsilon_j$, there is a state $(-u_j^*(x)\;\;v_j^*(x))^T$ with energy $-\epsilon_j$ (see Sec.~\ref{stationary solution}). Therefore the terms involving derivatives in Eq.~\eqref{Emf1} vanish when summed over all states. Further, $(v_j^*(x)\;\;u_j^*(x))^T$ is also a state with energy $\epsilon_j$, which lets us write
\begin{equation}
2\sum\nolimits_j^{\prime}\int\hspace{-0.05cm} d x\hspace{0.05cm} |v_j(x)|^2 = 2\sum\nolimits_j^{\prime}\int\hspace{-0.05cm} d x \hspace{0.05cm}|u_j(x)|^2 = \sum\nolimits_j^{\prime}\int\hspace{-0.05cm }d x \hspace{0.05cm} \big(|u_j(x)|^2 + |v_j(x)|^2\big) = \sum\nolimits_j^{\prime} 1\hspace{0.05cm}.
\label{Emfsimplify}
\end{equation}
Using these results in Eq.~\eqref{Emf1}, we find
\begin{equation}
E = \sum\nolimits_j^{\prime} \Big(2 (\epsilon_j + h)\hspace{0.05cm} n_{\mbox{\tiny{F}}} (\epsilon_j + h) - h\Big) - g_{\mbox{\tiny{1D}}}^{-1} \int \hspace{-0.1cm} dx \hspace{0.05cm} \big(\Delta_0(x)\big)^2\hspace{0.05cm}.
\label{Emfsimplified1}
\end{equation}
One can show that $E$ is an even function of $h$, using the identity $n_{\mbox{\tiny{F}}} (-\epsilon) = 1 - n_{\mbox{\tiny{F}}} (\epsilon)$ and the fact that the spectrum is symmetric for positive and negative energies. Thus we can write
\begin{equation}
E = \sum\nolimits_j^{\prime} \Big(2 (|h| - \epsilon_j)\hspace{0.05cm} n_{\mbox{\tiny{F}}} (|h| - \epsilon_j) - |h|\Big) - g_{\mbox{\tiny{1D}}}^{-1} \int \hspace{-0.1cm} dx \hspace{0.05cm} \big(\Delta_0(x)\big)^2\hspace{0.05cm}.
\label{Emfsimplified2}
\end{equation}
In the limit of zero temperature and large system size ($L\to\infty$), Eq.~\eqref{Emfsimplified2} gives an energy density
\begin{equation}
\mathcal{E}\equiv\frac{E}{L} = -2 \int _{|h|}^{\epsilon_{k_c}} \hspace{-0.05cm}\epsilon \hspace{0.05cm}\rho(\epsilon)\hspace{0.05cm} d\epsilon - \frac{|h|}{\pi} \int_{\epsilon_k < |h|}\hspace{-0.05cm} dk - \frac{g_{\mbox{\tiny{1D}}}^{-1}}{L} \int\hspace{-0.05cm} dx \hspace{0.05cm}(\Delta_0(x))^2\hspace{0.05cm},
\label{energydensity}
\end{equation}
where $\rho(\epsilon)$ represents the density of states, and we have assumed that $\Delta_0(x)$ is either uniform or periodic, such that the spectrum can be labeled by quasimomenta $k$, with $\epsilon_k \geq 0\;\forall\;k$.

Next we apply Eq.~\eqref{energydensity} to calculate energy densities of different stationary states. We start with the Normal state where $\Delta_0(x) = 0$. The spectrum is given by $\epsilon_k = k_{\text{F}} |k|$ (see Eq.~\eqref{uniformAndreev}). Hence $\rho(\epsilon) \equiv (1/\pi) |dk/d\epsilon| = 1/(\pi k_{\text{F}})$. Using these expressions in Eq.~\eqref{energydensity} yields an energy density
\begin{align}
\mathcal{E}_{\text{N}} &= -\frac{1}{\pi k_{\text{F}}} \big(k_{\text{F}}^2 k_c^2 + h^2\big) = -\frac{1}{\pi k_{\text{F}}} \Big[\Delta_{\text{BCS}}^2 \sinh^2 \Big(\frac{\pi}{2} k_{\text{F}} a_{\mbox{\tiny{1D}}}\Big) + h^2\Big]\hspace{0.05cm},
\label{ENormal}\\
\text{or}\quad \bar{\mathcal{E}}_{\text{N}} &\equiv \frac{\mathcal{E}_{\text{N}}}{k_{\text{F}} \epsilon_{\text{F}}} = -\frac{1}{2\pi} \Big[(\Delta_{\text{BCS}}/\epsilon_{\text{F}})^2 \sinh^2 \Big(\frac{\pi}{2} k_{\text{F}} a_{\mbox{\tiny{1D}}}\Big) + (h/\epsilon_{\text{F}})^2\Big]\hspace{0.05cm},
\label{EtildeNormal}
\end{align}
where we have used Eq.~\eqref{kc} for $k_c$, and defined a rescaled energy density $\bar{\mathcal{E}} \equiv \mathcal{E}/(k_{\text{F}} \epsilon_{\text{F}})$. The parameter $\Delta_{\text{BCS}}/\epsilon_{\text{F}}$ is a function of $k_{\text{F}} a_{\mbox{\tiny{1D}}}$ as given in Eq.~\eqref{DeltaBCS}. Note that the dependence of $\bar{\mathcal{E}}_{\text{N}}$ on the interaction strength $k_{\text{F}} a_{\mbox{\tiny{1D}}}$ is a consequence of the finite cutoff $k_c$ in the Andreev approximation. In the full model, the normal state has an energy density $\bar{\mathcal{E}}_{\text{N}}^{\text{Full}} = -(2/3\pi)\big[(1+h/\epsilon_{\text{F}})^{3/2}+(1-h/\epsilon_{\text{F}})^{3/2}\big] = -(1/6\pi)[8+3(h/\epsilon_{\text{F}})^2] - \mathcal{O}((h/\epsilon_{\text{F}})^4)$ for $|h|\leq\epsilon_{\text{F}}$. Thus we see that the introduction of the cutoff in the Andreev approximation renormalizes the energy of the Normal state, such that $\smash{\bar{\mathcal{E}}^{\text{Full}}=\bar{\mathcal{E}} + \Lambda}$, where the shift $\Lambda$ depends on $k_{\text{F}} a_{\mbox{\tiny{1D}}}$. We find that for $k_{\text{F}} a_{\mbox{\tiny{1D}}} \gtrsim 1$ and $|h|\ll \epsilon_{\text{F}}$, the energy of a uniform superfluid experiences the same renormalization, and $\Lambda$ is irrelevant for comparing the energies of different states with one another. Therefore we will drop this term in the following. 

For a uniform superfluid phase with $\Delta_0(x) = \Delta_u \geq 0$, the spectrum is given by $\epsilon_k = \sqrt{(k_{\text{F}} k)^2 + \Delta_u^2}$. Thus,
\begin{equation}
\rho(\epsilon) \equiv \frac{1}{\pi} \bigg|\frac{d k}{d\epsilon}\bigg| = \frac{1}{\pi k_{\text{F}}} \frac{\epsilon}{\sqrt{\epsilon^2 - \Delta_u^2}} \hspace{0.05cm}\Theta(\epsilon - \Delta_u)\hspace{0.05cm}.
\label{uniformdos}
\end{equation}
Substituting this result in Eq.~\eqref{energydensity} and using $g_{\mbox{\tiny{1D}}} = -2/a_{\mbox{\tiny{1D}}}$, we find
\begin{align}
\hspace{-0.15cm}\mathcal{E}_u &= -\frac{2}{\pi k_{\text{F}}} \int_{\text{Max}(|h|,\Delta_u)}^{\sqrt{(k_{\text{F}} k_c)^2 + \Delta_u^2}} \frac{\epsilon^2 d\epsilon}{\sqrt{\epsilon^2 - \Delta_u^2}} - \frac{|h|}{\pi} \int_{\sqrt{(k_{\text{F}} k)^2 + \Delta_u^2} < |h|} dk + \frac{\Delta_u^2 a_{\mbox{\tiny{1D}}}}{2}
\label{uniformenergyintegral}\\
&= -\frac{1}{\pi k_{\text{F}}} \bigg[k_{\text{F}} k_c \sqrt{(k_{\text{F}} k_c)^2 + \Delta_u^2} + \Delta_u^2 \hspace{0.05cm}\sinh^{-1} \frac{k_{\text{F}} k_c}{\Delta_u} +\hspace{-0.05cm} \bigg(\hspace{-0.05cm }|h| \sqrt{h^2 - \Delta_u^2} -\Delta_u^2 \cosh^{-1} \frac{|h|}{\Delta_u}\bigg)\hspace{0.02cm} \Theta(|h|-\Delta_u) \bigg] \hspace{-0.05cm}+ \frac{\Delta_u^2 a_{\mbox{\tiny{1D}}}}{2}\hspace{0.05cm},
\label{uniformenergy}
\end{align}
where $k_{\text{F}} k_c = \Delta_{\text{BCS}} \sinh(\pi k_{\text{F}} a_{\mbox{\tiny{1D}}}/2)$ from Eq.~\eqref{kc}. The stationary states correspond to local extrema of $\mathcal{E}_u$, i.e., $d\mathcal{E}_u/d\Delta_u = 0$. For $|h| < (1-e^{-\pi k_{\text{F}} a_{\mbox{\tiny{1D}}}}) \Delta_{\text{BCS}}/2$, there is a maximum at $\Delta_u = 0$ (the Normal state) and a minimum at $\Delta_u = \Delta_{\text{BCS}}$ (the `BCS' state). For $|h| > (1-e^{-\pi k_{\text{F}} a_{\mbox{\tiny{1D}}}}) \Delta_{\text{BCS}}/2$, the Normal state turns into a minimum, and a new maximum appears at $\Delta_u = (h^2 - (\Delta_{\text{BCS}} - |h|)^2 \tanh^2 (\pi k_{\text{F}} a_{\mbox{\tiny{1D}}}/2))^{1/2}$, which represents the unstable Sarma phase. As $|h|$ is increased, the Sarma maximum approaches the BCS minimum, and the two annihilate at $|h| = \Delta_{\text{BCS}}$. For larger values of $|h|$, only the Normal phase minimum survives. The energy of the `BCS' state can be obtained by setting $\Delta_u = \Delta_{\text{BCS}}>|h|$ in Eq.~\eqref{uniformenergy}, yielding
\begin{equation}
\mathcal{E}_{\text{BCS}} = -\frac{1}{2\pi k_{\text{F}}} \Delta_{\text{BCS}}^2 \sinh(\pi k_{\text{F}} a_{\mbox{\tiny{1D}}})\hspace{0.05cm},\quad\text{or}\quad \bar{\mathcal{E}}_{\text{BCS}} = -\frac{1}{4\pi} (\Delta_{\text{BCS}}/\epsilon_{\text{F}})^2 \sinh(\pi k_{\text{F}} a_{\mbox{\tiny{1D}}})\hspace{0.05cm}.
\label{BCSenergy}
\end{equation}
Note that $\Delta_{\text{BCS}}/\epsilon_{\text{F}}$ only depends on $k_{\text{F}} a_{\mbox{\tiny{1D}}}$ [Eq.~\eqref{DeltaBCS}], so $\bar{\mathcal{E}}_{\text{BCS}}$ is independent of $h$. Comparing Eqs.~\eqref{ENormal} and \eqref{BCSenergy} we find that $\mathcal{E}_N < \mathcal{E}_{\text{BCS}}$ for $h^2>(1-e^{-\pi k_{\text{F}} a_{\mbox{\tiny{1D}}}}) \Delta_{\text{BCS}}^2/2$.

For a soliton train phase with period $2\pi/k_0$ and sharpness parameter $k_1$ (see Sec.~\ref{stationary solution}),
\begin{equation}
\Delta_0(x) = (2 k_{\text{F}} k_0 k_1 K(k_1) /\pi) \hspace{0.1cm}\text{sn}\big(2 K(k_1) k_0 x/\pi, k_1\big)\hspace{0.05cm}.
\label{Delta0x}
\end{equation}
The quasiparticle spectrum and the density of states are given in Eqs.~\eqref{quasiparticlespectrum} and \eqref{DOS}. In terms of rescaled quantities we defined earlier, the energy density in Eq.~\eqref{energydensity} can be expressed as
\begin{equation}
\bar{\mathcal{E}}_{\text{ST}} = -4 \hspace{0.05cm}\frac{k_0^2}{k_{\text{F}}^2} \bigg[\int_{|\tilde{h}|}^{\tilde{\epsilon}_c} \tilde{\epsilon} \hspace{0.05cm}\tilde{\rho}(\tilde{\epsilon}) \hspace{0.05cm} d\tilde{\epsilon} + \frac{1}{\pi}\hspace{0.05cm}\tilde{k}_{|\tilde{h}|} |\tilde{h}| - \frac{k_{\text{F}} a_{\mbox{\tiny{1D}}}}{8 \pi}\hspace{0.05cm} k_0 \hspace{-0.05cm}\int_{-\pi\hspace{-0.03cm}/\hspace{-0.03cm}k_0}^{\pi\hspace{-0.03cm}/\hspace{-0.03cm}k_0}\hspace{-0.03cm}d x\hspace{0.05cm} (\tilde{\Delta}_0 (x))^2 \bigg].
\label{solitontrainenergyintegral}
\end{equation}
Here $\bar{\mathcal{E}}_{\text{ST}} \equiv \mathcal{E}_{\text{ST}}/(k_{\text{F}}\epsilon_{\text{F}})$, $\tilde{k} \equiv k/k_0$, $(\tilde{\epsilon}, \tilde{h}, \tilde{\Delta}_0(x)) \equiv (\epsilon, h, \Delta_0(x))/(k_{\text{F}} k_0)$, $\tilde{\epsilon}_c \equiv \tilde{\epsilon}_{\tilde{k}_c}$, and $k_{|h|}$ denotes the non-negative quasimomentum such that $\epsilon_{k_{|h|}} = |h|$. From Eq.~\eqref{quasiparticlespectrum} we see that
\begin{equation}
\tilde{k}_{|h|} = \frac{|\tilde{h}|}{\pi \tilde{\epsilon}_+} \text{Re} \hspace{-0.05cm} \left[ \sqrt{\frac{\tilde{\epsilon}_-^2 - \tilde{h}^2}{\tilde{\epsilon}_+^2 - \tilde{h}^2}} \hspace{0.1cm}\Pi \hspace{-0.05cm} \left(\frac{\tilde{\epsilon}_+^2 - \tilde{\epsilon}_-^2}{\tilde{\epsilon}_+^2 - \tilde{h}^2}, \sqrt{1-\frac{\tilde{\epsilon}_-^2}{\tilde{\epsilon}_+^2}}\right) \right],
\label{khtilde}
\end{equation}
where $\tilde{\epsilon}_{\pm} \equiv (1\pm k_1) K(k_1)/\pi$. Using $\tilde{\rho}(\tilde{\epsilon})$ from Eq.~\eqref{DOS} we find
\begin{equation}
\hspace{-0.1cm}\int_{|\tilde{h}|}^{\tilde{\epsilon}_c} \tilde{\epsilon} \hspace{0.05cm}\tilde{\rho}(\tilde{\epsilon}) \hspace{0.05cm} d\tilde{\epsilon} = \frac{1}{2\pi} \text{Re}\Bigg[ \big(\tilde{\epsilon}_+^2 + \tilde{\epsilon}_-^2 - 2\tilde{\epsilon}_g^2\big) \ln\hspace{-0.05cm} \left(\hspace{-0.05cm}\frac{(\tilde{\epsilon}_c^2 - \tilde{\epsilon}_-^2)^{\frac{1}{2}} + (\tilde{\epsilon}_c^2 - \tilde{\epsilon}_+^2)^{\frac{1}{2}}}{(\tilde{\epsilon}_-^2 - \tilde{h}^2)^{\frac{1}{2}} + (\tilde{\epsilon}_+^2 - \tilde{h}^2)^{\frac{1}{2}}}\hspace{-0.05cm}\right) + ((\tilde{\epsilon}_c^2 - \tilde{\epsilon}_-^2) (\tilde{\epsilon}_c^2 - \tilde{\epsilon}_+^2))^{\frac{1}{2}} + (\tilde{\epsilon}_-^2 - \tilde{h}^2)^{\frac{1}{2}} (\tilde{\epsilon}_+^2 - \tilde{h}^2)^{\frac{1}{2}}\Bigg],
\label{DOSintegral}
\end{equation}
where $\tilde{\epsilon}_g^2 \equiv \tilde{\epsilon}_+^2 \hspace{0.05cm} E\big((1-\tilde{\epsilon}_-^2/\tilde{\epsilon}_+^2)^{\frac{1}{2}}\big) / K\big((1-\tilde{\epsilon}_-^2/\tilde{\epsilon}_+^2)^{\frac{1}{2}}\big)$. Finally, integrating over $(\Delta_0(x))^2$ we find
\begin{equation}
k_0 \hspace{-0.05cm}\int_{-\pi\hspace{-0.03cm}/\hspace{-0.03cm}k_0}^{\pi\hspace{-0.03cm}/\hspace{-0.03cm}k_0}\hspace{-0.03cm}d x\hspace{0.05cm} (\tilde{\Delta}_0 (x))^2 = \frac{8}{\pi}\hspace{0.05cm} K(k_1) \hspace{0.05cm}\big(K(k_1) - E(k_1)\big)\hspace{0.05cm}.
\label{Delta0xsqintegral}
\end{equation}
Substituting Eqs.~\eqref{khtilde}, \eqref{DOSintegral}, and \eqref{Delta0xsqintegral} into Eq.~\eqref{solitontrainenergyintegral} yields the energy density of a soliton train phase. The stationary states with a given soliton spacing ($k_0/k_{\text{F}}$) are obtained by extremizing $\bar{\mathcal{E}}_{\text{ST}}$ with respect to $k_1$, or equivalently, by solving Eq.~\eqref{selfconsistencyfinal} (see Fig.~\ref{stationarysolutions}). For the C-FFLO phase with $\tilde{\epsilon}_- < |\tilde{h}| < \tilde{\epsilon}_+$, the expression for $\bar{\mathcal{E}}_{\text{ST}}$ simplifies to
\begin{equation}
\bar{\mathcal{E}}_{\text{C-FFLO}} = \bar{\mathcal{E}}_0 - (k_0/\pi k_{\text{F}}) (|h|/\epsilon_{\text{F}})\hspace{0.05cm},
\label{energycfflo}
\end{equation}
where $\bar{\mathcal{E}}_0$ is independent of $h$ (as is $k_1$, see Eq.~\eqref{selfconsistencyfinal}),
\begin{equation}
\bar{\mathcal{E}}_0 = \frac{2}{\pi} \frac{k_0^2}{k_{\text{F}}^2} \Bigg[\frac{2}{\pi} k_{\text{F}} a_{\mbox{\tiny{1D}}}\hspace{0.05cm} K(k_1) \hspace{0.05cm}\big(K(k_1) - E(k_1)\big) - \big(\tilde{\epsilon}_+^2 + \tilde{\epsilon}_-^2 - 2\tilde{\epsilon}_g^2\big) \ln\hspace{-0.05cm} \left(\hspace{-0.05cm}\frac{(\tilde{\epsilon}_c^2 - \tilde{\epsilon}_-^2)^{\frac{1}{2}} + (\tilde{\epsilon}_c^2 - \tilde{\epsilon}_+^2)^{\frac{1}{2}}}{(\tilde{\epsilon}_+^2 - \tilde{\epsilon}_-^2)^{\frac{1}{2}}}\hspace{-0.05cm}\right) - ((\tilde{\epsilon}_c^2 - \tilde{\epsilon}_-^2) (\tilde{\epsilon}_c^2 - \tilde{\epsilon}_+^2))^{\frac{1}{2}}\Bigg].
\end{equation}
For a given value of $k_0/k_{\text{F}}$, both the balanced soliton train ($n_s=0$) and the C-FFLO phase ($n_s = 1$) represent local minima of $\bar{\mathcal{E}}_{\text{ST}}(k_1)$, whereas the phase with $n_s > 1$ represents a maximum lying between the C-FFLO and the Normal phase minima, thus forming an analog of the unstable Sarma phase. The incommensurate FFLO phases with $0<n_s<1$ come in both varieties (maximum/minimum). By comparing the energies of these phases with the uniform states [Eqs.~\eqref{EtildeNormal}, \eqref{BCSenergy}, and \eqref{solitontrainenergyintegral}], we arrive at the phase diagram shown in Fig. 4 of the main article.

If $k_0/k_{\text{F}}$ is allowed to vary, only the `BCS', Normal, and C-FFLO phases remain as local energy minima in the higher-dimensional space. A direct comparison of their energies reveal that the ground state changes from `BCS' for $|h|<(2/\pi)\Delta_{\text{BCS}}$ to C-FFLO for $|h|>(2/\pi)\Delta_{\text{BCS}}$ via a second-order phase transition \cite{smertsching1981incommensurate, smachida1984superconductivity}. As $|h|$ is increased further, more nodes are introduced in the C-FFLO ground state to host the excess fermions. In an exact Bethe ansatz calculation, the ground state eventually changes from FFLO to a fully polarized state for $|h| \gtrsim \epsilon_{\text{F}}$ \cite{sorso2007attractive}. However, in the Andreev approximation, which is valid for $|h| \ll \epsilon_{\text{F}}$, a fully polarized state is never the ground state. It has an energy $\mathcal{E}_P = -(\Delta_{\text{BCS}} \sinh(\pi k_{\text{F}} a_{\mbox{\tiny{1D}}}/2) + |h|)^2 / (2\pi k_{\text{F}}) \geq \mathcal{E}_N$.

\section{\label{integralequation}Integral equations for collective modes}
Here we derive a pair of integral equations describing the collective modes of the order parameter in the Andreev approximation (Eqs. (4) and (5) in the main article) by linearizing the dynamics about the stationary solution.

We start from the time-dependent BdG equations (see Sec.~\ref{stationary solution})
\begin{align}
i \partial_t \left(\hspace{-0.1cm}\begin{array}{c} U^{\pm}_j (x,t)\\ V^{\pm}_j (x,t) \end{array}\hspace{-0.1cm}\right) &= 
\begin{pmatrix}  \mp i k_{\mbox{\tiny{F}}} \partial_x + h & \Delta(x,t) \\ \Delta^*(x,t) & \pm i k_{\mbox{\tiny{F}}} \partial_x + h\end{pmatrix} 
\left(\hspace{-0.1cm}\begin{array}{c} U^{\pm}_j (x,t) \\ V^{\pm}_j (x,t) \end{array}\hspace{-0.1cm}\right) \hspace{0.05cm},
\label{tBdG1}\\
\text{with}\quad\Delta(x,t) &= g_{\mbox{\tiny{1D}}} \sideset{}{'} \sum_{s=\pm,j} n_{\mbox{\tiny{F}}} (\epsilon_j + h) \hspace{0.05cm} U^{s}_j (x,t) V^{s *}_j (x,t)\hspace{0.05cm},
\label{tBdG2}
\end{align}
where the prime on the summation imposes the high-energy cutoff. We substitute $\Delta(x,t) = \Delta_0 (x) +\delta\Delta(x,t)$ and $(U_j^{\pm}(x,t), V_j^{\pm}(x,t)) = \big(u^{\pm}_j(x) + \delta u^{\pm}_j (x,t), v^{\pm}_j(x) + \delta v^{\pm}_j (x,t)\big) \hspace{0.05cm} e^{-i(\epsilon_j + h)t}$ into Eqs.~\eqref{tBdG1} and \eqref{tBdG2}, and retain terms which are linear in the fluctuations, yielding
\begin{align}
i \partial_t \left(\hspace{-0.1cm}\begin{array}{c} \delta u^{\pm}_j (x,t)\\ \delta v^{\pm}_j (x,t) \end{array}\hspace{-0.1cm}\right) &= 
\begin{pmatrix}  \mp i k_{\mbox{\tiny{F}}} \partial_x - \epsilon_j & \Delta_0 (x) \\ \Delta_0 (x) & \pm i k_{\mbox{\tiny{F}}} \partial_x - \epsilon_j\end{pmatrix} 
\left(\hspace{-0.1cm}\begin{array}{c} \delta u^{\pm}_j (x,t)\\ \delta v^{\pm}_j (x,t) \end{array}\hspace{-0.1cm}\right) + \left(\hspace{-0.1cm}\begin{array}{c} \delta \Delta (x,t) \hspace{0.05cm} v^{\pm}_j (x)\\ \delta \Delta^* (x,t) \hspace{0.05cm} u^{\pm}_j (x) \end{array}\hspace{-0.1cm}\right) \hspace{0.05cm},
\label{linearizedtBdG1}\\
\text{and}\quad \delta\Delta (x,t) &= g_{\mbox{\tiny{1D}}} \sideset{}{'}\sum_{s=\pm,j} n_{\mbox{\tiny{F}}} (\epsilon_j + h)\hspace{0.05cm} \big(u^s_j (x) \hspace{0.05cm}\delta v^{s *}_j (x,t) + v^{s *}_j (x) \hspace{0.05cm}\delta u^s_j (x,t)\big)\hspace{0.05cm}.
\label{linearizedtBdG2}
\end{align}
Next we decouple the fluctuations into frequency components by writing $\delta\Delta(x,t) = e^{-\eta t} \big(\delta\Delta_+ (x) \hspace{0.05cm}e^{i\omega t} + \delta\Delta_- (x) \hspace{0.05cm} e^{-i\omega t}\big)$, $\delta u^s_j (x,t) = e^{-\eta t} \big(\delta u^s_{j,+} (x)\hspace{0.05cm} e^{i \omega t} + \delta u^s_{j,-} (x) \hspace{0.05cm}e^{-i\omega t}\big)$, and $\delta v^s_j (x,t) = e^{-\eta t} \big(\delta v^s_{j,+} (x)\hspace{0.05cm} e^{i \omega t} + \delta v^s_{j,-} (x) \hspace{0.05cm}e^{-i\omega t}\big)$ where $\eta,\omega\in\mathbb{R}$. Using these expressions in Eqs.~\eqref{linearizedtBdG1} and \eqref{linearizedtBdG2}, we find
\begin{align}
\begin{pmatrix}  - i s k_{\mbox{\tiny{F}}} \partial_x - \epsilon_j \pm \omega + i\eta & \Delta_0 (x) \\ \Delta_0 (x) & i s k_{\mbox{\tiny{F}}} \partial_x - \epsilon_j \pm\omega + i\eta \end{pmatrix} 
\left(\hspace{-0.1cm}\begin{array}{c} \delta u^{s}_{j,\pm} (x)\\ \delta v^{s}_{j,\pm} (x) \end{array}\hspace{-0.1cm}\right) + \left(\hspace{-0.1cm}\begin{array}{c} \delta \Delta_{\pm} (x) \hspace{0.05cm} v^{s}_j (x)\\ \delta \Delta_{\mp}^* (x) \hspace{0.05cm} u^{s}_j (x) \end{array}\hspace{-0.1cm}\right)& = 0 \quad\text{where}\; s=\pm\hspace{0.05cm},
\label{linearizedfrequencycomponents1}\\
\text{and} \quad \delta\Delta_{\pm}(x) = g_{\mbox{\tiny{1D}}} \sideset{}{'}\sum_{s=\pm,j} n_{\mbox{\tiny{F}}} (\epsilon_j + h)\hspace{0.05cm} \big(u^s_j (x) \hspace{0.05cm}\delta v^{s *}_{j,\mp} (x) + v^{s *}_j (x) \hspace{0.05cm}\delta u^s_{j,\pm} (x)\big)&\hspace{0.05cm}.
\label{linearizedfrequencycomponents2}
\end{align}
We then use the completeness of the stationary wavefunctions to express $\delta u^{s}_{j,\pm} (x)$ and $\delta v^{s}_{j,\pm} (x)$ in terms of $\delta \Delta_{\pm} (x)$ from Eq.~\eqref{linearizedfrequencycomponents1}. Specifically, we write $(\delta u^{s}_{j,\pm} (x), \delta v^{s}_{j,\pm} (x)) = \sum_{j^{\prime}} c^s_{j j^{\prime},\pm} (u^s_{j^{\prime}}(x), v^s_{j^{\prime}}(x))$, and use Eq.~\eqref{stationaryBdGAndreev} to obtain
\begin{align}
\sum_{j^{\prime}} c^s_{j j^{\prime},\pm} (\epsilon_{j^{\prime}} - \epsilon_j &\pm \omega + i\eta) \left(\hspace{-0.1cm}\begin{array}{c} u^s_{j^{\prime}}(x)\\ v^s_{j^{\prime}}(x) \end{array}\hspace{-0.1cm}\right) + \left(\hspace{-0.1cm}\begin{array}{c} \delta \Delta_{\pm} (x) \hspace{0.05cm} v^{s}_j (x)\\ \delta \Delta_{\mp}^* (x) \hspace{0.05cm} u^{s}_j (x) \end{array}\hspace{-0.1cm}\right) = 0\hspace{0.05cm},
\label{cjjprime1}\\
\text{or}\quad c^s_{j j^{\prime},\pm} = \frac{1}{\epsilon_j - \epsilon_{j^{\prime}} \mp \omega -i\eta} &\int dx \hspace{0.05cm}\big(v^s_j (x) \hspace{0.05cm} u^{s *}_{j^{\prime}} (x) \hspace{0.05cm} \delta\Delta_{\pm}(x) + u^s_j (x) \hspace{0.05cm} v^{s *}_{j^{\prime}} (x) \hspace{0.05cm} \delta\Delta_{\mp}^* (x)\big)\hspace{0.05cm}, \quad\text{for}\; s=\pm\hspace{0.05cm}.
\label{cjjprime2}
\end{align}
We have taken the inner product with $(u^{s *}_{j^{\prime}} (x)\;\;v^{s *}_{j^{\prime}} (x))$ from the left on Eq.~\eqref{cjjprime1}, and used the orthonormality of the stationary wavefunctions to arrive at Eq.~\eqref{cjjprime2}. Substituting the expression for $c^s_{j j^{\prime},\pm}$ into the self-consistency condition in Eq.~\eqref{linearizedfrequencycomponents2} yields a pair of coupled homogeneous integral equations for $\delta\Delta_{\pm}(x)$,
\begin{align}
\delta\Delta_{\pm}(x) &= g_{\mbox{\tiny{1D}}} \sum_{s=\pm}\sideset{}{'}\sum_j\sum_{j^{\prime}} n_{\mbox{\tiny{F}}} (\epsilon_j + h) \hspace{0.05cm}\big(u^s_j (x)\hspace{0.05cm} v^{s *}_{j^{\prime}}(x) \hspace{0.05cm} c^{s *}_{j j^{\prime},\mp} + v^{s *}_j (x) \hspace{0.05cm}u^s_{j^{\prime}} (x) \hspace{0.05cm}c^s_{j j^{\prime},\pm} \big)
\label{integraleqndeltaDeltaplusminusintermediate}\\
\nonumber &= \int \hspace{-0.05cm} d x^{\prime} \bigg[ g_{\mbox{\tiny{1D}}} \sideset{}{'}\sum_{s=\pm,j}\sum_{j^{\prime}} n_{\mbox{\tiny{F}}} (\epsilon_j + h) \left(\frac{u^s_j (x) v^{s *}_{j^{\prime}}(x) u^{s *}_{j} (x^{\prime}) v^s_{j^{\prime}} (x^{\prime})}{\epsilon_j - \epsilon_{j^{\prime}} \pm \omega + i\eta} + \frac{v^{s *}_j (x) u^{s}_{j^{\prime}}(x) v^{s}_{j} (x^{\prime}) u^{s *}_{j^{\prime}} (x^{\prime})}{\epsilon_j - \epsilon_{j^{\prime}} \mp \omega - i\eta} \right)\hspace{-0.05cm}\bigg] \hspace{0.05cm}\delta\Delta_{\pm}(x^{\prime})
\\
&+ \int \hspace{-0.05cm} d x^{\prime} \bigg[ g_{\mbox{\tiny{1D}}} \sideset{}{'}\sum_{s=\pm,j}\sum_{j^{\prime}} n_{\mbox{\tiny{F}}} (\epsilon_j + h) \left(\frac{u^s_j (x) v^{s *}_{j^{\prime}}(x) v^{s *}_{j} (x^{\prime}) u^s_{j^{\prime}} (x^{\prime})}{\epsilon_j - \epsilon_{j^{\prime}} \pm \omega + i\eta} + \frac{v^{s *}_j (x) u^{s}_{j^{\prime}}(x) u^{s}_{j} (x^{\prime}) v^{s *}_{j^{\prime}} (x^{\prime})}{\epsilon_j - \epsilon_{j^{\prime}} \mp \omega - i\eta} \right)\hspace{-0.05cm}\bigg] \hspace{0.05cm}\delta\Delta_{\mp}^* (x^{\prime}) \hspace{0.05cm}.
\label{integraleqndeltaDeltaplusminus}
\end{align}
Next we use two symmetries of the stationary wavefunctions: (i) if $(u^s_j (x)\;\;v^s_j (x))^T$ is an eigenstate with energy $\epsilon_j$, then $(v^{s *}_j (x)\;\;u^{s *}_j (x))^T$ is also an eigenstate with energy $\epsilon_j$, and (ii) $(u^-_j (x),v^-_j (x)) = (u^+_j (x),v^+_j (x))^*$ (see Sec.~\ref{stationary solution}). These two symmetries let us write Eq.~\eqref{integraleqndeltaDeltaplusminus} as
\begin{align}
\delta\Delta_+ (x) &= \int d x^{\prime} \big(\mathcal{M}_1 (x,x^{\prime};\Omega) \hspace{0.05cm}\delta\Delta_+ (x^{\prime}) + \mathcal{M}_2 (x,x^{\prime};\Omega)  \hspace{0.05cm} \delta\Delta_-^* (x^{\prime})\big)\hspace{0.05cm},
\label{integraleqndeltaDeltaplus}\\
\text{and}\quad \delta\Delta_-^* (x) &= \int d x^{\prime} \big(\mathcal{M}_2 (x,x^{\prime};\Omega) \hspace{0.05cm}\delta\Delta_+ (x^{\prime}) + \mathcal{M}_1 (x,x^{\prime};\Omega)  \hspace{0.05cm} \delta\Delta_-^* (x^{\prime})\big)\hspace{0.05cm},
\label{integraleqndeltaDeltaminus}
\end{align}
where $\Omega \equiv \omega + i \eta$, and
\begin{align}
\mathcal{M}_1(x,x^{\prime};\Omega) = g_{\mbox{\tiny{1D}}} \sideset{}{'}\sum_{j}\sum_{j^{\prime}} n_{\mbox{\tiny{F}}} (\epsilon_j + h) \hspace{0.05cm} \frac{2 (\epsilon_j - \epsilon_{j^{\prime}})}{(\epsilon_j - \epsilon_{j^{\prime}})^2 - \Omega^2} \hspace{0.05cm} \big(u_j^* (x) v_{j^{\prime}}(x) u_{j} (x^{\prime}) v^*_{j^{\prime}} (x^{\prime}) + v_j^* (x) u_{j^{\prime}}(x) v_{j} (x^{\prime}) u^*_{j^{\prime}} (x^{\prime})\big)\hspace{0.05cm},
\label{M1}\\
\mathcal{M}_2(x,x^{\prime};\Omega) = g_{\mbox{\tiny{1D}}} \sideset{}{'}\sum_{j}\sum_{j^{\prime}} n_{\mbox{\tiny{F}}} (\epsilon_j + h) \hspace{0.05cm} \frac{2 (\epsilon_j - \epsilon_{j^{\prime}})}{(\epsilon_j - \epsilon_{j^{\prime}})^2 - \Omega^2} \hspace{0.05cm} \big(u^*_j (x) v_{j^{\prime}}(x) v_{j} (x^{\prime}) u^*_{j^{\prime}} (x^{\prime}) + v^*_j (x) u_{j^{\prime}}(x) u_{j} (x^{\prime}) v^*_{j^{\prime}} (x^{\prime})\big)\hspace{0.05cm}.
\label{M2}
\end{align}
Here all the wavefunctions refer to the right-moving branch. We can express Eqs.~\eqref{integraleqndeltaDeltaplus} and \eqref{integraleqndeltaDeltaminus} in a simpler form by defining $\delta_{p,a} (x) \equiv \delta\Delta_+(x) \mp \delta\Delta_-^* (x)$. Note that the fluctuation of the order parameter is given by
\begin{equation}
\delta\Delta(x,t) = e^{-\eta t} \big(\delta\Delta_+ (x) \hspace{0.05cm}e^{i\omega t} + \delta\Delta_- (x) \hspace{0.05cm} e^{-i\omega t}\big) = \text{Re} \big(\delta_a (x) e^{i \Omega t}\big) + i\hspace{0.05cm} \text{Im} \big(\delta_p (x) e^{i\Omega t}\big)\hspace{0.05cm}.
\label{orderparamaterfluctuation}
\end{equation}
Since $\Delta_0 (x)$ is real, $\delta_p (x)$ and $\delta_a (x)$ describe the phase- and amplitude-fluctuations of the order parameter respectively. From Eqs.~\eqref{integraleqndeltaDeltaplus} and \eqref{integraleqndeltaDeltaminus} we see that the phase and amplitude fluctuations decouple, with
\begin{equation}
\delta_{p,a}(x) = -g_{\mbox{\tiny{1D}}} \int d x^{\prime} \mathcal{M}^{\pm} (x,x^{\prime};\Omega) \hspace{0.05cm} \delta_{p,a}(x^{\prime})\hspace{0.05cm},\quad\text{where}\quad \mathcal{M}^{\pm} \equiv -g_{\mbox{\tiny{1D}}}^{-1}(\mathcal{M}_1 \mp \mathcal{M}_2) \hspace{0.05cm}.
\label{integraleqndeltapa}
\end{equation}
Using Eqs.~\eqref{M1} and \eqref{M2} we find
\begin{align}
\mathcal{M}^{\pm} (x,x^{\prime};\Omega) &= - \sideset{}{'}\sum_{j}\sum_{j^{\prime}} \frac{2 \hspace{0.05cm} n_{\mbox{\tiny{F}}} (\epsilon_j + h) \hspace{0.05cm} (\epsilon_j - \epsilon_{j^{\prime}})}{(\epsilon_j - \epsilon_{j^{\prime}})^2 - \Omega^2} \hspace{0.05cm} \big(u^*_j (x) v_{j^{\prime}}(x) \mp v^*_j (x) u_{j^{\prime}}(x)\big) \big(u_{j} (x^{\prime}) v^*_{j^{\prime}} (x^{\prime}) \mp v_{j} (x^{\prime}) u^*_{j^{\prime}} (x^{\prime})\big)\hspace{0.05cm},
\label{Mpa1}\\
&= \sideset{}{'}\sum_{j}\sum_{j^{\prime}} \frac{2 \hspace{0.05cm} n_{\mbox{\tiny{F}}} (h - \epsilon_j) \hspace{0.05cm} (\epsilon_j + \epsilon_{j^{\prime}})}{(\epsilon_j + \epsilon_{j^{\prime}})^2 - \Omega^2} \hspace{0.05cm} \big(u^*_j (x) u_{j^{\prime}}(x) \pm v^*_j (x) v_{j^{\prime}}(x)\big) \big(u_{j} (x^{\prime}) u^*_{j^{\prime}} (x^{\prime}) \pm v_{j} (x^{\prime}) v^*_{j^{\prime}} (x^{\prime})\big)\hspace{0.05cm}.
\label{Mpa2}
\end{align}
In Eq.~\eqref{Mpa2} we have used the symmetry that for any eigenstate $(u_j (x)\;\;v_j(x))^{T}$ with energy $\epsilon_j$, there is a corresponding eigenstate $(-v_j (x)\;\;u_j(x))^{T}$ with energy $-\epsilon_j$ (see Sec.~\ref{stationary solution}). At zero temperature, $n_{\mbox{\tiny{F}}} (h - \epsilon_j) = \Theta(\epsilon_j - h)$. Then Eq.~\eqref{Mpa2} reduces to
\begin{equation}
\mathcal{M}^{\pm} (x,x^{\prime};\Omega) = \sideset{}{'}\sum_{j}\sum_{j^{\prime}} \frac{2 \hspace{0.05cm} (\epsilon_j + \epsilon_{j^{\prime}})}{(\epsilon_j + \epsilon_{j^{\prime}})^2 - \Omega^2} \hspace{0.05cm}\big(u^*_j (x) u_{j^{\prime}}(x) \pm v^*_j (x) v_{j^{\prime}}(x)\big) \big(u_{j} (x^{\prime}) u^*_{j^{\prime}} (x^{\prime}) \pm v_{j} (x^{\prime}) v^*_{j^{\prime}} (x^{\prime})\big)\hspace{0.05cm},
\label{MpazeroT}
\end{equation}
where the prime on the $j$-summation now stands for the condition $h\leq\epsilon_j<\epsilon_c$, $\epsilon_c$ being the high-energy cutoff. Eqs.~\eqref{integraleqndeltapa} and \eqref{MpazeroT} describe the collective modes of the order parameter about any stationary solution in the Andreev approximation. In particular, they apply to both a soliton train state and a uniform state.

\section{\label{goldstonehiggs}Goldstone and `Higgs' modes}
A soliton train spontaneously breaks both gauge- and translational-symmetries. Therefore, it has two gapless Goldstone modes: a phase mode described by $\delta_p(x) \propto \Delta_0 (x)$ and an amplitude mode described by $\delta_a (x) \propto \Delta_0^{\prime}(x)$ at zero energy. On the other hand, a uniform state breaks only gauge symmetry, and has only one Goldstone mode described by $\delta_p (x) \propto 1$ at zero energy. Below we show that more generally, $\delta_p (x) \propto \Delta_0 (x) \hspace{0.05cm} e^{i q x}$ is a collective mode with dispersion $\omega(q) = k_{\text{F}} q$, describing Anderson-Bogoliubov phonons traveling at the Fermi velocity.

We let $R$ represent the right-hand side of Eq.~\eqref{integraleqndeltapa}. For $\delta_p (x) = \Delta_0 (x) \hspace{0.05cm} e^{i q x}$ and $\Omega = k_{\text{F}} q$, with Eq.~\eqref{Mpa2},
\begin{equation}
R \hspace{-0.02cm}=\hspace{-0.05cm} - g_{\mbox{\tiny{1D}}}\hspace{-0.05cm} \sideset{}{'}\sum_j \sum_{j^{\prime}} n_{\mbox{\tiny{F}}} (h - \epsilon_j)\big(u^*_j (x) u_{j^{\prime}}(x) + v^*_j (x) v_{j^{\prime}}(x)\big) \frac{2 \hspace{0.05cm} (\epsilon_j + \epsilon_{j^{\prime}})}{(\epsilon_j + \epsilon_{j^{\prime}})^2 \hspace{-0.05cm}-\hspace{-0.05cm} (k_{\text{F}} q)^2} \hspace{-0.05cm}\int \hspace{-0.1cm}d x^{\prime} \big(u_{j} (x^{\prime}) u^*_{j^{\prime}} (x^{\prime}) + v_{j} (x^{\prime}) v^*_{j^{\prime}} (x^{\prime})\big) \Delta_0 (x^{\prime}) e^{i q x^{\prime}}\hspace{-0.1cm}.
\label{integraleqnphononmode}
\end{equation}
However, using the BdG equations [Eq.~\eqref{stationaryBdGAndreevright}] we can write
\begin{align}
\nonumber &\int d x^{\prime} \big(u_{j} (x^{\prime}) u^*_{j^{\prime}} (x^{\prime}) + v_{j} (x^{\prime}) v^*_{j^{\prime}} (x^{\prime})\big) \Delta_0 (x^{\prime}) \hspace{0.05cm} e^{i q x^{\prime}} \\
\nonumber &= \int d x^{\prime} u^{*}_{j^{\prime}} (x^{\prime}) \hspace{0.05cm}\big(\epsilon_{j} v_{j} (x^{\prime}) - i k_{\text{F}} \partial_{x^{\prime}} v_{j} (x^{\prime}) \big) \hspace{0.05cm} e^{i q x^{\prime}} + v_{j} (x^{\prime}) \hspace{0.05cm} \big(\epsilon_{j^{\prime}} u_{j^{\prime}}^* (x^{\prime}) - i k_{\text{F}} \partial_{x^{\prime}} u_{j^{\prime}}^* (x^{\prime})\big) \hspace{0.05cm} e^{i q x^{\prime}}\\
\nonumber &= (\epsilon_j + \epsilon_{j^{\prime}} - k_{\text{F}} q) \int d x^{\prime} u^{*}_{j^{\prime}} (x^{\prime}) v_{j} (x^{\prime}) \hspace{0.05cm} e^{i q x^{\prime}} - i k_{\text{F}} \int d x^{\prime} \partial_{x^{\prime}} \big(u^{*}_{j^{\prime}} (x^{\prime}) v_{j} (x^{\prime})  \hspace{0.05cm} e^{i q x^{\prime}}\big) \\
&= (\epsilon_j + \epsilon_{j^{\prime}} - k_{\text{F}} q) \int d x^{\prime} u^{*}_{j^{\prime}} (x^{\prime}) v_{j} (x^{\prime}) \hspace{0.05cm} e^{i q x^{\prime}}. \qquad\text{(using periodic boundary conditions)}
\label{phononmodeproofBdG1}
\end{align}
Similarly, one can show that
\begin{equation}
\int d x^{\prime} \big(u_{j} (x^{\prime}) u^*_{j^{\prime}} (x^{\prime}) + v_{j} (x^{\prime}) v^*_{j^{\prime}} (x^{\prime})\big) \Delta_0 (x^{\prime}) \hspace{0.05cm} e^{i q x^{\prime}} = (\epsilon_j + \epsilon_{j^{\prime}} + k_{\text{F}} q) \int d x^{\prime} v^{*}_{j^{\prime}} (x^{\prime}) u_{j} (x^{\prime}) \hspace{0.05cm} e^{i q x^{\prime}}.
\label{phononmodeproofBdG2}
\end{equation}
Using Eqs.~\eqref{phononmodeproofBdG1} and \eqref{phononmodeproofBdG2} in Eq.~\eqref{integraleqnphononmode}, we get
\begin{equation}
R = - g_{\mbox{\tiny{1D}}} \sum\nolimits_j^{\prime} n_{\mbox{\tiny{F}}} (h - \epsilon_j) \sum\nolimits_{j^{\prime}} \int d x^{\prime} \big(u^*_j (x) u_{j^{\prime}}(x) + v^*_j (x) v_{j^{\prime}}(x)\big) \big(u^{*}_{j^{\prime}} (x^{\prime}) v_{j} (x^{\prime}) + v^{*}_{j^{\prime}} (x^{\prime}) u_{j} (x^{\prime})\big) \hspace{0.05cm} e^{i q x^{\prime}}.
\label{R2}
\end{equation}
Next we substitute the completeness relations $\sum_{j^{\prime}} u_{j^{\prime}} (x) u^*_{j^{\prime}} (x^{\prime}) = v_{j^{\prime}} (x) v^*_{j^{\prime}} (x^{\prime}) = \delta(x-x^{\prime})$ and $\sum_{j^{\prime}} u_{j^{\prime}} (x) v^*_{j^{\prime}} (x^{\prime}) = 0$ in Eq. \eqref{R2}, and use the self-consistency of the stationary solution [Eq.~\eqref{stationaryBdGAndreevright}] to obtain
\begin{equation}
R = - g_{\mbox{\tiny{1D}}} \sum\nolimits_j^{\prime} n_{\mbox{\tiny{F}}} (h - \epsilon_j) \hspace{0.05cm} \big(u^*_j (x) v_j (x) + v^*_j (x) u_j (x)\big) \hspace{0.05cm} e^{i q x} = \Delta_0 (x) \hspace{0.05cm} e^{i q x}.
\label{phononmodeendproof}
\end{equation}
This result shows that the proposed collective mode indeed satisfies the integral equation [Eq.~\eqref{integraleqndeltapa}]. Note that in applying completeness, we use the fact that the sum over $j^{\prime}$ in Eq.~\eqref{R2} is unrestricted, i.e., it includes states beyond the high-energy cutoff. This is important for obtaining the correct Goldstone modes.

\begin{figure}[h]
\includegraphics[width=0.95\textwidth]{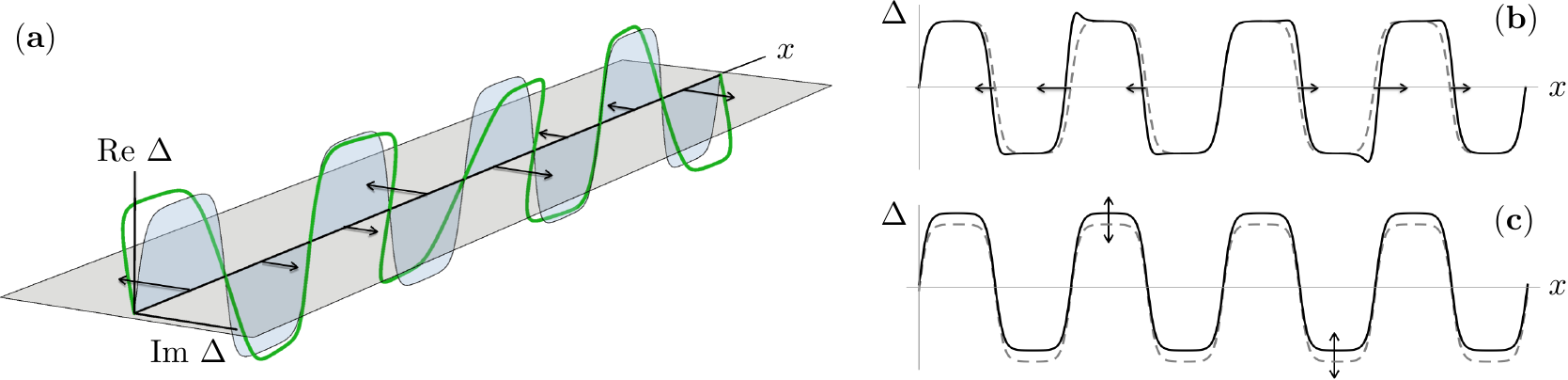}
\caption{\label{goldstonehiggsfig}(Color online.) Goldstone and `Higgs' modes of a fermionic soliton train with sharpness parameter $k_1 = 0.999$: {\bf (a)} Goldstone phase mode with wave-vector $q=k_0/4$, {\bf (b)} Goldstone amplitude mode with $q = k_0/4$, {\bf (c)} `Higgs' mode with $q=k_0$.}
\vspace{-0.05cm}
\end{figure}

Similarly, we can show that $\delta_a (x) \propto \Delta_0^{\prime}(x)$ represents a zero-energy collective mode for any non-uniform state. More generally, we find from numerics on the soliton train and the uniform state that $\delta_a (x) \propto u_{q/2}(x) v_{q/2}(x)$ describes a collective mode with dispersion $\omega(q) = 2 \epsilon (q/2)$, where $\epsilon(k)$ denotes the single-particle spectrum. For the uniform state, this mode describes sinusoidal oscillations of the order-parameter amplitude, also known as the Higgs mode. From Eq.~\eqref{uniformAndreev} we see that the `Higgs' mode has dispersion $\omega(q) = 2\sqrt{(k_{\text{F}} q/2)^2 + \Delta_{\text{BCS}}^2}$. For the soliton train, the mode has two branches: a gapless branch with $q<k_0$, and a gapped branch with $q>k_0$. The gapless branch represents a Goldstone amplitude mode, approaching $\delta_a (x) \propto \Delta_0^{\prime}(x)$ as $q\to 0$, whereas the gapped branch represents the remnant of the `Higgs' mode. The dispersion can be found using the single-particle spectrum given in Eq.~\eqref{quasiparticlespectrum}. In particular, the dispersion is linear as $q\to 0$, with a group velocity $d\omega/d q = k_{\text{F}} (\epsilon_+ \epsilon_-/\epsilon_g^2)$, where $\epsilon_{\pm}$ and $\epsilon_{g}$ are defined in Sec.~\ref{stationary solution}. Figure~\ref{goldstonehiggsfig} shows the fluctuations described by the Goldstone and `Higgs' modes of a soliton train.

\section{\label{matrixequations}Matrix equations for collective mode spectrum of a soliton train}
Here we derive matrix equations describing the collective modes of a soliton train at zero temperature. We utilize features of the stationary solution outlined in Sec.~\ref{stationary solution} to find simplified expressions suited for numerics.

The collective modes have a Brillouin zone structure as $\Delta_0(x)$ is periodic. Further, the Brillouin zone has size $2 k_0$ because of the additional symmetry $\Delta_0(x+\pi/k_0) = -\Delta_0 (x)$ \cite{sedge2009signature}. Thus we can write $\delta_{p,a}(x) = e^{i q x} \sum_{n} C^{\pm}_{n} e^{i n k_0 x}$, where $n$ varies over odd integers, and $-k_0 \leq q \leq k_0$. Substituting this Fourier expansion into Eq.~\eqref{integraleqndeltapa}, we get
\begin{align}
C^{\pm}_{n} &= - g_{\mbox{\tiny{1D}}} \sum\nolimits_m M^{\pm}_{nm} (q,\Omega) \hspace{0.05cm} C^{\pm}_m\hspace{0.05cm},
\label{Cpamatrixeqn}\\
\text{where}\quad M^{\pm}_{nm} (q,\Omega) &\equiv \frac{k_0}{2\pi} \int_{-\pi\hspace{-0.03cm}/\hspace{-0.03cm}k_0}^{\pi\hspace{-0.03cm}/\hspace{-0.03cm}k_0} \hspace{-0.03cm}d x \int \hspace{-0.05cm}d x^{\prime} e^{-i(q+ n k_0) x} e^{i(q+m k_0) x^{\prime}} \mathcal{M}^{\pm}(x,x^{\prime};\Omega)\hspace{0.05cm}.
\label{Mpanmdefinition}
\end{align}
The collective modes represent non-trivial solutions to Eq.~\eqref{Cpamatrixeqn}. Therefore the task is to compute the matrices $M^{\pm}_{nm} (q,\Omega)$. Using the zero-temperature expression for $\mathcal{M}^{\pm}(x,x^{\prime};\Omega)$ [Eq.~\eqref{MpazeroT}] in Eq.~\eqref{Mpanmdefinition}, we find
\begin{align}
\nonumber M^{\pm}_{nm} (q,\Omega) = \frac{k_0}{\pi} \sum\nolimits_j^{\prime} \sum\nolimits_{j^{\prime}} \frac{\epsilon_j + \epsilon_{j^{\prime}}}{(\epsilon_j + \epsilon_{j^{\prime}})^2 - \Omega^2} &\int_{-\pi\hspace{-0.03cm}/\hspace{-0.03cm}k_0}^{\pi\hspace{-0.03cm}/\hspace{-0.03cm}k_0} \hspace{-0.03cm}d x \hspace{0.05cm} e^{-i(q+ n k_0) x} \big(u_{j^{\prime}}(x) u_j^*(x) \pm v_{j^{\prime}}(x) v_j^* (x) \big)\\
&\times \int \hspace{-0.05cm}d x^{\prime} e^{i(q+m k_0) x^{\prime}} \big(u_{j^{\prime}}^* (x^{\prime}) u_j(x^{\prime}) \pm v_{j^{\prime}}^* (x^{\prime}) v_j (x^{\prime}) \big)
\label{Mpanm1}\\
&\hspace{-4.95cm}= \frac{k_0}{\pi} \sum\nolimits_k^{\prime} \sum\nolimits_{k^{\prime}} \left[\frac{\epsilon_k + \epsilon_{k^{\prime}}}{(\epsilon_k + \epsilon_{k^{\prime}})^2 - \Omega^2} \hspace{0.05cm} I^{(1)}_{k k^{\prime}} + \frac{\epsilon_k - \epsilon_{k^{\prime}}}{(\epsilon_k - \epsilon_{k^{\prime}})^2 - \Omega^2} \hspace{0.05cm} I^{(2)}_{k k^{\prime}}\right],
\label{MpanmI}
\end{align}
where
\begin{align}
I^{(1)}_{k k^{\prime}} \equiv \int_{-\pi\hspace{-0.03cm}/\hspace{-0.03cm}k_0}^{\pi\hspace{-0.03cm}/\hspace{-0.03cm}k_0} \hspace{-0.03cm}d x \hspace{0.05cm} e^{-i(q+ n k_0) x} \big(u_{k^{\prime}}(x) u_k^*(x) \pm v_{k^{\prime}}(x) v_k^* (x) \big) \int \hspace{-0.05cm}d x^{\prime} e^{i(q+m k_0) x^{\prime}} \big(u_{k^{\prime}}^* (x^{\prime}) u_k(x^{\prime}) \pm v_{k^{\prime}}^* (x^{\prime}) v_k (x^{\prime}) \big) \hspace{0.05cm},
\label{I1}\\
I^{(2)}_{k k^{\prime}} \equiv \int_{-\pi\hspace{-0.03cm}/\hspace{-0.03cm}k_0}^{\pi\hspace{-0.03cm}/\hspace{-0.03cm}k_0} \hspace{-0.03cm}d x \hspace{0.05cm} e^{-i(q+ n k_0) x} \big(u_{k^{\prime}}(x) v_k^*(x) \mp v_{k^{\prime}}(x) u_k^* (x) \big) \int \hspace{-0.05cm}d x^{\prime} e^{i(q+m k_0) x^{\prime}} \big(u_{k^{\prime}}^* (x^{\prime}) v_k (x^{\prime}) \mp v_{k^{\prime}}^* (x^{\prime}) u_k (x^{\prime})\big) \hspace{0.05cm}.
\label{I2}
\end{align}
Here we have relabeled the sums over $j$ and $j^{\prime}$ in terms of quasimomenta $k$ and $k^{\prime}$ with $\epsilon_k, \epsilon_{k^{\prime}} \geq 0$, and used the symmetry that a negative energy state with quasimomentum $k$ has energy $-\epsilon_k$ and wavefunction $(-v_k (x)\;\;u_k(x))^T$. As before, the prime on the $k$ sum in Eq.~\eqref{MpanmI} stands for $h \leq \epsilon_k < \epsilon_c$, and we consider $h \geq 0$ without loss of generality. Next we use the symmetry $\epsilon_{-k} = \epsilon_{k}$ and $(u_{-k}(x),v_{-k}(x)) = (v_k^*(x), u_k^* (x))\hspace{0.1cm}\forall \hspace{0.05cm}k$ in Eqs.~\eqref{MpanmI} and \eqref{I2} to obtain
\begin{equation}
M^{\pm}_{nm} (q,\Omega) = \frac{k_0}{\pi} \sum\nolimits_k^{\prime} \sum\nolimits_{k^{\prime}} \left[\frac{\epsilon_k + \epsilon_{k^{\prime}}}{(\epsilon_k + \epsilon_{k^{\prime}})^2 - \Omega^2} \hspace{0.05cm} I^{(1)}_{k k^{\prime}} + \frac{\epsilon_k - \epsilon_{k^{\prime}}}{(\epsilon_k - \epsilon_{k^{\prime}})^2 - \Omega^2} \hspace{0.05cm} I^{(2)}_{k^{\prime} k}\right].
\label{MpanmI2}
\end{equation}
Combining Eqs.~\eqref{MpanmI} and \eqref{MpanmI2}, we can write
\begin{equation}
M^{\pm}_{nm} (q,\Omega) = \frac{k_0}{\pi} \sum\nolimits_{k,k^{\prime}}^{\scriptsize{\raisebox{.5pt}{\textcircled{\raisebox{-.6pt} {1}}}}} \frac{\epsilon_k + \epsilon_{k^{\prime}}}{(\epsilon_k + \epsilon_{k^{\prime}})^2 - \Omega^2} \hspace{0.05cm} I^{(1)}_{k k^{\prime}} + \frac{k_0}{\pi} \sum\nolimits_{k,k^{\prime}}^{\scriptsize{\raisebox{.5pt}{\textcircled{\raisebox{-.6pt} {2}}}}} \frac{\epsilon_k - \epsilon_{k^{\prime}}}{(\epsilon_k - \epsilon_{k^{\prime}})^2 - \Omega^2} \hspace{0.05cm} I^{(2)}_{k k^{\prime}}\hspace{0.05cm},
\label{MpanmIseparate}
\end{equation}
where {\small{\raisebox{.5pt}{\textcircled{\raisebox{-.6pt} {1}}}}} stands for the condition $\epsilon_k \in [h,\epsilon_c)$, and {\small{\raisebox{.5pt}{\textcircled{\raisebox{-.6pt} {2}}}}} stands for the condition $\epsilon_k \in [h,\epsilon_c) \land \epsilon_{k^{\prime}} \notin [h,\epsilon_c)$.

Next we note from Eqs.~\eqref{I1} and \eqref{I2} that both $I^{(1)}_{k k^{\prime}}$ and $I^{(2)}_{k k^{\prime}}$ vanish unless $k^{\prime} = k + q + r k_0$ where $r$ is an integer, in which case $I^{(1)}_{k k^{\prime}} = (2\pi/k_0 L) \hspace{0.05cm}\big(\xi^{k^{\prime}\hspace{-0.05cm},k}_{\pm,n}\big)^* \xi^{k^{\prime}\hspace{-0.05cm},k}_{\pm,m}$, and $I^{(2)}_{k k^{\prime}} = (2\pi/k_0 L) \hspace{0.05cm}\big(\chi^{k^{\prime}\hspace{-0.05cm},k}_{\pm,n}\big)^* \chi^{k^{\prime}\hspace{-0.05cm},k}_{\pm,m}$, where
\begin{align}
\xi^{k_1,k_2}_{\pm,m} &\equiv \frac{k_0 L}{2\pi} \int_{-\pi\hspace{-0.03cm}/\hspace{-0.03cm}k_0}^{\pi\hspace{-0.03cm}/\hspace{-0.03cm}k_0} \hspace{-0.03cm}d x \hspace{0.05cm} e^{i(q+m k_0) x} \big(u_{k_1}^* (x) u_{k_2}(x) \pm v_{k_1}^* (x) v_{k_2}(x) \big)\hspace{0.05cm},
\label{xi}\\
\text{and}\quad \chi^{k_1,k_2}_{\pm,m} &\equiv \frac{k_0 L}{2\pi} \int_{-\pi\hspace{-0.03cm}/\hspace{-0.03cm}k_0}^{\pi\hspace{-0.03cm}/\hspace{-0.03cm}k_0} \hspace{-0.03cm}d x \hspace{0.05cm} e^{i(q+m k_0) x} \big(u_{k_1}^* (x) v_{k_2}(x) \mp v_{k_1}^* (x) u_{k_2}(x) \big)\hspace{0.05cm}.
\label{chi}
\end{align}
Using this result in Eq.~\eqref{MpanmIseparate}, we obtain
\begin{align}
\nonumber &M^{\pm}_{nm} (q,\Omega) \\
&= \frac{2}{L} \sum\nolimits_{r,k}^{\scriptsize{\raisebox{.5pt}{\textcircled{\raisebox{-.6pt} {1}}}}} \frac{\epsilon_{k+q+r k_0}  \hspace{-0.1cm}+ \epsilon_{k}}{(\epsilon_{k+q+r k_0} \hspace{-0.1cm}+ \epsilon_{k})^2 - \Omega^2} \hspace{0.05cm} \big(\xi^{k+q+r k_0,k}_{\pm,n}\big)^* \xi^{k+q+r k_0,k}_{\pm,m} - \frac{2}{L} \sum\nolimits_{r,k}^{\scriptsize{\raisebox{.5pt}{\textcircled{\raisebox{-.6pt} {2}}}}} \frac{\epsilon_{k+q+r k_0}  \hspace{-0.1cm}- \epsilon_{k}}{(\epsilon_{k+q+r k_0}  \hspace{-0.1cm}- \epsilon_{k})^2 - \Omega^2} \hspace{0.05cm} \big(\chi^{k+q+r k_0,k}_{\pm,n}\big)^* \chi^{k+q+r k_0,k}_{\pm,m},
\label{MpanmIseparatexichi}
\end{align}
where $r$ varies over integers, {\small{\raisebox{.5pt}{\textcircled{\raisebox{-.6pt} {1}}}}} stands for $\epsilon_k \in [h,\epsilon_c)$ as before, and now {\small{\raisebox{.5pt}{\textcircled{\raisebox{-.6pt} {2}}}}} stands for $\epsilon_k \in [h,\epsilon_c) \land \epsilon_{k+q+r k_0} \notin [h,\epsilon_c)$. Taking the limit $L\to\infty$ in Eq.~\eqref{MpanmIseparatexichi}, we find
\begin{align}
\nonumber &\hspace{-0.4cm} - g_{\mbox{\tiny{1D}}} M^{\pm}_{nm} (q,\Omega) \\
&\hspace{-0.4cm}= \frac{1}{\pi k_{\text{F}} a_{\mbox{\tiny{1D}}}} \sum_r \bigg[\int_{\scriptsize{\raisebox{.5pt}{\textcircled{\raisebox{-.6pt} {1}}}}} \hspace{-0.05cm} d \tilde{k} \frac{\tilde{\epsilon}_{\tilde{k}+\tilde{q}+r}  \hspace{-0.1cm}+ \tilde{\epsilon}_{\tilde{k}}}{(\tilde{\epsilon}_{\tilde{k}+\tilde{q}+r} \hspace{-0.1cm}+ \tilde{\epsilon}_{\tilde{k}})^2 - \tilde{\Omega}^2} \hspace{0.05cm} \big(\xi^{\tilde{k}+\tilde{q}+r,\tilde{k}}_{\pm,n}\big)^* \xi^{\tilde{k}+\tilde{q}+r,\tilde{k}}_{\pm,m} -\hspace{-0.05cm}  \int_{\scriptsize{\raisebox{.5pt}{\textcircled{\raisebox{-.6pt} {2}}}}} \hspace{-0.05cm} d \tilde{k} \frac{\tilde{\epsilon}_{\tilde{k}+\tilde{q}+r}  \hspace{-0.1cm}- \tilde{\epsilon}_{\tilde{k}}}{(\tilde{\epsilon}_{\tilde{k}+\tilde{q}+r} \hspace{-0.1cm}- \tilde{\epsilon}_{\tilde{k}})^2 - \tilde{\Omega}^2} \hspace{0.05cm} \big(\chi^{\tilde{k}+\tilde{q}+r,\tilde{k}}_{\pm,n}\big)^* \chi^{\tilde{k}+\tilde{q}+r,\tilde{k}}_{\pm,m}\bigg].\hspace{-0.2cm}
\label{MpanmIseparatexichirescaled}
\end{align}
Here we have expressed the integrals in rescaled coordinates (see Sec.~\ref{stationary solution}), and used the relation $g_{\mbox{\tiny{1D}}} = -2/a_{\mbox{\tiny{1D}}}$ \cite{solshanii1998atomic, sbergeman2003atom}.

Fortunately, the $\xi^* \xi$ and $\chi^* \chi$ terms in Eq.~\eqref{MpanmIseparatexichirescaled} can be computed in closed form for the stationary wavefunctions in Eqs.~\eqref{freestatesuv} and \eqref{boundstatesuv}. First we note that $u_k(x)$ has only even Fourier components if $k$ is positive and only odd Fourier components if $k$ is negative. For $v_k(x)$, its the other way around. Therefore, Eq.~\eqref{xi} implies that $\xi^{k+q+r k_0, k}_{\pm,m}$ will be non-zero only if $k$ and $k+q+r k_0$ have the same sign when $r$ and $m$ have the same parity, and opposite sign when $r$ and $m$ have opposite parity. For $\chi^{k+q+r k_0, k}_{\pm,m}$, the conditions are reversed. A corollary of this result is that $M^{\pm}_{nm} = 0$ unless $n$ and $m$ have the same parity, which validates our premise of choosing both $n$ and $m$ to be odd. When the terms in Eq.~\eqref{MpanmIseparatexichirescaled} are non-zero, their computation involves evaluating the sums
\begin{equation}
S_{\pm}(\alpha;z_1,z_2) \equiv \sum_{n \text{ even}} \frac{1}{\sinh(n \alpha + z_1) \sinh(n\alpha + z_2)} \pm \sum_{n \text{ odd}} \frac{1}{\sinh(n \alpha + z_1) \sinh(n\alpha + z_2)}\hspace{0.05cm},
\label{Splusminusdefintion}
\end{equation}
where $\alpha \in \mathbb{R}$ and $z_1, z_2 \in \mathbb{C}$. These sums can be calculated in terms of the $q$-digamma function $\psi_q(z)$ as \cite{saskey1978q}
\begin{align}
S_{+}(\alpha;z_1,z_2) &= \frac{4(z_2 - z_1) + \psi_{e^{2\alpha}} (\frac{z_1}{\alpha}) - \psi_{e^{2\alpha}} (\frac{z_2}{\alpha}) + \psi_{e^{2\alpha}} (1- \frac{z_2}{\alpha}) - \psi_{e^{2\alpha}} (1- \frac{z_1}{\alpha})}{\alpha \sinh(z_1 - z_2)}\hspace{0.05cm},
\label{Splus}\\
\text{and}\quad S_{-}(\alpha;z_1,z_2) &= 2 S_{+}(2\alpha;z_1,z_2) - S_{+}(\alpha;z_1,z_2)\hspace{0.05cm},
\label{Sminus}
\end{align}
with the understanding that for $z_1 = z_2$ one has to take the limit $z_1 \to z_2$ in Eq.~\eqref{Splus}. When the $\xi^* \xi$ and $\chi^* \chi$ terms in Eq.~\eqref{MpanmIseparatexichirescaled} are non-zero, they are both given by the expression
\begin{equation}
\frac{S_{\pm}\hspace{-0.05cm} \left(\alpha; \hspace{0.05cm} (m-r)\alpha \hspace{-0.05cm}+\hspace{-0.05cm} \frac{1}{2} a_{\tilde{k}+\tilde{q}+r} \hspace{-0.07cm}+\hspace{-0.03cm} \phi_{\tilde{k}+\tilde{q}+r}, \hspace{0.05cm}\frac{1}{2} a_{\tilde{k}} \hspace{-0.05cm}+\hspace{-0.05cm} \phi_{\tilde{k}}\right) S_{\pm}\hspace{-0.05cm} \left(\alpha; \hspace{0.05cm} (n-r)\alpha \hspace{-0.05cm}+\hspace{-0.05cm} \frac{1}{2} a_{\tilde{k}+\tilde{q}+r} \hspace{-0.1cm}+\hspace{-0.05cm} \phi_{\tilde{k}+\tilde{q}+r}, \frac{1}{2} a_{\tilde{k}} + \phi_{\tilde{k}}\right)}{16 \left(\tilde{\epsilon}_g^2 - \tilde{\epsilon}_{\tilde{k}+\tilde{q}+r}^2\right) \left(\tilde{\epsilon}_g^2 - \tilde{\epsilon}_{\tilde{k}}^2\right)}\in\mathbb{R},\hspace{-0.1cm}
\label{beta}
\end{equation}
where $\phi_{\tilde{k}} = (i\pi/2) \hspace{0.05cm}\Theta(1/2-|\tilde{k}|)$, and $\tilde{\epsilon}_g$, $\tilde{\epsilon}_{\tilde{k}}$, $a_{\tilde{k}}$ are given in Eqs.~\eqref{DOS}, \eqref{freestatesparam} and \eqref{boundstatesparam}. From Eqs.~\eqref{MpanmIseparatexichirescaled} and \eqref{beta} we see that $M^{\pm}_{nm} = M^{\pm}_{mn}$, and $M^{\pm}(q,\Omega^*) = (M^{\pm}(q,\Omega))^*$. For computation purposes, it is most convenient to convert the integrals in Eq.~\eqref{MpanmIseparatexichirescaled} into integrals over the spectral parameter $a_{\tilde{k}}$, using the relation $d\tilde{k} = \big(\tilde{\epsilon}_g^2 - \tilde{\epsilon}_{\tilde{k}}^2\big) \hspace{0.05cm} d a_{\tilde{k}}$.

The expression in Eq.~\eqref{beta} can be further simplified using properties of the $q$-digamma function, so that the complexity of computing the matrix grows as $\mathcal{O}(n_{\text{max}})$, where $n,m$ take on values from $-n_{\text{max}}$ to $n_{\text{max}}$ in steps of 2. To speed up computation, we also truncate the infinite sums over $r$ in Eq.~\eqref{MpanmIseparatexichirescaled} to finite sums from $-r_{\text{max}}$ to $r_{\text{max}}$. We find that the sums converge rapidly for $r_{\text{max}} \gtrsim n_{\text{max}}$. To evaluate $M^{\pm} (q,\Omega)$ for real $\Omega$, we include a small imaginary part to avoid branch cut singularities arising from the particle-hole continua.

\section{\label{susceptibility}Relation between collective modes and pairing susceptibility}
Here we find a simple relation between the matrix derived in Sec.~\ref{matrixequations} and the pairing susceptibility. To calculate the susceptibility, we find the linear response to a small external pairing field, $\delta \hat{H} \hspace{-0.05cm}=\hspace{-0.05cm} \int\hspace{-0.05cm} dx \hspace{0.05cm} f(x,t) \hat{\Psi}_{\uparrow}^{\dagger}(x,t) \hat{\Psi}_{\downarrow}^{\dagger}(x,t)$ + H.c. In the time-dependent BdG dynamics (see Sec.~\ref{integralequation}), this pairing field changes $\Delta(x,t)$ to $\Delta(x,t) + f(x,t)$ in Eq.~\eqref{tBdG1} and $\delta\Delta(x,t)$ to $\delta\Delta(x,t) + f(x,t)$ in Eq.~\eqref{linearizedtBdG1}. We assume the drive to be at frequency $\omega\in\mathbb{R}$, i.e.,
\begin{equation}
f(x,t) = f_+(x) \hspace{0.05cm} e^{i \omega t} + f_-(x)\hspace{0.05cm} e^{-i\omega t} \equiv \text{Re} (f_a (x) \hspace{0.05cm}e^{i \omega t}) + i \hspace{0.05cm}\text{Im} (f_p(x) \hspace{0.05cm}e^{i \omega t})\hspace{0.05cm}.
\label{drive}
\end{equation}
In steady-state the order-parameter fluctuations oscillate at the same frequency $\omega$. Hence we set $\eta = 0$ or $\Omega = \omega$ in the remaining steps in Sec.~\ref{integralequation}. The inclusion of the driving terms changes $\delta\Delta_{\pm}(x)$ to $\delta\Delta_{\pm}(x) + f_{\pm}(x)$ in Eqs.~\eqref{linearizedfrequencycomponents1}, \eqref{cjjprime1}, and \eqref{cjjprime2}, and $\delta\Delta_{\pm}(x^{\prime})$ to $\delta\Delta_{\pm}(x^{\prime}) + f_{\pm}(x^{\prime})$ in the right-hand sides of Eqs.~\eqref{integraleqndeltaDeltaplusminus}, \eqref{integraleqndeltaDeltaplus}, and \eqref{integraleqndeltaDeltaminus}. The phase and amplitude modes still decouple. However, they are now governed by an inhomogeneous integral equation,
\begin{equation}
\delta_{p,a}(x) = -g_{\mbox{\tiny{1D}}} \int d x^{\prime} \mathcal{M}^{\pm} (x,x^{\prime};\omega) \hspace{0.05cm} \big(\delta_{p,a}(x^{\prime}) + f_{p,a}(x^{\prime})\big)\hspace{0.05cm}.
\label{integraleqnwithdrive}
\end{equation}
Expanding $\delta_{p,a}(x)$ and $f_{p,a} (x)$ into Fourier components as $\delta_{p,a}(x) = e^{i q x} \sum_{n} C^{\pm}_{n} e^{i n k_0 x}$ and $f_{p,a}(x) = e^{i q x} \sum_{n} F^{\pm}_{n} e^{i n k_0 x}$, and substituting in Eq.~\eqref{integraleqnwithdrive} yield
\begin{equation}
C^{\pm}_{n} = - g_{\mbox{\tiny{1D}}} \sum\nolimits_m M^{\pm}_{nm} (q,\omega) \hspace{0.05cm} \big(C^{\pm}_m + F^{\pm}_m\big)\hspace{0.05cm},
\label{matrixeqnwithdrive}
\end{equation}
where the matrices $M^{\pm} (q,\omega)$ were studied in detail in Sec.~\ref{matrixequations}. From Eq.~\eqref{matrixeqnwithdrive} we can write $C^{\pm}_{n} = \sum_m X^{\pm}_{nm} (q,\omega) \hspace{0.05cm} C^{\pm}_m$, where $X^{\pm} (q,\omega)$ represent the susceptibility matrices, given by
\begin{equation}
X^{\pm}(q,\omega) = - g_{\mbox{\tiny{1D}}}\hspace{0.05cm}(I+g_{\mbox{\tiny{1D}}} M^{\pm}(q,\omega))^{-1} M^{\pm}(q,\omega)\hspace{0.05cm}.
\label{susceptibilitymatrix}
\end{equation}
We define scalar susceptibilities $\chi^{\pm} (q,\omega) \equiv \text{Tr } X^{\pm}(q,\omega)$. In terms of the eigenvalues $\lambda_j^{\pm}$ of the matrices $- g_{\mbox{\tiny{1D}}} M^{\pm}(q,\omega)$, we can write $\smash{\chi^{\pm} = \sum_j \lambda_j^{\pm}/(1-\lambda_j^{\pm})}$. From Eq.~\eqref{Cpamatrixeqn} we see that the collective modes with real $\Omega$ represent zeros of the eigenvalues of $I + g_{\mbox{\tiny{1D}}} M^{\pm}(q,\omega)$. Thus they show up as isolated poles of $\chi^{\pm} (q,\omega)$. The branch cuts of $M^{\pm}(q,\omega)$, which originate from particle-hole excitations, show up as broad diffuse spectra in $\text{Im}\hspace{0.05cm} \chi^{\pm} (q,\omega)$.

An alternative derivation of this susceptibility involves writing
\begin{align}
X =
\begin{pmatrix} \frac{\delta \Delta(x,t)}{\delta f(x^{\prime},t^{\prime})} & \frac{\delta \Delta(x,t)}{\delta f^*\hspace{-0.03cm}(x^{\prime},t^{\prime})}\vspace{0.1cm}  \\
\frac{\delta \Delta^{\hspace{-0.05cm *}} \hspace{-0.03cm}(x,t)}{\delta f(x^{\prime},t^{\prime})} & \frac{\delta \Delta^{\hspace{-0.05cm *}} \hspace{-0.03cm}(x,t)}{\delta f^*\hspace{-0.03cm}(x^{\prime},t^{\prime})} \end{pmatrix} = 
&-g_{\mbox{\tiny{1D}}}
\begin{pmatrix} \frac{\delta \langle \hat{\Psi}_{\downarrow} (x,t) \hat{\Psi}_{\uparrow} (x,t) \rangle}{\delta f(x^{\prime},t^{\prime})} & \frac{\delta \langle \hat{\Psi}_{\downarrow} (x,t) \hat{\Psi}_{\uparrow} (x,t) \rangle}{\delta f^*\hspace{-0.03cm}(x^{\prime},t^{\prime})}\vspace{0.1cm}  \\
\frac{\delta \langle \hat{\Psi}_{\uparrow}^{\dagger} (x,t) \hat{\Psi}_{\downarrow}^{\dagger} (x,t) \rangle}{\delta f(x^{\prime},t^{\prime})} & \frac{\delta \langle \hat{\Psi}_{\uparrow}^{\dagger} (x,t) \hat{\Psi}_{\downarrow}^{\dagger} (x,t) \rangle}{\delta f^*\hspace{-0.03cm}(x^{\prime},t^{\prime})} \end{pmatrix}_{\hspace{-0.1cm} 0} \\
&-g_{\mbox{\tiny{1D}}} \int\hspace{-0.1cm} d x^{\prime\prime} d t^{\prime\prime}
\begin{pmatrix} \frac{\delta \langle \hat{\Psi}_{\downarrow} (x,t) \hat{\Psi}_{\uparrow} (x,t) \rangle}{\delta \Delta(x^{\prime\prime},t^{\prime\prime})} & \frac{\delta \langle \hat{\Psi}_{\downarrow} (x,t) \hat{\Psi}_{\uparrow} (x,t) \rangle}{\delta \Delta^{\hspace{-0.05cm *}}\hspace{-0.03cm}(x^{\prime\prime},t^{\prime\prime})}\vspace{0.1cm}  \\
\frac{\delta \langle \hat{\Psi}_{\uparrow}^{\dagger} (x,t) \hat{\Psi}_{\downarrow}^{\dagger} (x,t) \rangle}{\delta \Delta(x^{\prime\prime},t^{\prime\prime})} & \frac{\delta \langle \hat{\Psi}_{\uparrow}^{\dagger} (x,t) \hat{\Psi}_{\downarrow}^{\dagger} (x,t) \rangle}{\delta \Delta^{\hspace{-0.05cm *}}\hspace{-0.03cm}(x^{\prime\prime},t^{\prime\prime})} \end{pmatrix}
\begin{pmatrix} \frac{\delta \Delta(x^{\prime\prime},t^{\prime\prime})}{\delta f(x^{\prime},t^{\prime})} & \frac{\delta \Delta(x^{\prime\prime},t^{\prime\prime})}{\delta f^*\hspace{-0.03cm}(x^{\prime},t^{\prime})}\vspace{0.1cm}  \\
\frac{\delta \Delta^{\hspace{-0.05cm *}} \hspace{-0.03cm}(x^{\prime\prime},t^{\prime\prime})}{\delta f(x^{\prime},t^{\prime})} & \frac{\delta \Delta^{\hspace{-0.05cm *}} \hspace{-0.03cm}(x^{\prime\prime},t^{\prime\prime})}{\delta f^*\hspace{-0.03cm}(x^{\prime},t^{\prime})} \end{pmatrix},
\label{susceptibility1}
\end{align}
where the first term on the right is the response neglecting self-consistency, and the second term gives the correction from self-consistency. We define
\begin{equation}
M \equiv \begin{pmatrix} \frac{\delta \langle \hat{\Psi}_{\downarrow} (x,t) \hat{\Psi}_{\uparrow} (x,t) \rangle}{\delta f(x^{\prime},t^{\prime})} & \frac{\delta \langle \hat{\Psi}_{\downarrow} (x,t) \hat{\Psi}_{\uparrow} (x,t) \rangle}{\delta f^*\hspace{-0.03cm}(x^{\prime},t^{\prime})}\vspace{0.1cm}  \\
\frac{\delta \langle \hat{\Psi}_{\uparrow}^{\dagger} (x,t) \hat{\Psi}_{\downarrow}^{\dagger} (x,t) \rangle}{\delta f(x^{\prime},t^{\prime})} & \frac{\delta \langle \hat{\Psi}_{\uparrow}^{\dagger} (x,t) \hat{\Psi}_{\downarrow}^{\dagger} (x,t) \rangle}{\delta f^*\hspace{-0.03cm}(x^{\prime},t^{\prime})} \end{pmatrix}_{\hspace{-0.1cm} 0} = 
\begin{pmatrix} \frac{\delta \langle \hat{\Psi}_{\downarrow} (x,t) \hat{\Psi}_{\uparrow} (x,t) \rangle}{\delta \Delta(x^{\prime},t^{\prime})} & \frac{\delta \langle \hat{\Psi}_{\downarrow} (x,t) \hat{\Psi}_{\uparrow} (x,t) \rangle}{\delta \Delta^{\hspace{-0.05cm *}}\hspace{-0.03cm}(x^{\prime},t^{\prime})}\vspace{0.1cm}  \\
\frac{\delta \langle \hat{\Psi}_{\uparrow}^{\dagger} (x,t) \hat{\Psi}_{\downarrow}^{\dagger} (x,t) \rangle}{\delta \Delta(x^{\prime},t^{\prime})} & \frac{\delta \langle \hat{\Psi}_{\uparrow}^{\dagger} (x,t) \hat{\Psi}_{\downarrow}^{\dagger} (x,t) \rangle}{\delta \Delta^{\hspace{-0.05cm *}}\hspace{-0.03cm}(x^{\prime},t^{\prime})} \end{pmatrix}.
\label{defineM}
\end{equation}
In the Fourier domain Eq.~\eqref{susceptibility1} has the same form as Eq.~\eqref{susceptibilitymatrix}. A straightforward application of linear response theory \cite{splischke2006equilibrium} confirms that in the Andreev approximation $\mathcal{M} = \begin{pmatrix} \mathcal{M}_1 & \mathcal{M}_2 \\ \mathcal{M}_2 & \mathcal{M}_1 \end{pmatrix}$ where $\mathcal{M}_1$ and $\mathcal{M}_2$ are defined in Eqs.~\eqref{M1} and \eqref{M2}.

\section{\label{figures}Collective mode spectra at different interactions and spin imbalance}
In this section we present additional figures showing the variation of the collective mode spectrum of a soliton train with spin imbalance and interaction strength.

\begin{figure}[t]
\vspace{-0.1cm}
\includegraphics[scale=1]{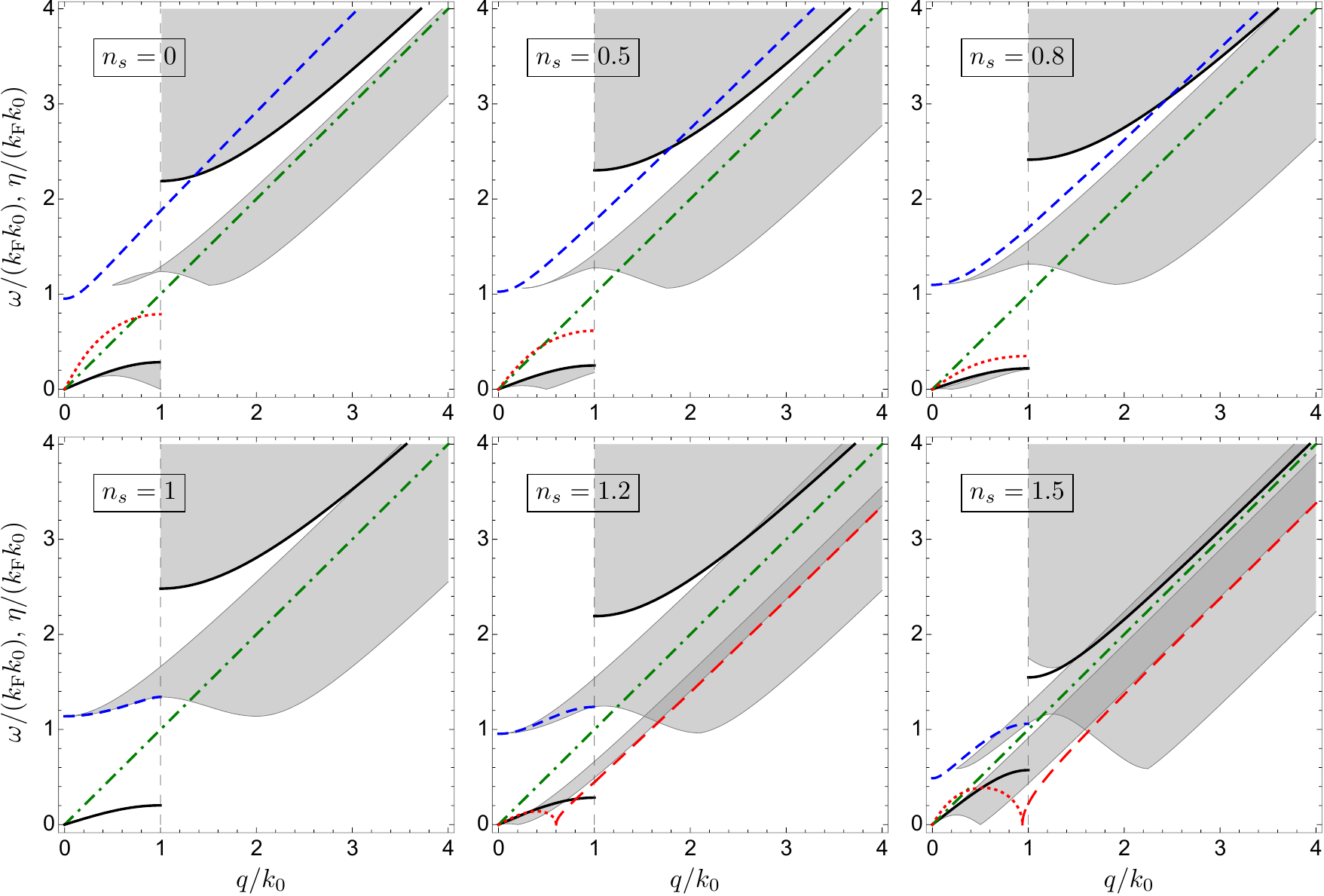}
\vspace{-0.5cm}
\caption{\label{spectravsimbalance}(Color online.) Collective excitation spectra of a fermionic soliton train at different spin imbalance in the extended zone representation, for $k_0/k_{\text{F}} = 0.05$ and $k_{\text{F}} a_{\mbox{\tiny{1D}}} = 2.65$. Gray regions show particle-hole continua, dot-dashed (green) curves show the Goldstone phase (phonon) mode, while dashed (blue) curves represent the doubly-degenerate `core' modes. The Goldstone amplitude (`elastic') mode with $q\leq k_0$, as well as the `Higgs' mode with $q\geq k_0$ are shown by solid (black) lines. The dotted (red) curves give the growth rate $\eta$ of the two-fold degenerate unstable modes. For $n_s>1$, the unstable modes are connected via a cusp to a new branch of stable modes, shown as wide-dashed (red) curves.}
\vspace{-0.3cm}
\end{figure}

Figure~\ref{spectravsimbalance} shows how the spectrum changes as the number of unpaired fermions per soliton ($n_s$) is varied. As $n_s$ is increased from 0, the instability rate falls toward zero, and vanishes for the C-FFLO phase with $n_s=1$. Since a change in $n_s$ comes about from altering the chemical potentials, it also affects the allowed particle-hole excitations, so the two-particle continua are modified. In the limit $n_s \to 1$, the degenerate `core' modes merge with the neighboring continuum, and the Goldstone amplitude mode becomes undamped. For $n_s \geq 1$, the `core' modes (dashed, blue lines in Fig.~\ref{spectravsimbalance}) are confined to wave-vectors $q\leq k_0$. This is different from the $n_s<1$ case, when the `core' modes extend to arbitrarily large $q$. As described in the main article, the amplitude and phase branches of the `core' modes are related as $\delta_a(x) \propto \delta_p^{\prime}(x)\;\forall\;q$. At $q=0$, they are described by $\delta_p(x) \propto \text{cn}(\Delta_1 x/k_{\text{F}}, k_1)$ and $\omega = \Delta_1 k_1$. For $n_s\geq 1$, the endpoint of the `core' spectrum at $q=k_0$ also has a simple form: there we find $\delta_p(x) \propto \text{dn}(\Delta_1 x/k_{\text{F}}, k_1)$ and $\omega = \Delta_1$ (Note that $\Delta_1 \equiv 2 k_{\text{F}} k_0 K(k_1)/\pi$). The state with $n_s>1$ is an analog of the unstable Sarma phase (see Sec.~\ref{energy}). It has zero-energy particle-hole excitations, and we again find dynamical instabilities. As before, there are two degenerate unstable modes which reduce to $\delta_p(x) \propto \Delta_0 (x)$ and $\delta_a(x) \propto \Delta_0^{\prime} (x)$ at $q=0$, shown as dotted red lines in Fig.~\ref{spectravsimbalance}. However, the maximum instability occurs at a wave-vector $q<k_0$, and the unstable modes are connected, via a cusp at $q=q_0 \leq k_0$, to a doubly degenerate stable mode which extends to large wave-vectors, shown as dashed red lines. As $n_s$ is increased, the maximum instability rate grows, and $q_0$ moves toward $k_0$. In addition, more of the low-energy `Higgs' modes become undamped. A soliton train solution exists as long as $n_s < n_s^{\text{max}} = (1+(4\hspace{0.03cm}\tilde{\epsilon}_c^2 - 1)\hspace{0.05cm} e^{-\pi k_{\text{F}} a_{\mbox{\tiny{1D}}}})^{1/2}$. In the limit $n_s \to n_s^{\text{max}}$, the soliton train reduces to a sinusoid of vanishingly small amplitude.

\begin{figure}[H]
\vspace{0.15cm}
\includegraphics[scale=1]{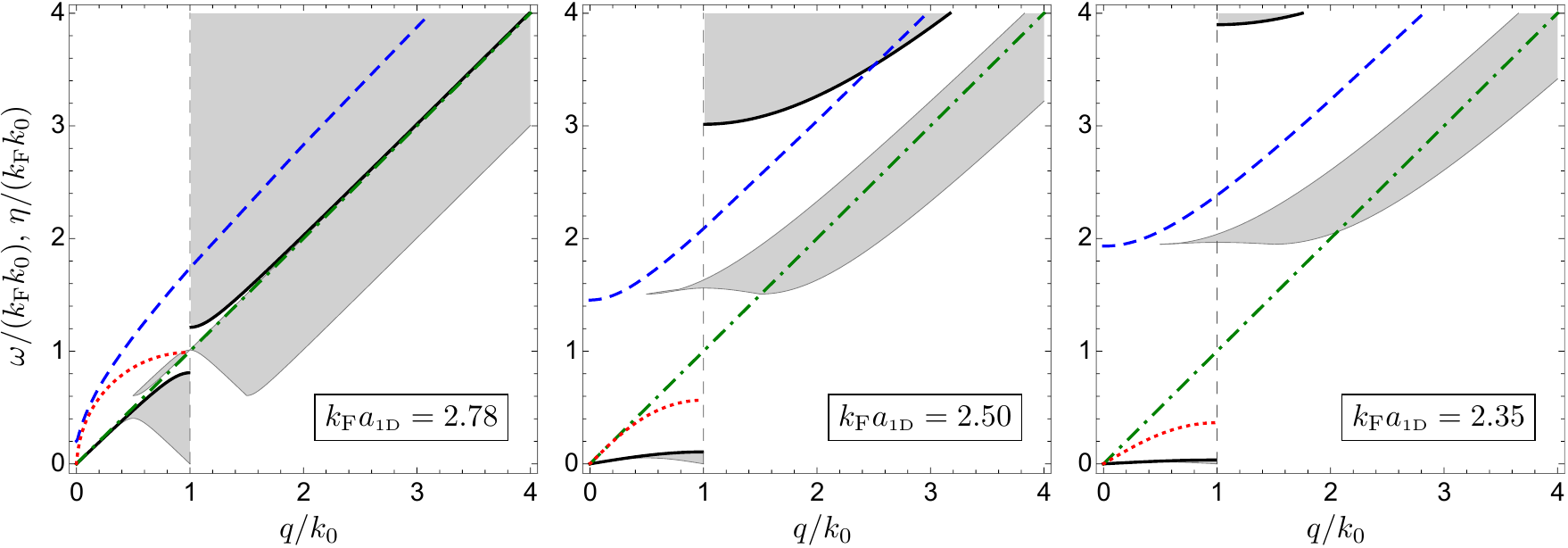}
\caption{\label{spectravsinteraction}(Color online.) Collective excitation spectra of a fermionic soliton train at different interaction strengths in the extended zone representation, for $k_0/k_{\text{F}} = 0.05$ and $n_s = 0$. Conventions for the curves/regions are the same as in Fig.~\ref{spectravsimbalance}.}
\end{figure}

Figure~\ref{spectravsinteraction} shows the variation in the spectrum with interaction strength. Stronger interactions (smaller $k_{\text{F}} a_{\mbox{\tiny{1D}}}$) increase the band-gap between the bound states and free states in the single-particle spectrum (see Fig.~1 of the main article), which leads to a larger separation of energy scales in the collective mode spectrum. In particular, the `Higgs' mode and the `core' modes move to higher frequencies, while the Goldstone amplitude mode moves to a lower frequency. Stronger interactions also produce sharp, well-separated solitons, thus reducing the instability rate. In the next section, we present simulations of the full time-dynamics of a two-soliton system without the Andreev apprximation, which show that the instabilities become unnoticeable for $k_{\text{F}} a_{\mbox{\tiny{1D}}} \lesssim 2$.

\section{\label{simulation} Simulation of the full dynamics of a two-soliton system}

\begin{figure}[b]
\includegraphics[scale=1]{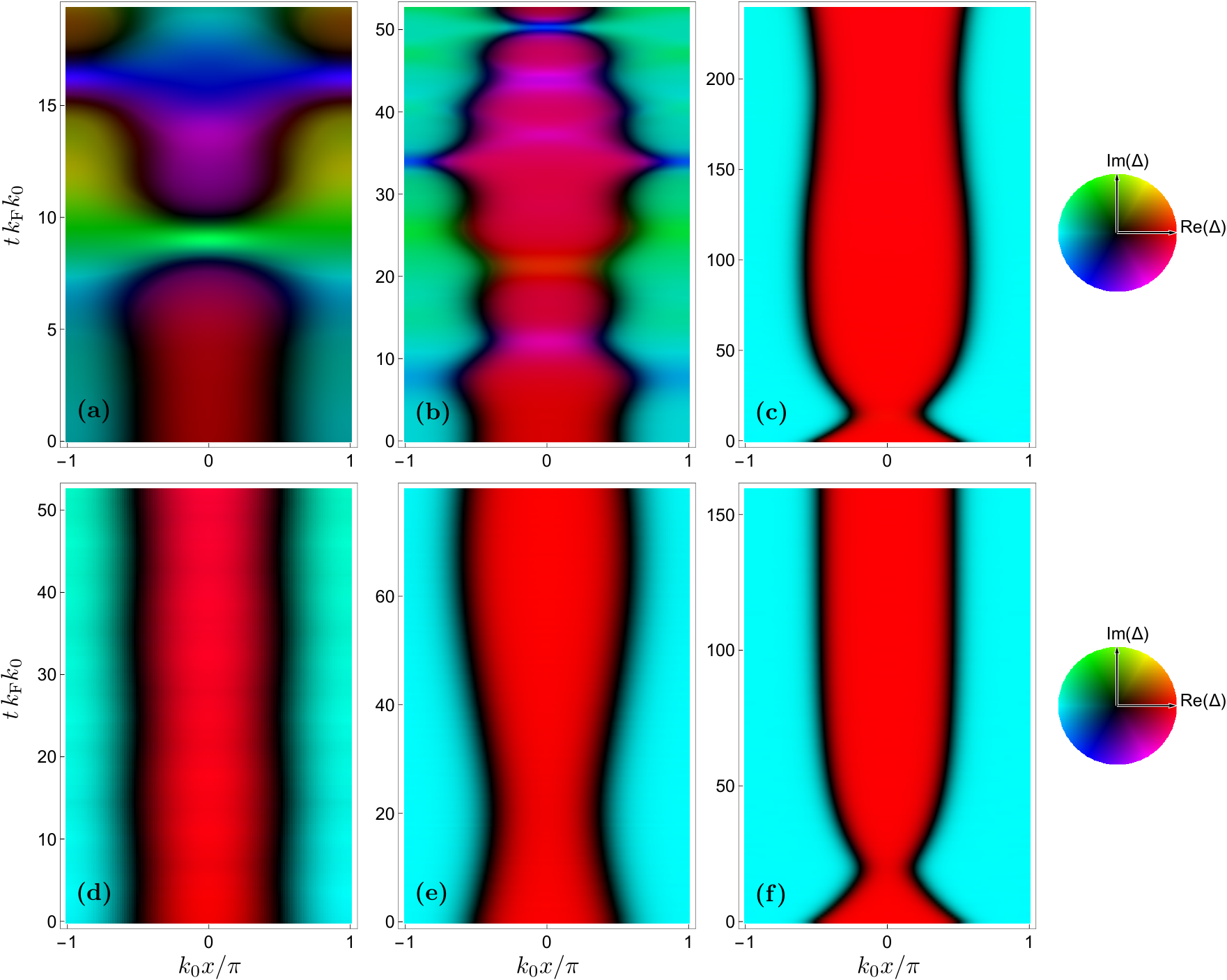}
\caption{\label{simulationarrayplots}(Color online.) Full dynamics of two solitons with periodic boundary conditions for $k_0/k_{\text{F}} = 0.05$. The absolute magnitude of $\Delta$ is proportional to the brightness, so dark bands represent the location of domain walls. The color wheels on the right show how the phase of $\Delta$ is encoded. Upper row: no spin imbalance, with (a) $k_{\text{F}} a_{\mbox{\tiny{1D}}} = 2.75$, (b) $k_{\text{F}} a_{\mbox{\tiny{1D}}} = 2.4$, and (c) $k_{\text{F}} a_{\mbox{\tiny{1D}}} = 2$. Lower row: with spin imbalance, and (d) $k_{\text{F}} a_{\mbox{\tiny{1D}}} = 2.75$, (e) $k_{\text{F}} a_{\mbox{\tiny{1D}}} = 2.4$, and (f) $k_{\text{F}} a_{\mbox{\tiny{1D}}} = 2$.}
\end{figure}

We simulate the full BdG dynamics [Eq.~\eqref{heisenbergeom}] of two solitons on a ring without the Andreev approximation, using a split-step algorithm. We start from the stationary solution $\Delta_0(x)$, with two domain walls located at $x=\pm \pi/(2 k_0)$, and apply a small perturbation $\smash{\delta \hat{H} \hspace{-0.05cm}=\hspace{-0.05cm} \int\hspace{-0.05cm} dx \hspace{0.05cm} f(x,t) \hat{\Psi}_{\uparrow}^{\dagger}(x,t) \hat{\Psi}_{\downarrow}^{\dagger}(x,t)}$ + H.c. where $f(x,t) = -(0.01/k_0) |\Delta_0^{\prime} (x)|\hspace{0.02cm} e^{-5 t k_{\text{F}} k_0}\Theta(t)$. The perturbation is short-lived compared to the dynamical timescales. However, it causes the two solitons to start moving toward each other. Figure~\ref{simulationarrayplots} shows the ensuing dynamics for different interaction strengths and spin imbalance. The dark bands represent the location of the domain walls (where $\Delta(x,t)=0$) as a function of time. For weak interactions and no spin imbalance, the system has dynamical instabilities: the solitons approach one another and annihilate, then repeatedly reform and annihilate. At stronger interactions, it takes longer for the solitons to merge, during which they exhibit a rich oscillatory motion. For very strong interactions ($k_{\text{F}} a_{\mbox{\tiny{1D}}} \lesssim 2$ for $k_0/k_{\text{F}} = 0.05$), we no longer see a merging of the solitons. Instead, they collide elastically off one another, and the domain walls undergo slow oscillations. This increase in stability with interaction strength is in agreement with the variation of the collective-mode spectrum of a soliton train in Fig.~\ref{spectravsinteraction}. Figures~\ref{simulationarrayplots}(d)-(f) show the dynamics in the presence of spin imbalance, such that the chemical potential difference $|h|$ is larger then the energy of the Andreev bound states, but smaller than the energy of all free quasiparticle states. This situation corresponds to the C-FFLO state of a soliton train, which we found earlier to be dynamically stable (see Fig.~\ref{spectravsimbalance}). Similarly we find here that the domain walls undergo stable oscillations. The oscillation timescale corresponds to the frequency of the `elastic' modes in Fig~\ref{spectravsimbalance}, which become slower at stronger interactions. The initial response to the perturbation is larger for stronger interactions.

\section{\label{GPE}Collective modes of a soliton train in the Gross-Pitaevskii equation}
Here we find the collective modes of a soliton train in a 1D Bose superfluid modeled by the Gross-Pitaevskii (GP) equation. We first set up the equations for a general stationary state, then consider the special case of a soliton train.

The GP equation describes the dynamics of the macroscopic wavefunction of the superfluid, $\psi(x,t)$, as
\begin{equation}
i \hspace{0.05cm}\partial_t\hspace{0.02cm} \psi(x,t) = \big(\hspace{-0.05cm}-\partial_x^2/2 - \mu + g\hspace{0.05cm} |\psi(x,t)|^2\big)\hspace{0.02cm} \psi(x,t)\hspace{0.05cm},
\label{GPE}
\end{equation}
where $\mu$ denotes the chemical potential, and $g$ is the coupling constant. Thus, a stationary state $\psi_0(x)\in\mathbb{R}$ satisfies
\begin{equation}
\big(\hspace{-0.05cm}-\partial_x^2/2 - \mu + g\hspace{0.05cm} (\psi_0(x))^2\big)\hspace{0.02cm} \psi_0(x) = 0\hspace{0.05cm}.
\label{GPEstationary}
\end{equation}
To find the collective modes, we linearize Eq.~\eqref{GPE} for small fluctuations $\delta \psi(x,t)$ about $\psi_0(x)$, obtaining
\begin{equation}
i \hspace{0.05cm}\partial_t\hspace{0.02cm} \delta\psi(x,t) = (-\partial_x^2/2 - \mu) \hspace{0.03cm}\delta\psi(x,t) + g \hspace{0.04cm}(\psi_0(x))^2 \hspace{0.03cm}(2\hspace{0.02cm}\delta\psi(x,t) + \delta\psi^* (x,t))\hspace{0.05cm}.
\label{linearizedGPE}
\end{equation}
Next we decompose $\delta\psi$ into amplitude and phase fluctuations, $\delta\psi(x,t) = \text{Re} \big(\delta \psi_a (x)\hspace{0.05cm} e^{i\Omega t}\big) + i\hspace{0.05cm} \text{Im} \big(\delta\psi_p (x) \hspace{0.05cm}e^{i\Omega t}\big)$ where $\delta \psi_a (x), \delta \psi_p (x), \Omega \in \mathbb{C}$. Comparing real and imaginary parts in Eq.~\eqref{linearizedGPE}, we find
\begin{align}
-\Omega\hspace{0.05cm} \delta\psi_p(x) &= \big(\hspace{-0.05cm}-\partial_x^2/2 - \mu + g\hspace{0.05cm} (\psi_0(x))^2\big)\hspace{0.02cm} \delta\psi_a(x)\hspace{0.05cm},
\label{collectivemodesGPE1}\\
\text{and}\quad -\Omega\hspace{0.05cm} \delta\psi_a(x) &= \big(\hspace{-0.05cm}-\partial_x^2/2 - \mu + 3 g\hspace{0.05cm} (\psi_0(x))^2\big)\hspace{0.02cm} \delta\psi_p(x)\hspace{0.05cm}.
\label{collectivemodesGPE2}
\end{align}
Equations~\eqref{collectivemodesGPE1} and \eqref{collectivemodesGPE2} describe the collective modes. Note that the amplitude and phase modes are coupled, unlike the fermionic case [Eq.~\eqref{integraleqndeltapa}]. We also see that $\Omega^2$ is an eigenvalue of a Hermitian operator. Thus $\Omega^2$ must be real, which means $\Omega$ is either real or imaginary. For a soliton train phase (as well as a uniform state), we find $\Omega$ is real for all collective modes, i.e., there is no dynamical instability.

A soliton train solution to Eq.~\eqref{GPEstationary} with period $2\pi/k_0$ exists for $0<\kappa<1$, where $\kappa \equiv k_0^2/(2\mu)$ is a measure of the ratio of kinetic energy to interaction energy. The soliton train profile is given by
\begin{equation}
\psi_0(x) = \sqrt{2\kappa n_0} \hspace{0.05cm}(2 k_1 K(k_1) /\pi) \hspace{0.1cm}\text{sn}\big(2 K(k_1)\hspace{0.02cm} k_0 x/\pi, k_1\big)\hspace{0.05cm},
\label{GPEpsi0x}
\end{equation}
where $n_0 \equiv \mu/g$, and $k_1$ is a sharpness parameter set by the equation $2\hspace{0.05cm}(1+k_1^2)^{1/2} K(k_1) = \pi/\sqrt{\kappa}$. For $\kappa \to 0$, $k_1 \to 1$, and $\psi_0(x)$ describes an array of sharp domain walls separating uniform regions with $\psi_0 = \pm \sqrt{n_0}$. Conversely, for $\kappa\to 1$, $k_1\to 0$, and $\psi_0(x)$ reduces to a sinusoid of vanishing amplitude. Note that $\psi_0(x)$ in Eq.~\eqref{GPEpsi0x} has the same spatial variation as a soliton train in a Fermi superfluid [Eq.~\eqref{Delta0x}].

To obtain the collective mode spectrum, we write $\delta\psi_{p,a}(x) = e^{i q x} \sum_n A^{p,a}_n e^{i n k_0 x}$ in Eqs.~\eqref{collectivemodesGPE1} and \eqref{collectivemodesGPE2}, and use
\begin{equation}
(\psi_0(x))^2 = 2 \kappa n_0 \Big[\left(2 K(k_1)/\pi\right)^2 \big(1-E(k_1)/K(k_1)\big) - 4\sum\nolimits_{n=1}^{\infty} n \cos(2 n k_0 x)\Big/\hspace{-0.05cm}\sinh\hspace{-0.05cm}\left(\hspace{-0.05cm}n \pi K\big((1-k_1^2)^{\frac{1}{2}}\big) \Big/ K(k_1)\right)\Big]
\label{fourierseries}
\end{equation}
to yield a matrix equation for the coefficients $A^{p,a}_n$. The eigenvalues of this matrix give the frequencies of collective oscillations. The collective modes are characterized by the parameter $\kappa$. Figure 2(c) in the main article shows the spectrum for $\kappa = 0.7$. The spectrum only contains two gapless Goldstone modes which arise from the spontaneous breaking of gauge and translational symmetry. In contrast, the fermionic soliton train has a much richer spectrum [Figs.~\ref{spectravsimbalance} and \ref{spectravsinteraction}]. In the limit $\kappa \to 0$ (sharp, well-separated solitons) and $q \to 0$, the mode with unbounded spectrum is described by $(\delta\psi_p (x),\delta\psi_a (x)) \propto (\psi_0(x),0)$ and $\Omega = \sqrt{\mu}\hspace{0.05cm}q$, whereas the mode with bounded spectrum is described by $(\delta\psi_p (x),\delta\psi_a (x)) \propto (0,\psi_0^{\prime}(x))$ and $\Omega \approx 0$. In other words, the two modes become pure phase (phonon) and pure amplitude (`elastic') modes respectively. In the opposite limit $\kappa \to 1$, the phase and amplitude oscillations are strongly mixed. For $q\to 0$, both modes are given by $(\delta\psi_p (x),\delta\psi_a (x)) \propto (\psi_0(x),i\psi_0^{\prime}(x))$ and $\Omega = \sqrt{2\mu}\hspace{0.05cm}q$.

\section{\label{KG}Collective modes of a soliton train in a nonlinear Klein-Gordon equation}
Here we study the collective modes of an `order parameter' $\phi(x,t)$ governed by the nonlinear Klein-Gordon equation
\begin{equation}
- (\partial_t^2/2)\hspace{0.02cm} \phi(x,t) = v^2\big(\hspace{-0.05cm}-\partial_x^2/2 - \mu + g\hspace{0.05cm} |\phi(x,t)|^2\big)\hspace{0.02cm} \phi(x,t)\hspace{0.05cm},
\label{KGequation}
\end{equation}
where $\mu$, $g$, and $v$ are phenamenological parameters playing the role of chemical potential, interaction strength, and speed of sound respectively. The motivation for studying such an equation is twofold. First, it presents a simple `fermionic analog' of the GP equation [Eq.~\eqref{GPE}]. The stationary solutions of the GP equation and Eq.~\eqref{KGequation} are identical. However, Eq.~\eqref{KGequation} is second-order in time, which causes phase and amplitude collective modes to decouple, as is the case in a Fermi superfluid [Eq.~\eqref{integraleqndeltapa}]. Further, Eq.~\eqref{KGequation} reproduces several features of the collective modes in a Fermi superfluid, in particular, analogs of Goldstone and `Higgs' modes, as well as soliton `core' modes and dynamical instabilities. This similarity suggests that the features could arise more generically in mesoscopic nonlinear systems, which might allow their broad characterization in terms of effective, coarse-grained models.

To study the collective modes, we linearize Eq.~\eqref{KGequation} about a stationary solution $\phi_0(x)\in\mathbb{R}$, yielding
\begin{equation}
\big(\partial_t^2/(2v^2)-\partial_x^2/2 - \mu\big) \hspace{0.03cm}\delta\phi(x,t) + g \hspace{0.04cm}(\phi_0(x))^2 \hspace{0.03cm}(2\hspace{0.02cm}\delta\phi(x,t) + \delta\phi^* (x,t)) = 0\hspace{0.05cm}.
\label{linearizedNKG}
\end{equation}
Substituting $\delta\phi(x,t) = \text{Re} \big(\delta \phi_a (x)\hspace{0.05cm} e^{i\Omega t}\big) + i\hspace{0.05cm} \text{Im} \big(\delta\phi_p (x) \hspace{0.05cm}e^{i\Omega t}\big)$ and comparing real and imaginary parts, we find
\begin{align}
(\Omega/v)^2\hspace{0.05cm} \delta\phi_p(x) &= 2\hspace{0.05cm}\big(\hspace{-0.05cm}-\partial_x^2/2 - \mu + g\hspace{0.05cm} (\phi_0(x))^2\big)\hspace{0.02cm} \delta\phi_p(x)\hspace{0.05cm},
\label{collectivemodesNKG1}\\
\text{and}\quad (\Omega/v)^2\hspace{0.05cm} \delta\phi_a(x) &= 2\hspace{0.05cm}\big(\hspace{-0.05cm}-\partial_x^2/2 - \mu + 3 g\hspace{0.05cm} (\phi_0(x))^2\big)\hspace{0.02cm} \delta\phi_a(x)\hspace{0.05cm}.
\label{collectivemodesNKG2}
\end{align}
Thus the amplitude and phase oscillations decouple, and $\Omega^2$ is given by an eigenvalue of a Hermitian operator. Thus, $\Omega$ must be either real or imaginary for any collective mode.

For a uniform stationary solution, $\phi_0(x) = (\mu/g)^{1\hspace{-0.02cm}/\hspace{-0.015cm}2}$, there is a Goldstone phase mode arising from the spontaneous breaking of gauge symmetry, described by $\delta \phi_p(x) \propto e^{i q x}$ and $\Omega = v \hspace{0.02cm} q$. It is similar to the Anderson-Bogoliubov phonon mode of a uniform Fermi superfluid. There is also an amplitude mode described by $\delta \phi_a (x) \propto e^{i q x}$ and $\Omega = v\hspace{0.03cm} (q^2 + 4 \mu)^{1\hspace{-0.02cm}/\hspace{-0.015cm}2}$, which is analogous to the `Higgs' mode in a uniform Fermi superfluid (see Sec.~\ref{goldstonehiggs}).

\begin{figure}[h]
\includegraphics[scale=1]{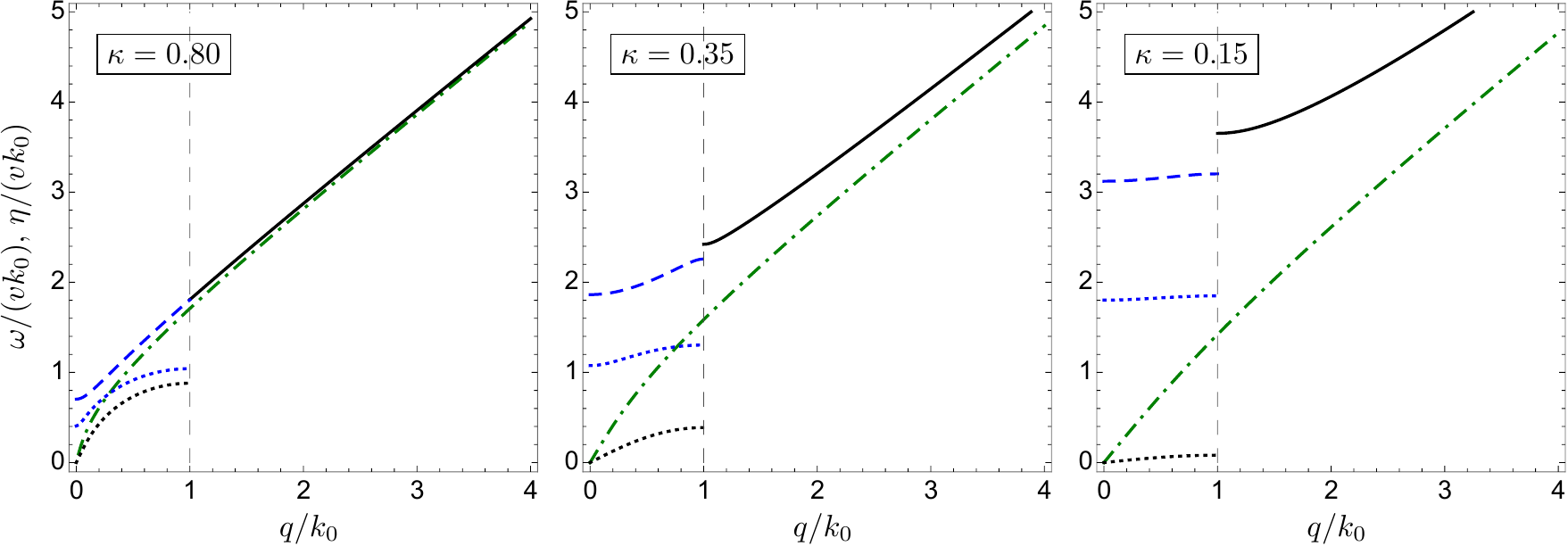}
\caption{\label{NKGspectra}(Color online.) Collective mode spectra of a soliton train described by a nonlinear Klein-Gordon equation [Eq.~\eqref{KGequation}] for different values of $\kappa \equiv k_0^2/(2\mu)$. There are three stable modes: a Goldstone phase mode, a Higgs-like amplitude mode, and a `core' mode describing width oscillations of solitons. These are represented by dot-dashed (green), solid (black), and dashed (blue) curves respectively. The dotted curves give the growth rate of two unstable modes: a gapless Goldstone amplitude mode (shown in black), and a `core' mode describing grayness oscillations of solitons (shown in blue).}
\end{figure}

A stationary soliton train has the same profile as in Eq.~\eqref{GPEpsi0x}, $\phi_0(x) = \sqrt{2 \kappa n_0}\hspace{0.05cm} k_1 \nu\hspace{0.08cm} \text{sn}(\nu k_0 x, k_1)$ where $\nu \equiv 2 K(k_1)/\pi$, $\kappa \equiv k_0^2/(2 \mu)$, and $n_0 \equiv (\mu/g)^{1\hspace{-0.02cm}/\hspace{-0.015cm}2}$. Its collective modes are characterized by $\kappa$. Figure~\ref{NKGspectra} shows the spectrum for different values of $\kappa$ in the extended zone scheme. Comparing with Figs.~\ref{spectravsimbalance} and \ref{spectravsinteraction}, we find a number of similarities, as well as some differences, with the spectrum in a Fermi superfluid. Like a fermionic soliton train, we find a Goldstone phase mode with an unbounded spectrum, which in the limit $q\to 0$, is given by $\delta \phi_p (x) \propto \phi_0(x)$. However, the Goldstone amplitude mode here is unstable with imaginary frequency. At $q=0$, the mode describes a uniform translation, $\delta \phi_a (x) \propto \phi_0^{\prime} (x)$, whereas at $q=k_0$, it describes a fluctuation $\delta\phi_a(x) \propto 1-\varepsilon_- \text{sn}^2 (\nu k_0 x, k_1)$ growing at a rate $\eta = \Gamma_-$, where $\varepsilon_{\pm} \equiv \big(1 \pm (1-3 \kappa^2 k_1^2 \nu^4)^{1\hspace{-0.02cm}/\hspace{-0.015cm}2}\big)/(\kappa\hspace{0.05cm} \nu^2)$ and $\Gamma_{\pm} \equiv v k_0 \big(1\pm 2\hspace{0.03cm} (1-3 \kappa^2 k_1^2 \nu^4)^{1\hspace{-0.02cm}/\hspace{-0.015cm}2}\hspace{0.02cm}\big)\hspace{-0.02cm}{}^{1\hspace{-0.02cm}/\hspace{-0.015cm}2}/\sqrt{\kappa}$. This instability is similar to the one we had found in the fermionic soliton train, where neighboring solitons approach and annihilate one another (see Fig.~4(f) in the main article). There we also had an instability where the entire soliton train moves off into the complex plane. Such an instability is only present here in the limit $\kappa \to 1$. In general, the maximum instability comes from a gapped unstable phase mode at $q=k_0$, where a fluctuation $\delta\phi_p(x) \propto \text{dn}(\nu k_0 x, k_1)$ grows at a rate $\eta_{\text{max}} = v k_0 (1/\kappa - k_1^2 \nu^2)^{1\hspace{-0.02cm}/\hspace{-0.015cm}2}$. In the limit $\kappa \to 1$, $\text{dn}(\nu k_0 x, k_1) \to 1$ and $\eta_{\text{max}}=v k_0$. At $q=0$, the unstable phase mode describes a fluctuation $\delta\phi_p(x) \propto \text{cn}(\nu k_0 x, k_1)$ growing at a rate $\eta_{\text{max}} = v k_0 (1/\kappa - \nu^2)^{1\hspace{-0.02cm}/\hspace{-0.015cm}2}$. This fluctuation, in fact, has the same functional form as the `core' mode which described grayness oscillations of each soliton in the fermionic soliton train (Fig.~4(e) in the main article). Thus, the phase branch of the `core' modes has turned into an unstable mode. In contrast, the amplitude branch, describing oscillations in the soliton widths, is still present, but only for $q\leq k_0$. At $q=0$, it is given by $\delta\phi_a(x)\propto \text{sn}(\nu k_0 x, k_1)\hspace{0.04cm}\text{dn}(\nu k_0 x, k_1)$ and $\Omega = (3 k_1 \nu)^{1\hspace{-0.02cm}/\hspace{-0.015cm}2}$, whereas at $q=k_0$, it is given by $\delta\phi_a(x)\propto \text{sn}(\nu k_0 x, k_1)\hspace{0.04cm}\text{cn}(\nu k_0 x, k_1)$ and $\Omega = (3 \nu)^{1\hspace{-0.02cm}/\hspace{-0.015cm}2}$. 
In addition, we have the analog of the `Higgs' mode for $q \geq k_0$. At $q=k_0$, it describes an amplitude oscillation similar to the one in a fermionic soliton train shown in Fig.~\ref{goldstonehiggsfig}(c), given by $\delta\phi_a(x) \propto 1-\varepsilon_+ \text{sn}^2 (\nu k_0 x, k_1)$ and $\Omega = \Gamma_+$. Since $\kappa$ measures the ratio of kinetic to interaction energy, decreasing $\kappa$ correspond to stronger interactions, or smaller values of $k_{\text{F}} a_{\mbox{\tiny{1D}}}$ in Fig.~\ref{spectravsinteraction}.

\end{document}